\documentclass[preprint]{elsarticle}

\usepackage{geometry}
\usepackage{tabulary}
\usepackage{fancyhdr}

\usepackage[latin1]{inputenc}
\usepackage{graphicx,epsfig,color}
\usepackage{wrapfig,rotating}
\usepackage{amssymb,amsmath,array}
\usepackage{latexsym, amsfonts, wasysym}
\usepackage{epstopdf}
\usepackage{lscape}
\usepackage[]{lineno}

\usepackage[above,section]{placeins}

\usepackage[english]{babel}
\usepackage[T1]{fontenc}
\usepackage{subfig}

\begin{document}
\newcommand{\ecm}{\mathrm{\sqrt{s}}}
\newcommand{\eplus}{\mathrm{e}^+}
\newcommand{\eminus}{\mathrm{e}^-}
\newcommand{\epem}{\eplus\eminus}
\newcommand{\lplm}{l^+l^-}
\newcommand{\mpmm}{\mu^+\mu^-}
\newcommand{\tptm}{\tau^+\tau^-}
\newcommand{\eeX}{\epem X}
\newcommand{\mmX}{\mpmm X}
\newcommand{\ZH}{\mathrm{ZH}}
\newcommand{\Zzero}{\mathrm{Z}}
\newcommand{\Zo}{\mathrm{Z^0}}
\newcommand{\Wboson}{\mathrm{W}}
\newcommand{\WpWm}{\Wboson^+\Wboson^-}
\newcommand{\Higgs}{\mathrm{H}}
\newcommand{\qq}{\mathrm{q}\overline{\mathrm{q}}}
\newcommand{\uubar}{\mathrm{u}\overline{\mathrm{u}}}
\newcommand{\ddbar}{\mathrm{d}\overline{\mathrm{d}}}
\newcommand{\ssbar}{\mathrm{s}\overline{\mathrm{s}}}
\newcommand{\GammaZ}{\Gamma_\Zzero}
\newcommand{\GammaW}{\Gamma_\Wboson}
\newcommand{\mZ}{\mathrm{M_\Zzero}}
\newcommand{\mW}{\mathrm{M_\Wboson}}
\newcommand{\mH}{\mathrm{M_\Higgs}}
\newcommand{\Mdl}{\mathrm{M_{dl}}}
\newcommand{\Mrecoil}{\mathrm{M_{recoil}}}
\newcommand{\Ptdl}{\mathrm{P_{Tdl}}}
\newcommand{\Pdl}{\mathbf{P_{dl}}}
\newcommand{\roots}{\mathrm{\sqrt{s}}}
\newcommand{\invfb}{\mathrm{fb^{-1}}}
\newcommand{\GeV}{\mathrm{GeV}}
\newcommand{\MeV}{\mathrm{MeV}}
\newcommand{\polRL}{\mathrm{e^-_R e^+_L}}
\newcommand{\polLR}{\mathrm{e^-_L e^+_R}}
\newcommand{\rmsn}{\mathrm{rms}_{90}}
\newcommand{\cthadl}{\mathrm{cos\theta_{dl}}}
\newcommand{\Ptbal}{\mathrm{\Delta P_{Tbal.}}}
\newcommand{\Ptgamma}{\mathrm{P_{T\gamma}}}
\newcommand{\Pt}{\mathrm{P_{T}}}
\newcommand{\Naddtks}{\mathrm{N_{add.TK}}}
\newcommand{\dthattk}{\mathrm{|\Delta \theta_{2tk}|}}
\newcommand{\dthamin}{\mathrm{|\Delta \theta_{min}|}}
\newcommand{\cthamiss}{\mathrm{|cos\theta_{missing}|}}

\hyphenation{brems-strah-lung}

\begin{frontmatter}

\title{Testing Hadronic Interaction Models using a Highly Granular Silicon-Tungsten Calorimeter}


\author{\centering The CALICE Collaboration}

\author[add1]{B.\,Bilki\fnref{fn1}}
\author[add1]{J.\,Repond}
\author[add1]{J.\,Schlereth}
\author[add1]{L.\,Xia}
\address[add1]{Argonne National Laboratory, 9700 S.\ Cass Avenue, Argonne, IL 60439-4815, USA}
\fntext[fn1]{Also at University of Iowa}

\author[add2]{Z.\,Deng}
\author[add2]{Y.\,Li}
\author[add2]{Y.\,Wang} 
\author[add2]{Q.\,Yue}  
\author[add2]{Z.\,Yang}
\address[add2]{Tsinghua University, Department of Engineering Physics, Beijing, 100084, P.R. China}

\author[add3]{G.\,Eigen}
\address[add3]{University of Bergen, Inst.\, of Physics, Allegaten 55, N-5007 Bergen, Norway}

\author[add4]{Y.\,Mikami} 
\author[add4]{T.\,Price}
\author[add4]{N.\,K.\,Watson}
\address[add4]{University of Birmingham, School of Physics and Astronomy, Edgbaston, Birmingham B15 2TT, UK}

\author[add5]{M.\,A.\,Thomson} 
\author[add5]{D.\,R.\,Ward}
\address[add5]{University of Cambridge, Cavendish Laboratory, J J Thomson Avenue, CB3 0HE, UK}

\author[add6]{D.\,Benchekroun} 
\author[add6]{A.\,Hoummada} 
\author[add6]{Y.\,Khoulaki}
\address[add6]{Universit\'{e} Hassan II A\"{\i}n Chock, Facult\'{e} des sciences.\, B.P. 5366 Maarif, Casablanca, Morocco}

\author[add7]{C.\,C\^{a}rloganu}
\address[add7]{Clermont Universit\'{e}, Universit\'{e} Blaise Pascal, CNRS/IN2P3, LPC, BP 10448, F-63000, Clermont-Ferrand, France}

\author[add8]{S.\,Chang}
\author[add8]{A.\,Khan}
\author[add8]{D.\,H.\,Kim}
\author[add8]{D.\,J.\,Kong}
\author[add8]{Y.\,D.\,Oh}
\address[add8]{Department of Physics, Kyungpook National University, Daegu, 702-701, Republic of Korea}

\author[add9]{G.\,C.\,Blazey} 
\author[add9]{A.\,Dyshkant} 
\author[add9]{K.\,Francis} 
\author[add9]{J.\,G.\,R.\,Lima} 
\author[add9]{P.\,Salcido} 
\author[add9]{V.\,Zutshi}
\address[add9]{NICADD, Northern  Illinois University, Department of Physics, DeKalb, IL 60115, USA,}

\author[add10]{V.\,Boisvert}  
\author[add10]{B.\,Green} 
\author[add10]{A.\,Misiejuk} 
\author[add10]{F.\,Salvatore\fnref{fn2}}
\address[add10]{Royal Holloway University of London, Dept. of Physics, Egham, Surrey TW20 0EX, UK}
\fntext[fn2]{Now at University of Sussex, Physics and Astronomy Department, Brighton, Sussex, BN1 9QH, UK}

\author[add11]{K.\,Kawagoe}
\author[add11]{Y.\,Miyazaki} 
\author[add11]{Y.\,Sudo} 
\author[add11]{T.\,Suehara} 
\author[add11]{T.\,Tomita} 
\author[add11]{H.\,Ueno} 
\author[add11]{T.\,Yoshioka}
\address[add11]{Department of Physics, Kyushu University, Fukuoka 812-8581, Japan}

\author[add12]{J.\,Apostolakis}
\author[add12]{G.\,Folger}
\author[add12]{V.\,Ivantchenko}
\author[add12]{A.\,Ribon}
\author[add12]{V.\,Uzhinskiy}
\address[add12]{CERN, 1211 Gen\`{e}ve 23, Switzerland}

\author[add13]{S.\,Cauwenbergh} 
\author[add13]{M.\,Tytgat} 
\author[add13]{N.\,Zaganidis}
\address[add13]{Ghent University, Department of Physics and Astronomy, Proeftuinstraat 86, B-9000 Gent, Belgium}

\author[add14]{J.\,-Y.\,Hostachy} 
\author[add14]{L.\,Morin}
\address[add14]{Laboratoire de Physique Subatomique et de Cosmologie - Universit\'{e} Grenoble-Alpes, CNRS/IN2P3, Grenoble, France}

\author[add15]{K.\,Gadow} 
\author[add15]{P.\,G\"{o}ttlicher} 
\author[add15]{C.\,G\"{u}nter} 
\author[add15]{K.\,Kr\"{u}ger} 
\author[add15]{B.\,Lutz} 
\author[add15]{M.\,Reinecke}  
\author[add15]{F.\,Sefkow}
\address[add15]{DESY, Notkestrasse 85, D-22603 Hamburg, Germany}

\author[add16]{N.\,Feege\fnref{fn3}}
\author[add16]{E.\,Garutti}
\author[add16]{S.\,Laurien}
\author[add16]{S.\, Lu}
\author[add16,add9]{I.\,Marchesini}
\author[add16]{M.\,Matysek}
\author[add16]{M.\,Ramilli}
\address[add16]{Univ. Hamburg, Physics Department, Institut f\"ur Experimentalphysik,
Luruper Chaussee 149, 22761 Hamburg, Germany}
\fntext[fn3]{Now at Stony Brook University, Dept. of Physics \& Astronomy, Stony Brook, NY 11794-3800, USA}

\author[add17]{A.\,Kaplan}
\address[add17]{University of Heidelberg, Fakult\"at f\"ur Physik und Astronomie, Albert Uberle Str. 3-5 2.OG Ost, D-69120 Heidelberg, Germany}

\author[add18]{E.\,Norbeck} 
\author[add18]{D.\,Northacker}
\author[add18]{Y.\,Onel}
\address[add18]{University of Iowa, Dept. of Physics and Astronomy, 203 Van Allen Hall, Iowa City, IA 52242-1479, USA}

\author[add19]{E.\,J.\,Kim}
\address[add19]{Chonbuk National University, Jeonju, 561-756, South Korea}

\author[add20]{B.\,van\,Doren}
\author[add20]{G.\,W.\,Wilson}
\address[add20]{University of Kansas, Department of Physics and Astronomy, Malott Hall, 1251 Wescoe Hall Drive, Lawrence, KS 66045-7582, USA}

\author[add21,add15,add16]{M.\,Wing}
\address[add21]{Department of Physics and Astronomy, University College London, Gower Street, London WC1E 6BT, UK}

\author[add22,add23]{B.\,Bobchenko}
\author[add22,add23]{M.\,Chadeeva}
\author[add22]{R.\,Chistov}
\author[add22,add23]{M.\,Danilov\fnref{fn4}}
\author[add22,add23]{A.\,Drutskoy}
\author[add22]{A.\,Epifantsev}
\author[add22]{O.\,Markin}
\author[add22,add23]{R.\,Mizuk}
\author[add22]{E.Novikov}
\author[add22]{V.Popov}
\author[add22,add23]{V.\,Rusinov}
\author[add22,add23]{E.\,Tarkovsky}
\address[add22]{Institute of Theoretical and Experimental Physics, B. Cheremushkinskaya ul. 25, RU-117218 Moscow, Russia}
\fntext[fn4]{Also at Moscow Institute of Physics and Technology (State University)}

\author[add23]{D.\,Besson} 
\author[add23]{E.\,Popova}
\address[add23]{National Research Nuclear University MEPhI (Moscow Engineering Physics Institute) 31, Kashirskoye shosse, 115409 Moscow, Russia}

\author[add24]{M.\,Gabriel} 
\author[add24]{C.\,Kiesling} 
\author[add24]{F.\,Simon} 
\author[add24]{C.\,Soldner} 
\author[add24]{M.\,Szalay} 
\author[add24]{M.\,Tesar} 
\author[add24]{L.\,Weuste}
\address[add24]{Max Planck Inst. f\"ur Physik, F\"ohringer Ring 6, D-80805 Munich, Germany}

\author[add25]{M.\,S.\,Amjad\fnref{fn5}} 
\author[add25]{J.\,Bonis} 
\author[add25]{S.\,Callier}
\author[add25]{S.\, Conforti di Lorenzo} 
\author[add25]{P.\,Cornebise} 
\author[add25]{Ph.\,Doublet\fnref{fn6}}
\author[add25]{F.\,Dulucq} 
\author[add25]{M.\,Faucci-Giannelli\fnref{fn7}}
\author[add25]{J.\,Fleury}
\author[add25]{T.\,Frisson}
\author[add25]{B.\,K\'egl} 
\author[add25]{N.\,van\,der\,Kolk\fnref{fn8}\corref{cor1}}
\author[add25]{H.\,Li\fnref{fn9}} 
\author[add25]{G.\,Martin-Chassard} 
\author[add25]{F.\,Richard} 
\author[add25]{Ch.\,de la Taille} 
\author[add25]{R.\,P\"oschl} 
\author[add25]{L.\,Raux} 
\author[add25]{J.\,Rou\"en\'e} 
\author[add25]{N.\,Seguin-Moreau} 
\address[add25]{Laboratoire de l'Acc\'el\'erateur Lin\'eaire (LAL), Centre Scientifique d'Orsay, Universit\'e Paris-Sud XI, BP 34, B\^atiment 200, F-91898 Orsay CEDEX, France}
\fntext[fn5]{Now at COMSATS Institute of Information Technology, Islamabad, Pakistan}
\fntext[fn6]{Now at IUT d'Orsay (Universit\'e Paris-Sud)}
\fntext[fn7]{Now at Royal Holloway, University of London}
\fntext[fn8]{Also at LLR, through excellence cluster P2IO. Now at Max Planck Inst. f\"ur Physik, Munich, Germany}
\fntext[fn9]{Now at University of Virginia, Charlottesville, USA}
\cortext[cor1]{Corresponding author}

\author[add26]{M.\,Anduze}
\author[add26]{V.\,Balagura}
\author[add26]{E.\,Becheva}
\author[add26]{V.\,Boudry}
\author[add26]{J-C.\,Brient}
\author[add26]{R.\,Cornat}
\author[add26]{M.\,Frotin}
\author[add26]{F.\,Gastaldi}
\author[add26]{F.\,Magniette}
\author[add26]{A.\,Matthieu}
\author[add26]{P.\,Mora de Freitas}
\author[add26]{H.\,Videau}
\address[add26]{Laboratoire Leprince-Ringuet (LLR)  -- \'{E}cole Polytechnique, CNRS/IN2P3, Palaiseau, F-91128 France}

\author[add27]{J-E.\,Augustin} 
\author[add27]{J.\,David} 
\author[add27]{P.\,Ghislain} 
\author[add27]{D.\,Lacour} 
\author[add27]{L.\,Lavergne}
\address[add27]{Laboratoire de Physique Nucl\'eaire et de Hautes Energies (LPNHE), UPMC, UPD, CNRS/IN2P3, 4 Place Jussieu, 75005 Paris, France}

\author[add28]{J.\,Zacek} 
\address[add28]{Charles University, Institute of Particle \& Nuclear Physics, V Holesovickach 2, CZ-18000 Prague 8, Czech Republic}

\author[add29]{J.\,Cvach} 
\author[add29]{P.\,Gallus} 
\author[add29]{M.\,Havranek}  
\author[add29]{M.\,Janata}  
\author[add29]{J.\,Kvasnicka}  
\author[add29]{D.\,Lednicky}  
\author[add29]{M.\,Marcisovsky}   
\author[add29]{I.\,Polak}  
\author[add29]{J.\,Popule}  
\author[add29]{L.\,Tomasek}  
\author[add29]{M.\,Tomasek}  
\author[add29]{P.\,Ruzicka}  
\author[add29]{P.\,Sicho}  
\author[add29]{J.\,Smolik}  
\author[add29]{V.\,Vrba}  
\author[add29]{J.\,Zalesak}  
\address[add29]{Institute of Physics, Academy of Sciences of the Czech Republic, Na Slovance 2, CZ-18221 Prague 8, Czech Republic}

\author[add30]{D.\,Jeans}
\address[add30]{Department of Physics, Graduate School of Science, The University of Tokyo, 7-3-1 Hongo, Bunkyo-ku, Tokyo 113-0033, Japan}

\author[add31]{M.\,G\"otze}  
\address[add31]{Bergische Universit\"{a}t Wuppertal, Fachbereich 8 Physik, Gaussstrasse 20, D-42097 Wuppertal, Germany}

\begin{abstract}
A detailed study of hadronic interactions is presented using data recorded with the highly granular CALICE silicon-tungsten electromagnetic calorimeter.
Approximately 350,000 selected $\pi^-$ events at energies between 2 and 10 GeV have been studied.
The predictions of several physics models available within the {\sc Geant4} simulation tool kit are compared to this data.
A reasonable overall description of the data is observed; the Monte Carlo predictions are within 20\% of the data, and for many observables much closer.
The largest quantitative discrepancies are found in the longitudinal and transverse distributions of reconstructed energy.
\end{abstract}


\end{frontmatter}

\newpage

\tableofcontents

\newpage


\section{Introduction}\label{section:introduction}

The primary physics goals at a future high energy lepton collider require the precise measurement of the energy of hadronic jets~\cite{2013_TDR2}.
Particle flow algorithms ({\em PFA}) foreseen at future linear electron-positron colliders~\cite{2001_Brient, 2013_TDR4, 2012_CLIC} 
result in a jet energy resolution of 3--4\% for jets with an energy in the range from 40\,GeV to 400\,GeV~\cite{2009_Thomson}.

The PFA approach aims to reconstruct individually all particles in the final state of the $\epem$ collision. 
This requires highly segmented calorimeters to disentangle the contributions from showers created by different types of particles within a jet, i.e.\,from charged and neutral particles.
The CALICE collaboration\footnote{CALICE Collaboration web page: http://twiki.cern.ch/CALICE} designs, constructs and operates prototypes of calorimeters dedicated to the application of PFAs.

To develop realistic PFAs, the interactions of hadrons must be modelled reliably in Monte Carlo simulations 
and the detector response to hadrons must be well-understood. 
In view of this, highly granular calorimeter prototypes provide a unique means to test and to further develop models of hadronic cascades. 

In this paper, the response of a highly granular silicon-tungsten electromagnetic calorimeter prototype (Si-W ECAL)~\cite{2008_CALICE} is used to test hadronic shower models at low energies. 
The depth of the Si-W ECAL corresponds to approximately one interaction length ($\lambda_\mathrm{I}$), which means that, although the complete shower is not recorded, the first hadronic interaction can be studied in great detail because of the fine longitudinal and transversal sampling. 
The Si-W ECAL was operated in a test beam at Fermi National Accelerator Laboratory (FNAL) in 2008 with negatively charged pions ($\pi^-$) in the energy of range 2 -- 10\,GeV. 
The majority of charged pions and other hadrons within high energy jets have energies in this range and therefore it is of considerable interest to validate the performance of Monte Carlo simulations.
The high granularity of the Si-W ECAL permits a detailed measurement of hadronic interactions in terms of global observables describing both the longitudinal and transverse shower development.

This paper is organised as follows: the Si-W ECAL prototype is described in the following section, 
the data and Monte Carlo simulations, as well as the event selection criteria employed, are presented in Sect.~\ref{section:datasamples}. 
The algorithm used to identify interactions is described in Sect.~\ref{section:identifyinginteractingevents}. 
Results obtained using data taken by the prototype using a $\pi^-$ beam and comparisons with Monte Carlo are discussed in Sect.~\ref{section:results}. 
A summary, conclusions, and prospects for future studies are given in the last section.


\section{The Si-W ECAL prototype}\label{section:ecalprototype}

\begin{figure}[h!]
  {\centering  
    \includegraphics[height=8cm]{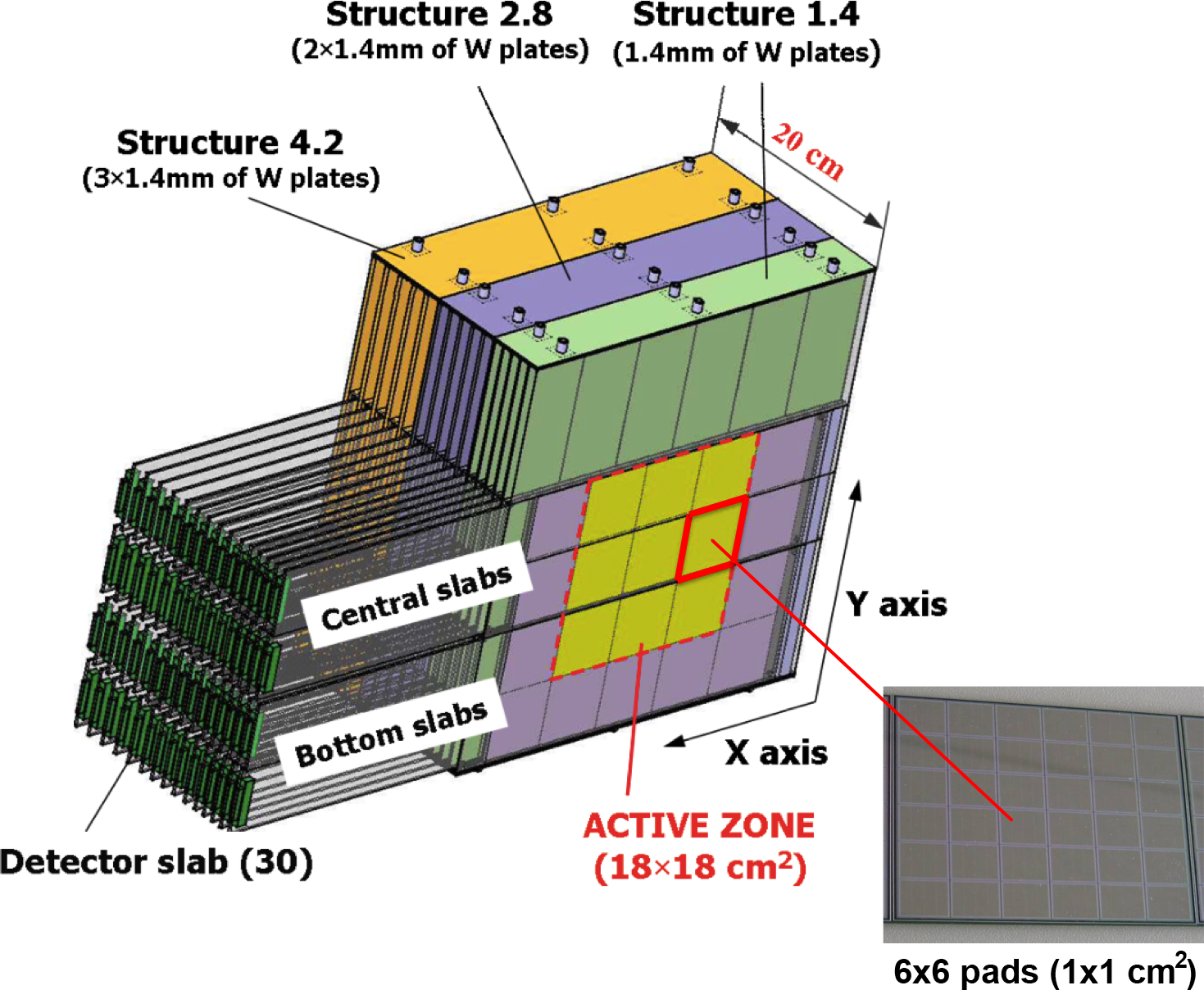}
    \caption{\sl Schematic view of the Si-W ECAL prototype} 
    \label{figure:ecalprototype}}
\end{figure}

The Si-W ECAL prototype consists of a sandwich structure of 30 layers of silicon as active material, alternating with tungsten as the absorber material.
The active layers are made of silicon wafers segmented into 1 $\times$ 1 cm$^{2}$ pixels (or pads). 
As shown in Fig.~\ref{figure:ecalprototype}, each wafer consists of a square of 6 $\times$ 6 pixels and each layer contains a 3 $\times$ 3 matrix of these wafers, 
resulting in an active zone of 18 $\times$ 18 cm$^2$.

The Si-W ECAL is divided into three modules of ten layers each. 
The tungsten thickness per layer is different in each module, increasing from 1.4 mm in the first module (layers 1--10), 
to 2.8 mm in the second (layers 11--20) and 4.2 mm in the third (layers 21--30). 
The total thickness corresponds to 24 radiation lengths (X$_{0}$) and approximately one interaction length. 
More than half of the hadrons traversing the Si-W ECAL prototype undergo a primary interaction within its volume. 


\section{Data samples}\label{section:datasamples}
	
Test beams were conducted in May and July of 2008 at the Fermilab Test Beam Facility\footnote{\label{note1}Fermilab Test Beam Facility web page: http://www-ppd.fnal.gov/MTBF-w} at FNAL.
The analysis presented in this paper uses data from runs with $\pi^{-}$ mesons at energies of 2, 4, 6, 8 and 10\,GeV. 
The Si-W ECAL was placed in front of two other CALICE prototypes: an analogue hadronic calorimeter (AHCAL)~\cite{2010_CALICE_2} and a Tail-Catcher and Muon Tracker (TCMT)~\cite{2012_CALICE}. 
Upstream of the Si-W ECAL the beam line was instrumented with two scintillator counters, covering an area of 10 $\times$ 10 cm$^{2}$, for triggering on incoming particles and two Cherenkov detectors for particle identification. 
The chosen coordinate system is right-handed with the z-axis pointing along the beam direction and the y-axis being vertical.

Monte Carlo simulations corresponding to the recorded test beam data have been produced using the simulation tool kit {\sc Geant4}~\cite{2003_Geant4}.
Version 9.6 patch 1 of {\sc Geant4} has been used as the default for this paper. 
The full geometry of the CALICE test beam set-up is taken into account in the simulation via the {\sc mokka} framework\footnote{Mokka web page: http://mokka.in2p3.fr} which provides the geometry interface to {\sc Geant4}. 
For a valid comparison of data and simulations realistic detector effects need to be present in the simulation.
Therefore a detailed digitization procedure is implemented that reproduces detector effects present in the data.
A detailed description of the detector simulation can be found in~\cite{2009_CALICE}.

\subsection{Simulation with various {\sc Geant4} physics lists}

Due to the complicated nature of hadronic interactions in material, it is difficult to achieve an accurate description of hadronic showers in simulations.
Several theory-driven and phenomenological hadronic interaction models are available~\cite{2010_Apostolakis} in {\sc Geant4}.
At higher energies theory-driven models are available, while for lower energies more approximate models are used.

At low energies,  where nucleons can be considered point-like in nature, two cascade models are implemented.
One is the Bertini cascade model, the other, the binary cascade model which is more theory based, is not relevant for this paper.
The Bertini cascade model simulates the initial interaction of the hadron with the nucleus, producing secondary particles which also collide with the nucleus in a so-called intra-nuclear cascade.
The particles are transported along straight lines through the nucleur medium and the interactions are modelled as free hadron-nucleon collisions.
The nuclear medium is approximated by several concentric shells of constant nucleon density.
In this process the nucleus is highly excited and evaporation models are included to de-excite the nucleus.

For medium to high energy hadronic interactions the theory-driven string parton models are implemented.
At these energies interactions between individual quarks in the projectile and the nucleons govern the scattering process.
There are two approaches, the Fritiof and the Quark-Gluon-String model.
In both approaches hadron-nucleus collisions are considered as a set of independent hadron-nucleon collisions.
In the Fritiof string model, diffractive scattering of the primary hadron with the nucleons is via momentum transfer alone, whereas in the Quark Gluon String model pomerons are exchanged.
An interaction results in several excited strings (and an excited nucleus) that are fragmented to produce secondary particles, which interact via a cascade model or a precompound model.
The fragmentation continues as long as the string energy is high enough for splitting.
The nucleus is de-excited by applying the precompound model and de-excitation models.
Additionally there are the Low Energy Parametrized (LEP) and High Energy Parametrized (HEP) models, which are based on fits to experimental data to predict the production of secondary particles.
Only the first hadron-nucleon collision is simulated in detail.
The remaining interactions within the nucleus are simulated by generating additional hadrons and assigning them as secondary particles from the initial interaction.
In these models energy is only conserved on average, not on an event-by-event basis.

These models are combined into {\it physics lists} within which they are applied in a specified energy range.
A number of reference physics lists are available with different tradeoffs between physics precision and speed.
Where two models are combined in a physics list, a smooth transition is achieved by randomly choosing the model on an event-by-event basis, with a probability that varies linearly with the energy in the interval.
The physics list {\sc qgsp\_bert}, for example, combines the Bertini model at low energies, $< 9.9$\,GeV, with the Low Energy Parametrized model at intermediate energies, 9.5 - 25\,GeV, and the Quark-Gluon-String Pre-compound model at high energies, $> 12$\,GeV.
Some models are only valid for certain hadrons, so within one physics list different models could be used for different hadrons.
The majority of the produced secondary particles in any hadron cascade are pions and thus the models used for pions dominate in general.

\begin{figure}[h!]
  {\centering 
    \includegraphics[width=0.8\textwidth]{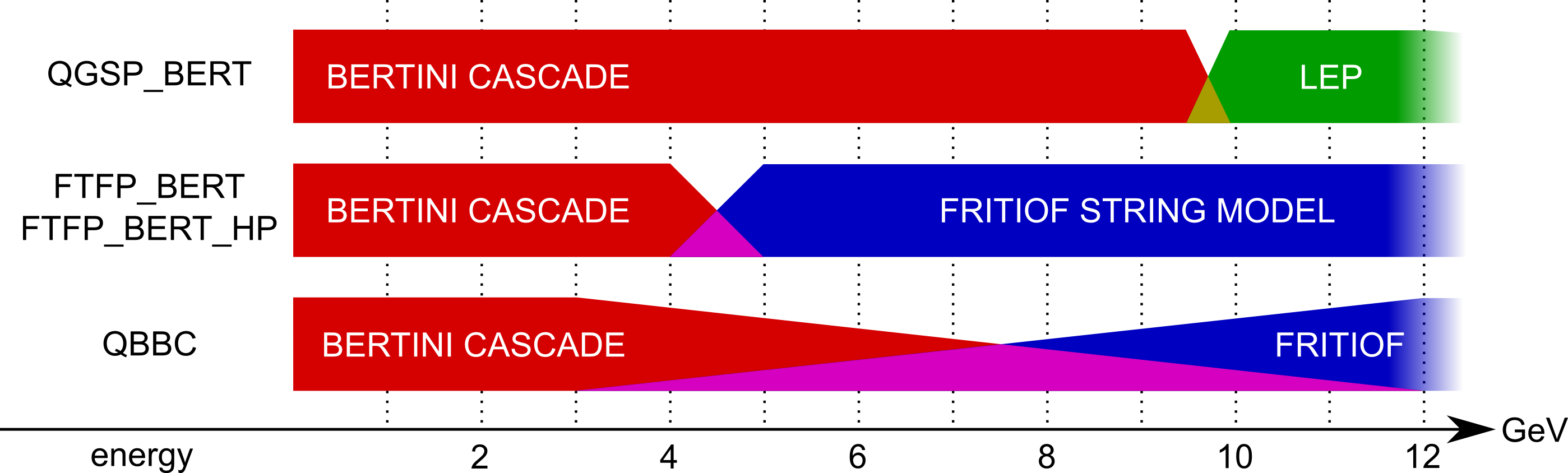} 
    \caption{\sl Model used for hadronic interactions of $\pi^-$ depending on the physics list and the energy of the interacting particle for the studied energy range.} 
    \label{figure:hadronicmodels}}
\end{figure}

In this paper four physics lists have been studied so as to be sensitive to differences between the hadronic interaction models and to the effect of the transitions between them.
The hadronic interaction models employed for pions by these physics lists in the studied energy range are illustrated in Fig.~\ref{figure:hadronicmodels}.
Electromagnetic processes for these physics lists all use the same, default underlying physics model.

The physics lists {\sc qgsp\_bert} and {\sc ftfp\_bert} allow the effect of the transition from the cascade to the string model to be studied, 
while {\sc qbbc} offers an alternative having a larger transition region between the two and by combining the Bertini and binary cascade models for neutrons and protons below 1.5\,GeV.
{\sc ftfp\_bert\_hp} is an extension of the {\sc ftfp\_bert} physics list which in addition employs a high precision treatment of neutrons with kinetic energies below 20\,MeV.
{\sc ftfp\_bert} is currently the recommended physics list for the simulation of LHC calorimeters~\cite{2011_Dotti} and is therefore used as the reference in this paper.

\subsection{Event selection}
\label{section:eventselection}

Data aquisition is triggered using the coincidence of the two scintillator counters upstream of the Si-W ECAL and $\pi^-$ mesons are identified with the help of two threshold Cherenkov counters.
The gas pressure in these counters is set such that for 2, 4 and 6\,GeV neither Cherenkov counters is triggered, while for 8 and 10 GeV only the first one is. 
The FNAL $\pi^{-}$ test beam is contaminated with $\mu^-$ and $e^-$, in particular at the lower energies where the beam is dominated by $e^-$.
At 2\,GeV the beam is estimated\footnote{\label{note1}Fermilab Test Beam Facility web page: http://www-ppd.fnal.gov/MTBF-w} to contain 5\% $\pi^-$ and 70\% $e^-$.
This contamination from $e^-$ is reduced significantly by the Cherenkov veto~\cite{2011_Feege}, however residual contamination remains.
The residual contamination is reduced by an additional event selection based on the position of the interaction of the incoming particle (see Sect.~\ref{section:identifyinginteractingevents}). 

Events are further selected as outlined below to guarantee a clean data set.
Data and simulation are subject to the same selection chain except where stated otherwise.
The {\sc ftfp\_bert} physics list is used as the default for background optimisation studies.

The response of the Si-W ECAL to charged particles has been calibrated with a $\mu^-$ beam~\cite{2008_CALICE, 2009_Li}.
Muons penetrate the whole detector volume with a (near) identical energy loss rate which is minimal for the beam energy used.
These muons are so-called minimum ionising particles (mip) and their mean energy loss in the active medium of a pad defines the energy unit MIP.
An energy threshold of 0.6 MIP on the reconstructed energy in an individual pixel (a hit) is applied to remove hits caused by detector noise.
Hits that are isolated (none of the 26 nearest-neighbour pixels in three dimensions contains a hit) are discarded in the analysis. 
This requirement removes 7 -- 10\% of the hits on average.

After this hit selection, events are selected that contain at least 25 hits.
This selection ensures that the incoming particle either passes through the Si-W ECAL as a mip or that it interacts inside the detector volume. 
Beam particles that enter the detector volume at an angle, due to multiple scattering in the material in the beam line, are in this way removed, as well as particles with a significant part of their trajectory in the inactive zones of the detector.
To avoid selecting events in which there may be significant lateral shower leakage, the lateral barycentres (energy weighted mean positions) $\bar{x}$ and $\bar{y}$ of the hits in an event are required to lie in the central part of the detector: $-50$\,mm < $\bar{x}$ < $50$\,mm and $-50$\,mm < $\bar{y}$ < $50$\,mm.
In addition events in the data in which instrumental noise (0.3\%) or spurious activity have been identified are excluded.

The contamination from $\mu^-$ in the data is reduced by a selection based on the number of hits in the TCMT ($N_{\mathrm{TCMT}}$). 
Based on the distribution of hits in a sample of simulated $\mu^-$ events, $\mu^-$s are identified as events where $N_{\mathrm{TCMT}} > 11$. 
At 2\,GeV, where the energy loss of $\mu^-$s in the HCAL is about 1.4\,GeV, the number of counts in the TCMT is reduced as the $\mu^-$s do not penetrate the full TCMT and the cut is changed to $N_{\mathrm{TCMT}} > 6$. 
The efficiency to reject $\mu^-$s is virtually 100\% for all studied energies. 
The loss of $\pi^-$ events due to the cut is 39\% at 2\,GeV and between 6\% and 10\% for 4 to 10\,GeV.
The efficiency to reject $\mu^-$s and the percentage of $\pi^-$ lost are based on samples of 500 k simulated $\mu^-$ and $\pi^-$ events.
Based on the fraction of events rejected by the muon selection in data, 
the FNAL $\pi^-$ beam is estimated to be contaminated with between 15\% of $\mu^-$ at 2\,GeV and 9\% at 10\,GeV.
The residual $\mu^-$ contamination in the data after the cuts are applied is negligible.
  
The $\pi^-$ beam is also contaminated with events in which two primary particles hit the Si-W ECAL simultaneously.
Events where a $\pi^-$ and $\mu^-$ are present are removed by the muon cut described above.
Events containing two $\pi^-$s are reduced by removing events in which two clusters of hits can be identified in the first eight layers of the Si-W ECAL.
Hits are clustered based on the distance (in three dimensions) between them and clusters are combined based on a cone algorithm.
Clusters containing at least 3 hits are accepted.
This selection can also reduce events where the $\pi^-$ has interacted upstream of the Si-W ECAL.
The efficiency of this selection to reduce multi-particle events has been estimated with the help of a sample of simulated $\pi^-$ events which were randomly overlaid with a second $\pi^-$ event. 
The efficiency is shown in Table~\ref{table:rejectionefficiencydoubleevents} together with the fraction of single $\pi^-$ events which are selected by this cut.

\begin{table}[h!]
  \caption{\sl Efficiencies to reject multi-particle events and to select single $\pi^-$ events based on the presence of two clusters of hits in the first eight layers of the Si-W ECAL for events which pass the selection described in the main text. The efficiency is estimated using Monte Carlo samples ({\sc ftfp\_bert}) in which $\pi^-$s were overlaid with other $\pi^-$s.}
  \begin{center}
    \begin{tabular}{| l | c | c | c | c | c |}
      \hline
      E (GeV) & 2 & 4 & 6 & 8 & 10 \\
      \hline
      \hline
      $\pi^-$ + $\pi^-$ event rejection efficiency & 0.76 & 0.79 & 0.80 & 0.78 & 0.77\\
      Single $\pi^-$ event selection efficiency & 0.87 & 0.85 & 0.84 & 0.84 & 0.84\\
      \hline
    \end{tabular}
  \end{center}
  \label{table:rejectionefficiencydoubleevents}
\end{table}

Events in which a $\pi^-$ and an $e^-$ are present are also rejected by this selection.
They are further reduced by rejecting events in which the incoming particle interacts in the beginning of the Si-W ECAL, a cut designed to reduce the fraction of $e^-$ events in the sample.
Details about this additional event selection are given in the next section.
The combination of these two cuts reduces the contamination due to events with a $\pi^-$ and an $e^-$ to a negligible level.

The estimated contamination of the FNAL data with double $\pi^-$ events is between 26\% at 2\,GeV and 5\% at 10\,GeV.
The residual contamination in the selected data sample is estimated based on the efficiencies found in the simulated samples and the number of events rejected in the data.
It is estimated to be between 8.8\% at 2\,GeV and 1.5\% at 10\,GeV.

The number of data events after the selection criteria are applied and the fraction of the total number of events that is selected are given in Table~\ref{table:numberofselectedevents}.
The size of the simulated event samples are of the order of 150 k events.
The applied selection cuts and the fraction of events that is sequentially removed are summarized in Table~\ref{table:cuts}.

\begin{table}[h!]
  \caption{\sl Number of events remaining after all selection criteria are applied to the FNAL $\pi^{-}$ data and the corresponding fraction of the original number of events.}
  \begin{center}
    \begin{tabular}{| l | c | c | c | c | c |}
      \hline
      E (GeV) & 2 & 4 & 6 & 8 & 10 \\
      \hline
      \hline
      Events & 8113 & 62431 & 40845 & 86934 & 154240 \\
      \hline
      Fraction & 0.12 & 0.29 & 0.39 & 0.37 & 0.40 \\ 
      \hline
    \end{tabular}
  \end{center}
  \label{table:numberofselectedevents}
\end{table}

\begin{table}[h!]
  \caption{\sl Summary of the applied selection criteria and the fraction of events removed by each criteria sequentially for 2 - 10\,GeV.}
  \begin{center}
    \begin{tabular}{| l | l |}
      \hline
      Selection criteria & Fraction \\
      \hline
      \hline
      {\bf Hit selection} & \\
      Energy threshold of 0.6 MIP & \\
      Isolated hits are discarded & \\
      \hline
      \hline
          {\bf Event selection} & \\
          Quality selection (Correct trigger, At least $25$ hits per event, & \\
          Event barycentres ($\bar{x}$, $\bar{y}$) are required in $-50 \,\mathrm{mm} \leq \bar{x} \leq 50 \,\mathrm{mm}$ & \\
          and $-50 \,\mathrm{mm} \leq \bar{y} \leq 50 \,\mathrm{mm}$,
          Events with instrumental noise & \\
          or spurious activity are rejected (data only)) & 0.64 - 0.29 \\
          $\mu^-$ rejection by requiring $N_{TCMT} > 11$  & 0.48 - 0.19 \\
          Double event rejection based on presence of two hit clusters & 0.29 - 0.19 \\
          $e^-$ rejection by requiring an interaction layer > 6 & 0.14 - 0.13 \\
      \hline
    \end{tabular}
  \end{center}
  \label{table:cuts}
\end{table}

\section{Identifying interacting events}\label{section:identifyinginteractingevents}

\begin{figure}[h!]
  {\centering 
    \includegraphics[width=0.6\textwidth]{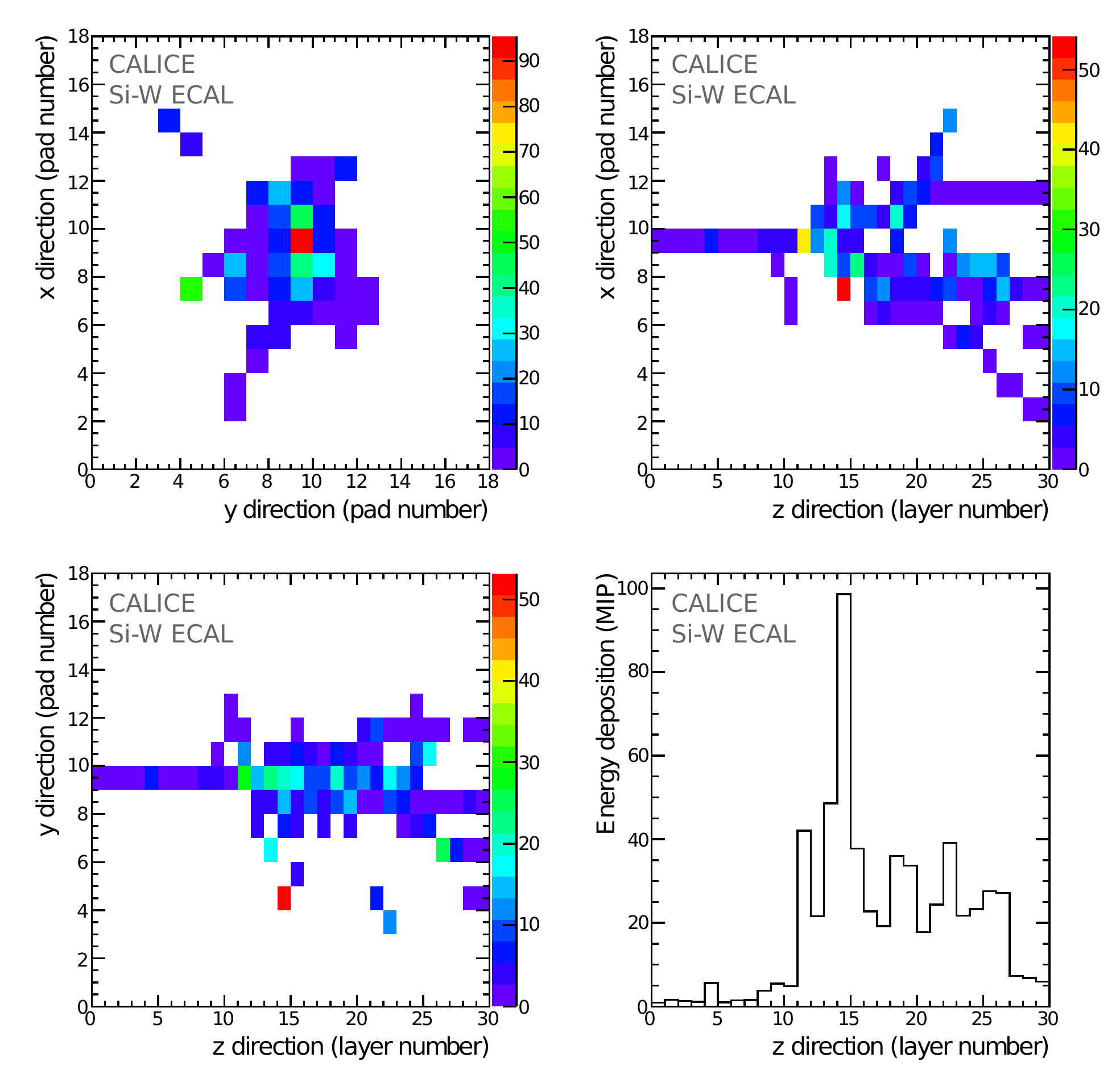} 
    \caption{\sl A hadronic interaction of a $\pi^-$ with an incident kinetic energy of 10\,GeV in the Si-W ECAL. 
      Top left: projection in the x-y plane of the reconstructed energy. 
      Top right: projection on the x-z plane of the reconstructed energy. 
      Bottom left: projection on the y-z plane of the reconstructed energy.
      Bottom right: the reconstructed energy in each layer of the Si-W ECAL.
      The energy unit is in MIPs.} 
    \label{figure:fireball}}
\end{figure}

A primary particle traversing the Si-W ECAL can either pass the detector material as an ionising particle or undergo interactions which lead to the creation of secondary particles.
In the latter case the ionising track in the first layers is followed by several secondary tracks after the interaction.
Figure~\ref{figure:fireball} shows a recorded event in which this can be seen. 
The bottom right histogram clearly illustrates that the reconstructed energy in consecutive layers increases significantly at the interaction point (here at layer 11).
This change in reconstructed energy can be used to identify the layer in which the interaction takes place.
Two criteria are applied: one based on the absolute energy increase, and one based on the relative energy increase~\cite{2012_Doublet}.

First a requirement is made on the reconstructed energy in each layer, $E_{i}$. 
If three consecutive layers have an energy higher than a threshold, $E_{\mathrm{cut}}$, the interaction layer is identified as the first of these (layer $i$).
This algorithm is not applicable for interactions occurring in the last two layers of the Si-W ECAL, and therefore has zero efficiency in this range. 
In this analysis the value of $E_{\mathrm{cut}}$ is chosen to be 8 MIP, which optimises for simulated 10 GeV events the interaction-finding efficiency and the standard deviation on the difference between the true and the reconstructed interaction layer.
The optimal value of $E_{\mathrm{cut}}$ varies by a maximum of one MIP between different Monte Carlo physics lists. 
 
This selection, based on absolute energy increase, is not efficient at lower beam energies, a particularly interesting region for hadronic modelling.  
Because at small hadron energies only a small number of low energy secondaries are produced, shower fluctuations are relatively strong making the interaction point less clearly defined. 
A second criterion based on the relative increase in reconstructed energy is applied to events without an interaction layer defined by the first criterion:
 
\begin{equation}
  \displaystyle \frac{E_{i}+E_{i+1}}{E_{i-1}+E_{i-2}} > F_{\mathrm{cut}}  \quad \text{and} \quad \frac{E_{i+1}+E_{i+2}}{E_{i-1}+E_{i-2}} > F_{\mathrm{cut}}~. 
  \label{equation:fcut}
\end{equation}

This measures a relative increase in energy before and after a given layer $i$. 
As two consecutive layers are grouped together the variables are less sensitive to local fluctuations in the reconstructed energy.
For a MIP-like energy deposit both fractions are around 1, while in case of a hadronic interaction they are larger.
The value of the threshold, $F_{\mathrm{cut}}$, for selecting interacting events, which minimises the contamination with non-interacting events is 6. 
This value is largely independent of energy and Monte Carlo physics list.
In cases where the relative increase continues over several layers one has to make sure that this increase is not an artefact caused by a backscattered particle that deposits energy several cells away from the incoming primary MIP track. 
To ensure that the increase is caused by the start of a hadronic interaction, the sum of the energies in the cell of the extrapolated primary MIP track (which is found by clustering hits in the first eight layers of the Si-W ECAL) and in the eight cells in the same layer ($i$) around it ($E_{\mathrm{around},i}$) should be at least half of the layer energy; $E_{\mathrm{around},i} > 0.5 E_{i}$.

\begin{figure}[h!]
  {\centering 
    \includegraphics[width=0.6\textwidth]{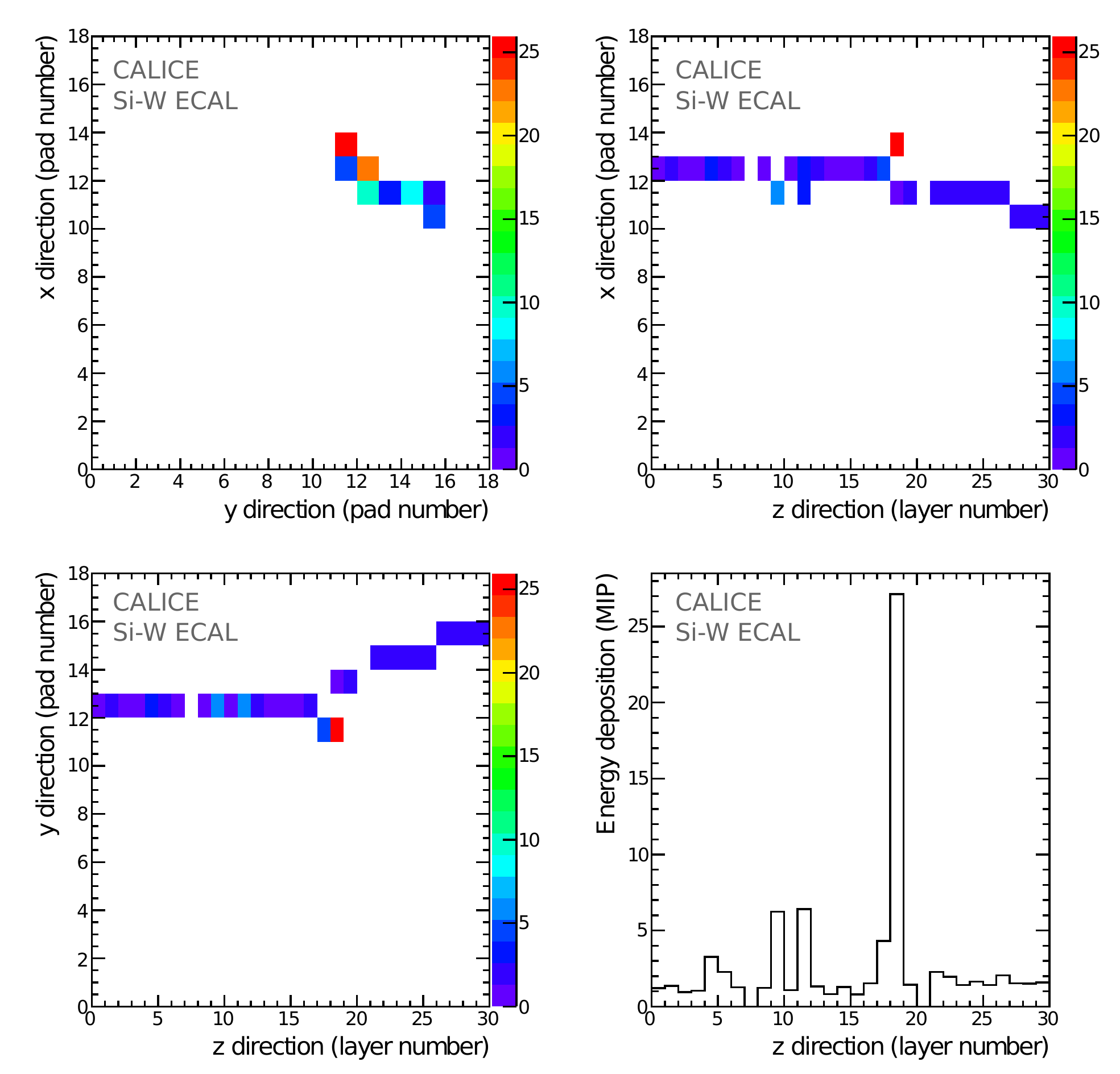} 
    \caption{\sl A hadronic interaction of a $\pi^-$ with an incident kinetic energy of 2\,GeV in the Si-W ECAL.
      Top left: projection in the x-y plane of the reconstructed energy. 
      Top right: projection on the x-z plane of the reconstructed energy. 
      Bottom left: projection on the y-z plane of the reconstructed energy.
      Bottom right: the reconstructed energy in each layer of the Si-W ECAL.
      The energy unit is in MIPs.} 
    \label{figure:pointlike}}
\end{figure}

The events with an interaction layer based on the second criterion show topologies with a small number of secondary particles.
An example is shown in Fig.~\ref{figure:pointlike}.
This event features a strong local increase in energy. 

As mentioned in Section~\ref{section:eventselection}, the contamination of the test beam data with $e^-$s is large even after the Cherenkov data selection is applied.
Therefor an additional event selection is applied based on the found interaction layer.
In simulated $e^-$ and $\pi^-$ events the rejection of events with an interaction found in the first six layers removes 84\% of $e^-$ events at 2\,GeV.
At 10\,GeV 98\% of $e^-$ events are removed.
The percentage of removed $\pi^-$ events is 20\% at all energies.
With this additional event selection the final contamination with $e^-$ is reduced from 15\% to 3\% at 2 GeV.
The contamination decreases quickly with energy and at 10\,GeV it is negligible. 
The estimate of the contamination is based on the rejection efficiency found in simulated events and the fraction of rejected events in the data when applying the selection cut.

The second selection criterion (Eq.~\ref{equation:fcut}) accepts a small fraction of delta rays.
This fraction is estimated to be between 2.2\% at 2 GeV and 3.2\% at 10 GeV. 
The estimate is based on the fraction of all events that are accepted as interacting by Eq.~\ref{equation:fcut} in a sample of 500\,k simulated $\mu^-$.
Because the mass of the $\mu^-$ and $\pi^-$ are very close, their behaviour in terms of electromagnetic interactions is very similar.

Table~\ref{table:fcutgain} shows the efficiency, $\eta$, to find an interaction inside the Si-W ECAL volume.
It is estimated from simulated data by comparing the found interaction layers to the Monte Carlo truth.
The efficiency is defined as the fraction of interacting events found by the algorithm described above, that are correctly classified as interacting according to the Monte Carlo truth. 
The efficiency increases with increasing energy. 
The efficiency found when only the absolute energy criterion is applied, $\eta_{Ecut}$, is lower than the total efficiency, $\eta$, where both criteria are applied, by 0.25 for 2 GeV and by 0.03 at 10 GeV.
Clearly at low beam energies the second criterion is needed.
The efficiency to identify the correct interaction layer with a maximum difference of one layer, $\eta_{\pm 1}$, and of two layers, $\eta_{\pm 2}$, with respect to the interaction layer given by the Monte Carlo truth are shown in the last two columns.

\begin{table}[h!]
  \caption{\sl The interaction-finding efficiency $\eta$, decomposed in the contribution of the absolute energy criteria only, $\eta_{Ecut}$, and the efficiency $\eta_{\pm 1(2)}$ to find interactions within $\pm$1(2) layer(s), measured in Monte Carlo events (\sc ftfp\_bert).}
  \begin{center}
    \begin{tabular}{| l | c | c || c | c |}
      \hline
      E (GeV) & $\eta$ & $\eta_{Ecut}$ & $\eta_{\pm 1}$ & $\eta_{\pm 2}$\\
      \hline
      \hline
      2 & 0.60 & 0.35 & 0.47 & 0.50\\
      4 & 0.81 & 0.67 & 0.67 & 0.69\\
      6 & 0.91 & 0.84 & 0.80 & 0.83\\
      8 & 0.92 & 0.88 & 0.82 & 0.85\\
      10 & 0.93 & 0.90 & 0.84 & 0.87\\
      \hline
    \end{tabular}
  \end{center}
  \label{table:fcutgain}
\end{table}

\begin{figure}[h]
  {\centering
    \includegraphics[width=0.6\textwidth]{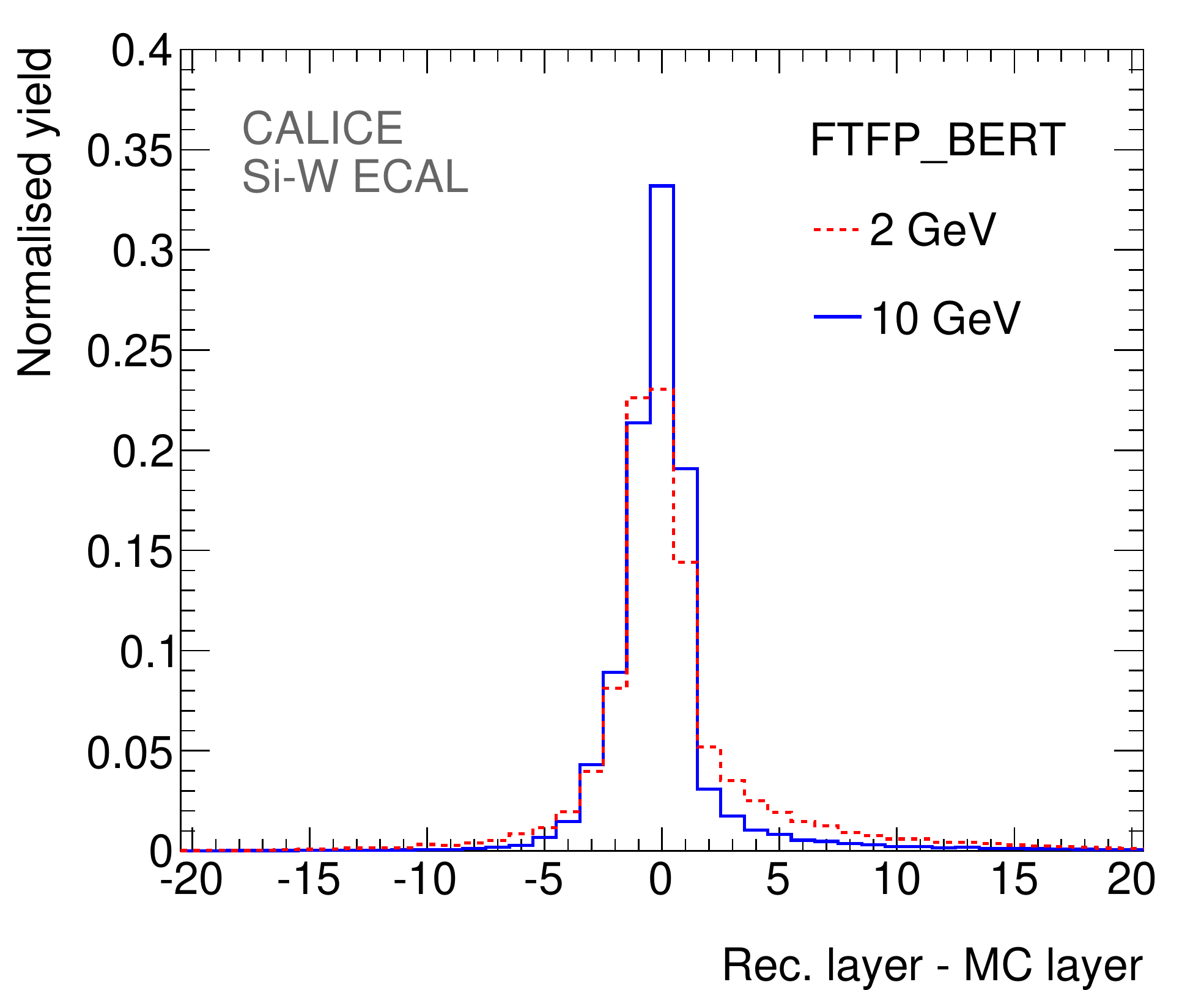}
    \caption{\sl The difference between the reconstructed and the true interaction layers found with the physics list {\sc ftfp\_bert} for $\pi^{-}$ of 2 and 10\,GeV.}
    \label{figure:interactionlayerdifference}
  }
\end{figure}

Figure~\ref{figure:interactionlayerdifference} shows the distribution of the difference between the reconstructed interaction layer and the true interaction layer.
The distribution is peaked around zero and slightly wider at 2\,GeV than at 10\,GeV.
The interaction is more often found in an earlier layer than the true interaction layer than in a later layer.

The interaction-finding efficiencies for the other studied Monte Carlo physics lists are consistent with those found for {\sc ftfp\_bert}, their maximum absolute difference is 0.03.

Events that are not identified by the criteria described above are considered as non-interacting events.
The event sample classified as interacting, however, contains a contamination with non-interacting events of between 2.4\% and 3.5\% for all energies and physics lists.
This contamination is defined as the fraction of events classified as interacting that are non-interacting according to the Monte Carlo truth.
It can be caused by e.g.\,backscattering from the AHCAL, delta rays or energy fluctuations.
	
\section{Comparing Monte Carlo models with data}\label{section:results}
	
Various Monte Carlo models are compared to the test beam data in terms of the fraction of interacting events and radial and longitudinal shower profiles of interacting events. 
The figures in the following sections show these comparisons for simulations based on the four studied Monte Carlo physics lists.

\subsection{Treatment of uncertainties and corrections to the data}\label{section:errortreatment}

The data are contaminated with $e^-$ (especially at low energies) and with events containing multiple interacting particles.
The contamination is reduced by applying triggers and selection cuts (see Section~\ref{section:eventselection}) and the data are corrected for the residual contamination.
The correction factors have been calculated based on Monte Carlo samples of $\pi^-$s mixed with $e^-$s, and mixed samples of single and double $\pi^-$ events.
These have been determined with the physics lists {\sc ftfp\_bert} and {\sc qgsp\_bert} and the average correction has been applied to the data for energies where these physics lists have a different model implementation, for the lowest energies the correction factor determined from {\sc ftfp\_bert} has been applied.
These correction factors are generally between 0.8 and 1.0 depending on the bin for the shower profiles, for the interaction fraction, shower energy, and the means and standard deviations of the shower profiles they are between 0.93 and 1.00.

The systematic uncertainty includes the effect of varying the cut values used to select interacting events, $E_{\mathrm{cut}}$ and $F_{\mathrm{cut}}$, by one unit, as well as the contamination with non-interacting events. 
The relative size of the systematic uncertainty has been estimated using simulated events ({\sc ftfp\_bert}).
The choice of the energy threshold of 0.6 MIP on the reconstructed energy per pixel in the Si-W ECAL has very little influence on the final analysis results: 
when changed to 0.4 MIP, mean results change by a maximum of 0.6\% and when changed to 0.8 MIP, the maximum change is 1.2\%. 
This contribution is small compared to other contributions and is therefore not included in the systematic uncertainty.
A change in the hit energy of 2\%, the estimated uncertainty in the calibration~\cite{2009_Li}, does not alter the results significantly,
nor does a change in the restriction on the event barycentre.

Systematic uncertainties related to the digitization procedure of the Monte Carlo data sets is estimated to be negligibly small.
The detailed digitisation procedure reproduces all effects in the data to a sufficient level.
For each cell the pedestal and the Gaussian noise recorded for that particular run are applied, the simulated hits in MeV are converted to ADC counts and calibration constants are then applied. The uncertainty in the calibration does not influence the results, as mentioned above.
Additionally the energy response of the Si-W ECAL is linear in the studied energy range.
As the signal to noise ratio of the Si-W ECAL is very good, $\approx 7.5$, the signal is well above the hit energy cut of 0.6 MIP, and there is little sensitivity to the noise spectrum itself. This is confirmed by the small effect a change in the hit energy threshold has on the final results.
Correlated noise was close to absent in the test beam periods at FNAL, due to proper grounding, and residual correlated noise is eliminated in the reconstruction procedure.

In each of the following figures the data has been corrected for residual contamination and the systematic uncertainty, determined as described above, is combined quadratically with the statistical uncertainty and is then visualised by a shaded band around the data.
In this combined uncertainty the systematic contribution is often dominant.
The stability of the mean and standard deviations of the studied observables have been estimated by performing the analysis on subsets of the data sets.
The maximum differences between the results of these subsets have been added to the systematic uncertainties for the means and standard deviations.
For figures~\ref{figure:interactionfraction}, \ref{figure:showerenergy}, \ref{figure:meansigmaradialdistance}, \ref{figure:meansigmaradialprofile}, \ref{figure:meansigmaz}, and \ref{figure:meansigmalongitudinalprofile} the systematic uncertainty is constructed such that it can be asymmetric; some contributions to the systematic uncertainty always increase the interaction fraction while others always decrease it. For the other figures this is not possible and the systematic uncertainty is symmetric.
No systematic uncertainty is assigned to the Monte Carlo data sets.

\subsection{Interaction fraction and reconstructed shower energy}

\begin{figure}[h]
  {\centering
    \includegraphics[width=0.6\textwidth]{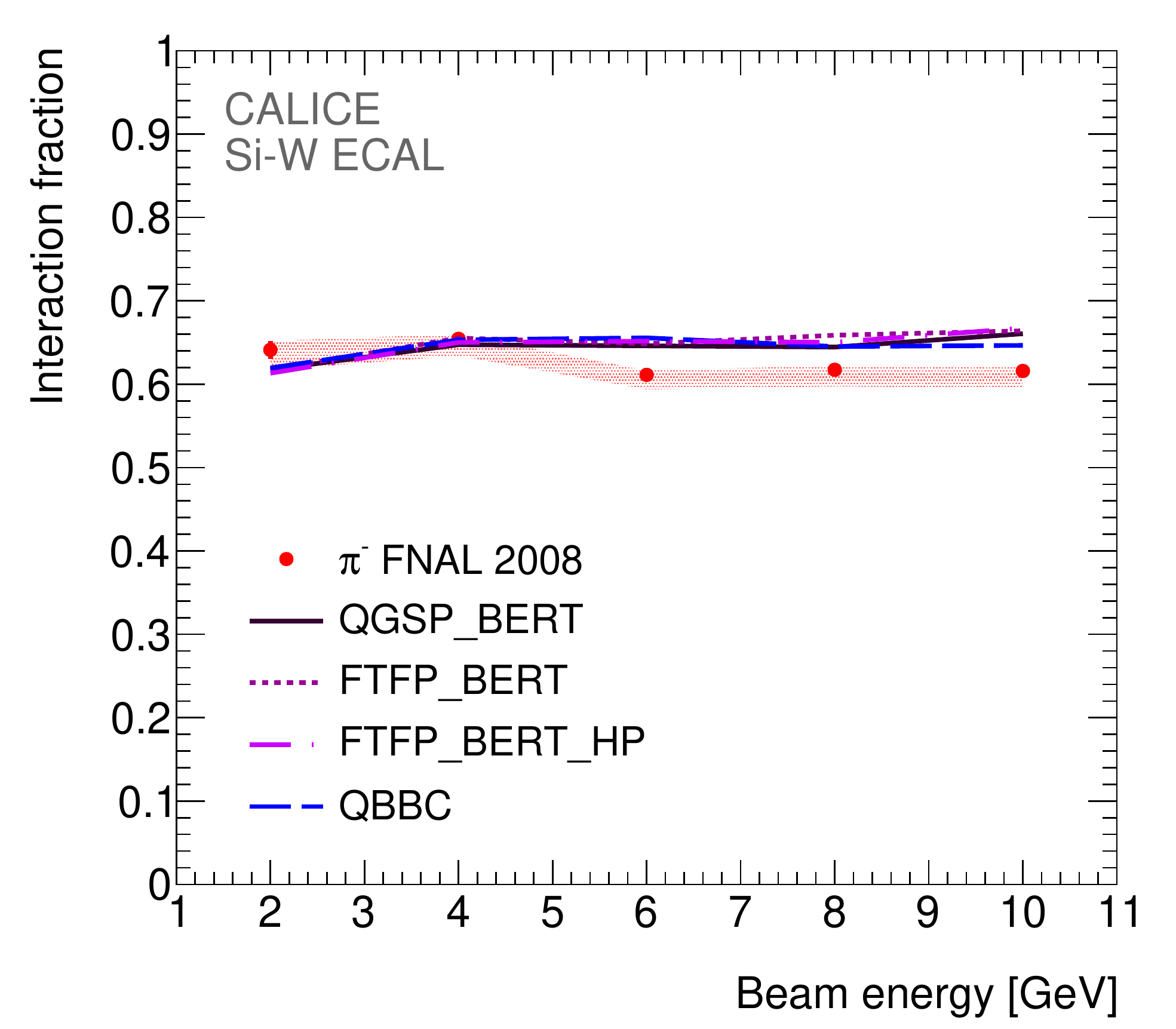}
    \caption{\sl Interaction fraction for $\pi^-$ in the Si-W ECAL for data and various Monte Carlo physics lists as a function of beam energy (2\,GeV to 10\,GeV).}
    \label{figure:interactionfraction}
  }
\end{figure}

The interaction fraction is the fraction of interacting events found among all events in an event sample according to the criteria described in Section~\ref{section:identifyinginteractingevents}, corrected by the interaction finding efficiency. 
For the test beam data the efficiency as given by the {\sc ftfp\_bert} physics list is used.
Figure~\ref{figure:interactionfraction} shows the interaction fraction as a function of the $\pi^-$ energy for data and the predictions of simulations using the physics lists {\sc qgsp\_bert}, {\sc ftfp\_bert}, {\sc ftfp\_bert\_hp} and {\sc qbbc}.

The interaction fraction is approximately independent of the beam energy and is consistent with the material budget of the Si-W ECAL (one interaction length).
For low beam energies the contribution from events with small energy transfer as well as events with high local energy transfer is highest, 
while at 10 GeV most of the events are selected by the absolute energy threshold criteria.
The physics lists are in good agreement with each other, and, at low energies, with the data.
At higher energies, all physics lists are found to overestimate the interaction fraction by about 7\%.

\begin{figure}[h]
  {\centering
    \includegraphics[width=0.6\textwidth]{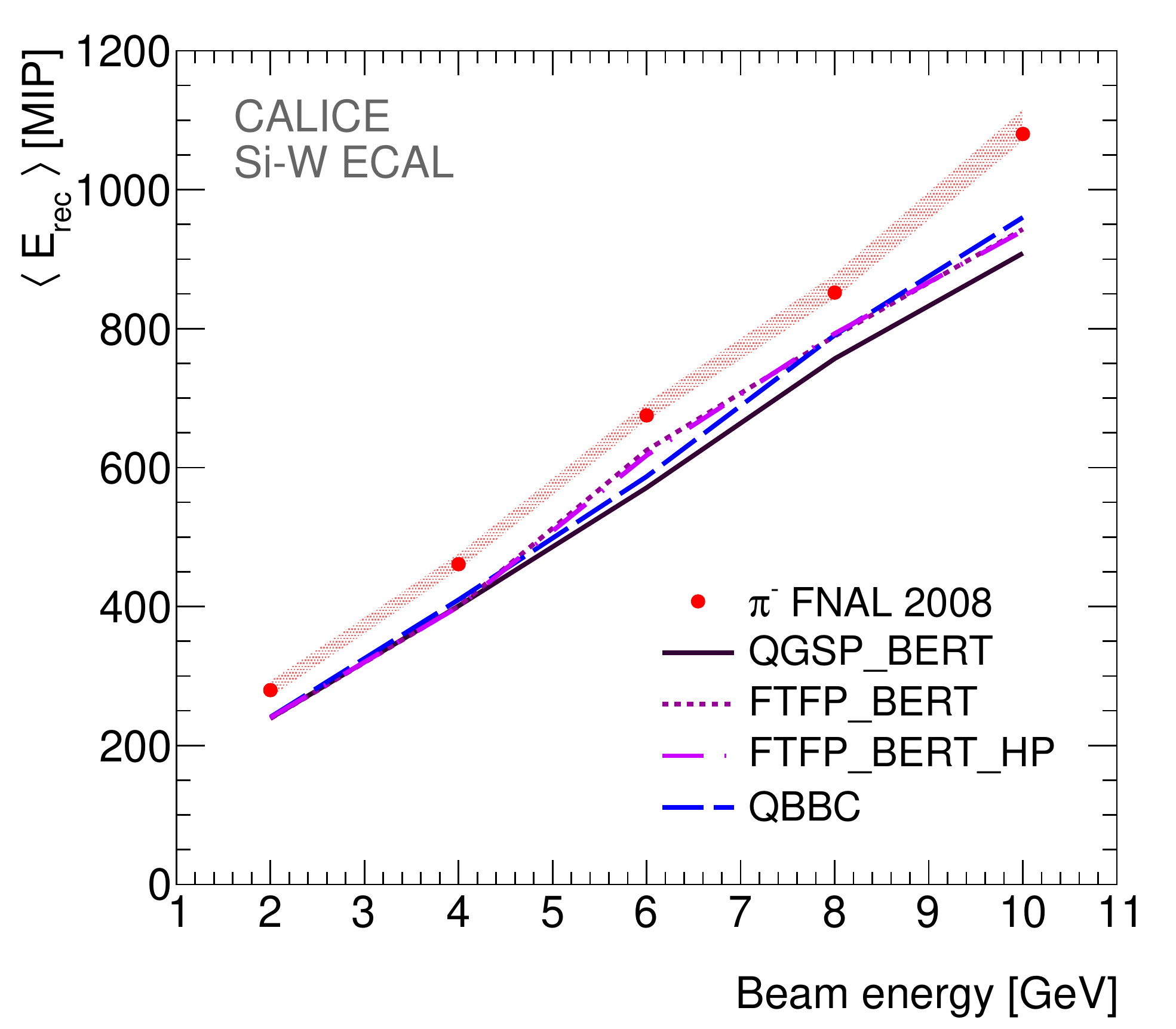}
    \caption{\sl Reconstructed $\pi^-$ shower energy in the Si-W ECAL for data and various Monte Carlo physics lists as a function of beam energy (2\,GeV to 10\,GeV).}
    \label{figure:showerenergy}
  }
\end{figure}

For the events identified as interacting Fig.~\ref{figure:showerenergy} shows the reconstructed energy of that part of the shower that is seen in the Si-W ECAL as a function of beam energy. This shower energy increases with the energy of the incoming $\pi^-$ and is on average 15\% higher in data than in simulation. This observation is true for all energies of the primary pions.

\subsection{Lateral shower extension}

\begin{figure}[h]
  {\centering
    \subfloat[]{\includegraphics[width=0.33\textwidth]{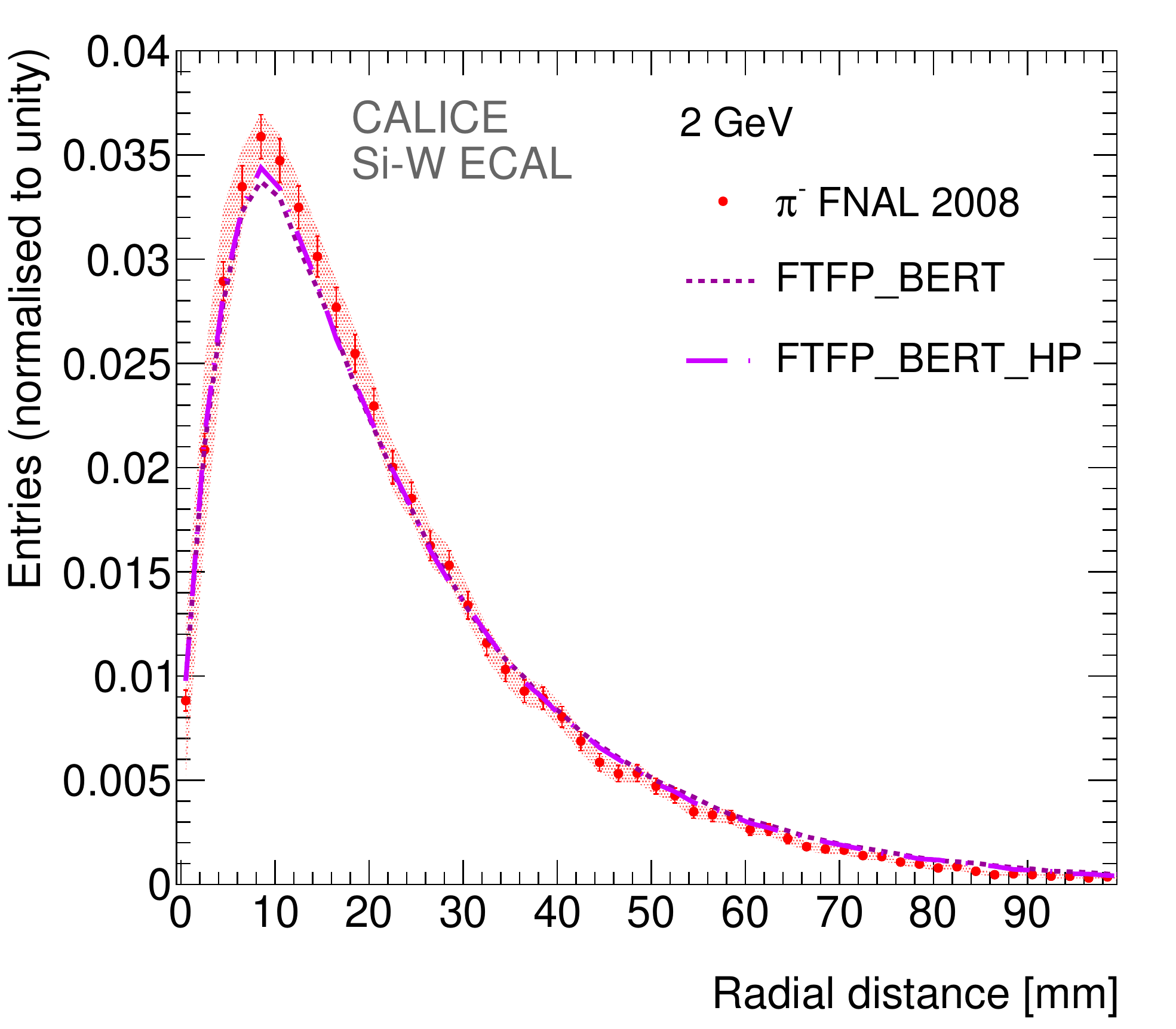}}
    \subfloat[]{\includegraphics[width=0.33\textwidth]{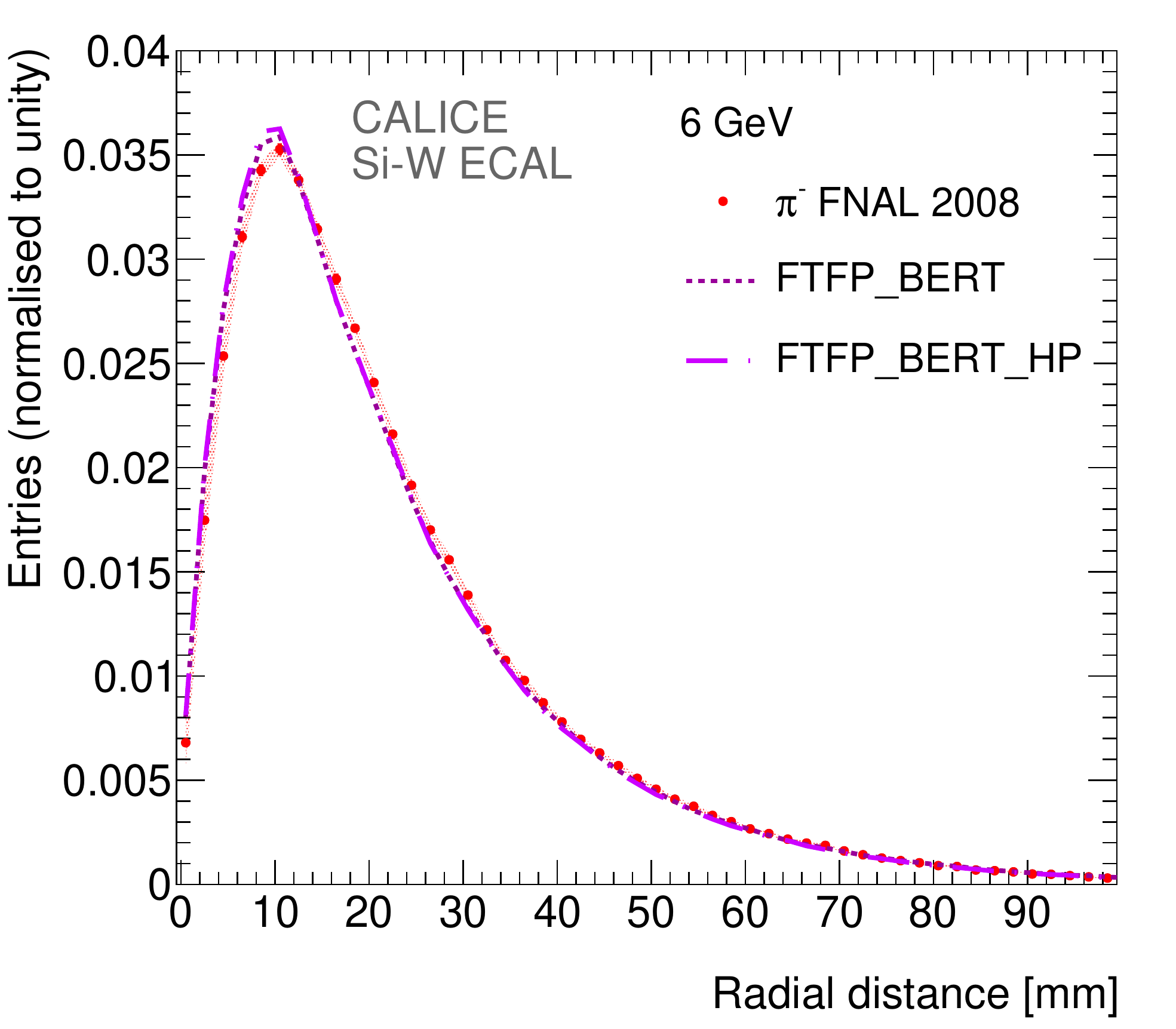}}
    \subfloat[]{\includegraphics[width=0.33\textwidth]{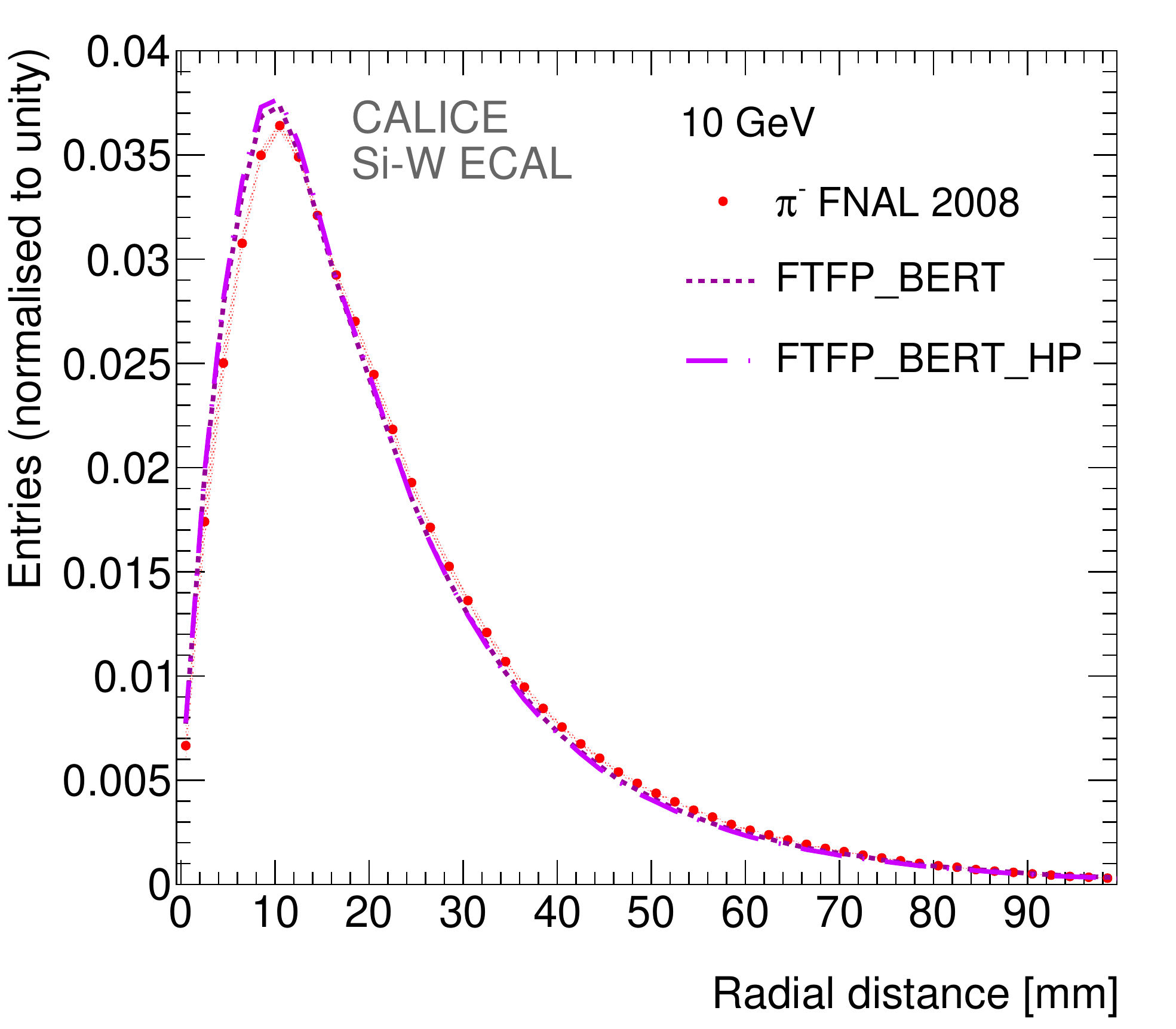}}
    \caption{\sl The radial distance from the shower centre of hits in the shower for interacting events at 2, 6 and 10\,GeV, for data and the Monte Carlo physics lists {\sc ftfp\_bert} and {\sc ftfp\_bert\_hp}. $\Delta r$ is 2 mm.}
    \label{figure:radialdistanceftfp}
  }
\end{figure}

\begin{figure}[h]
  {\centering
    \subfloat[]{\includegraphics[width=0.33\textwidth]{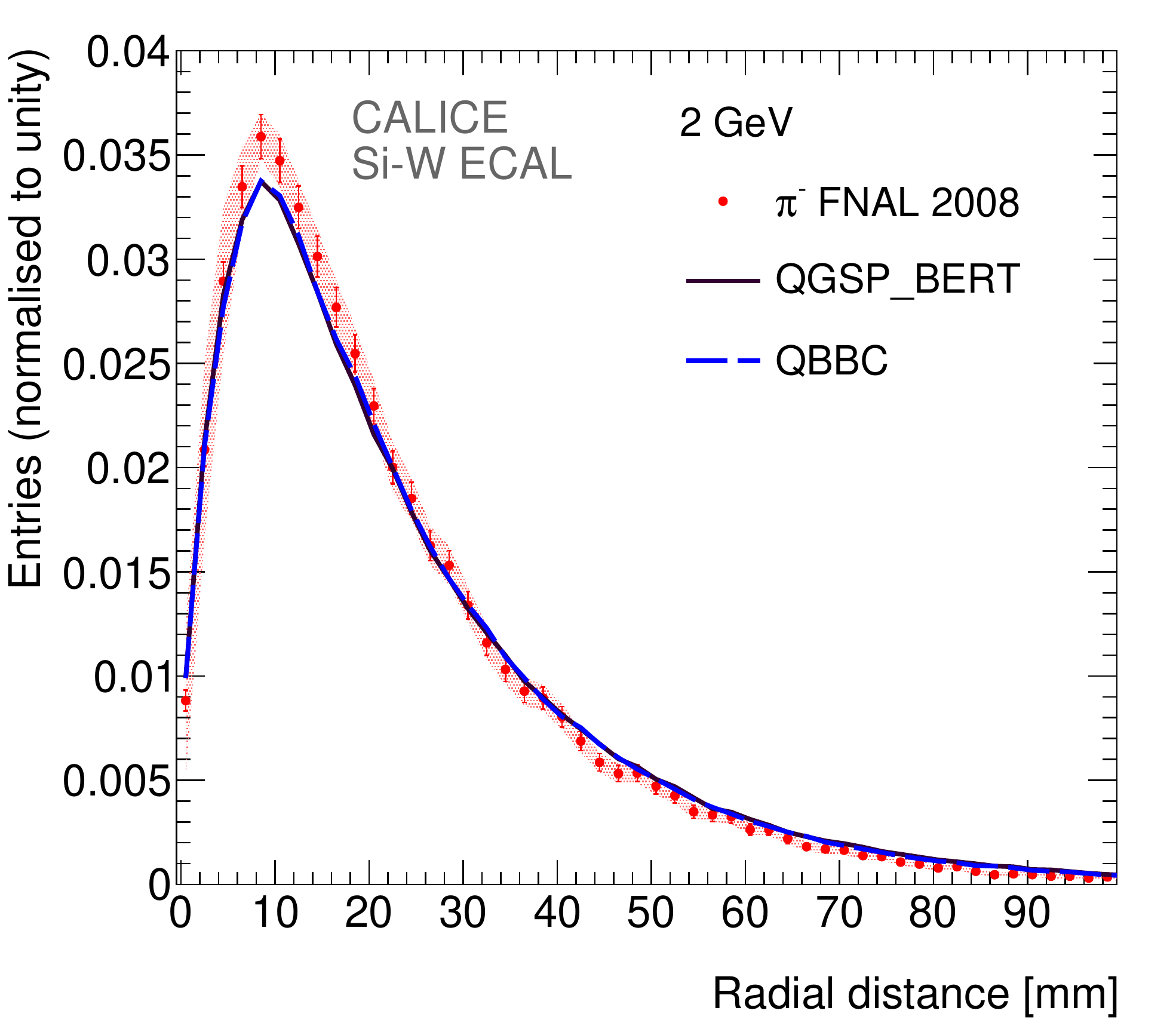}}
    \subfloat[]{\includegraphics[width=0.33\textwidth]{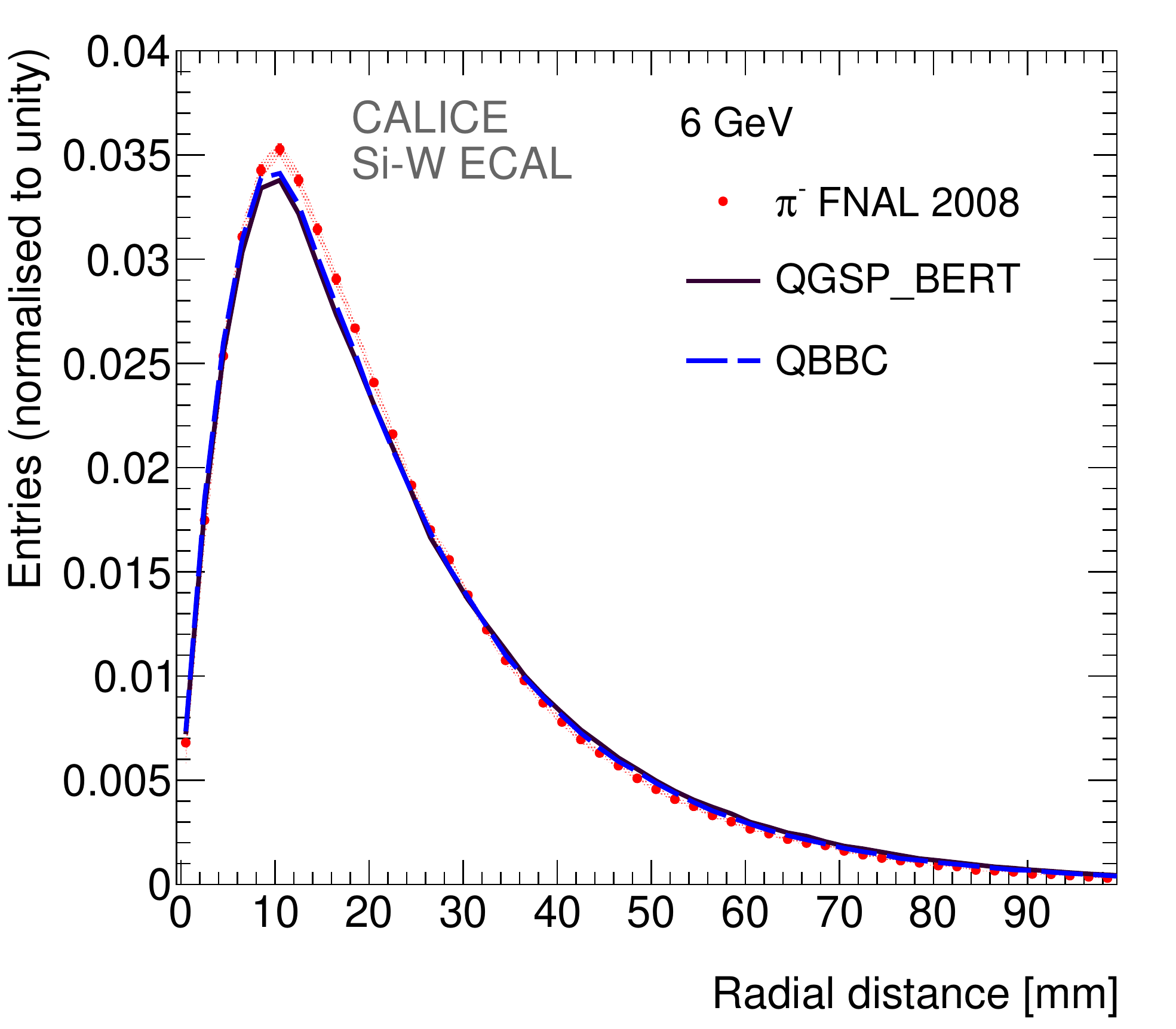}}
    \subfloat[]{\includegraphics[width=0.33\textwidth]{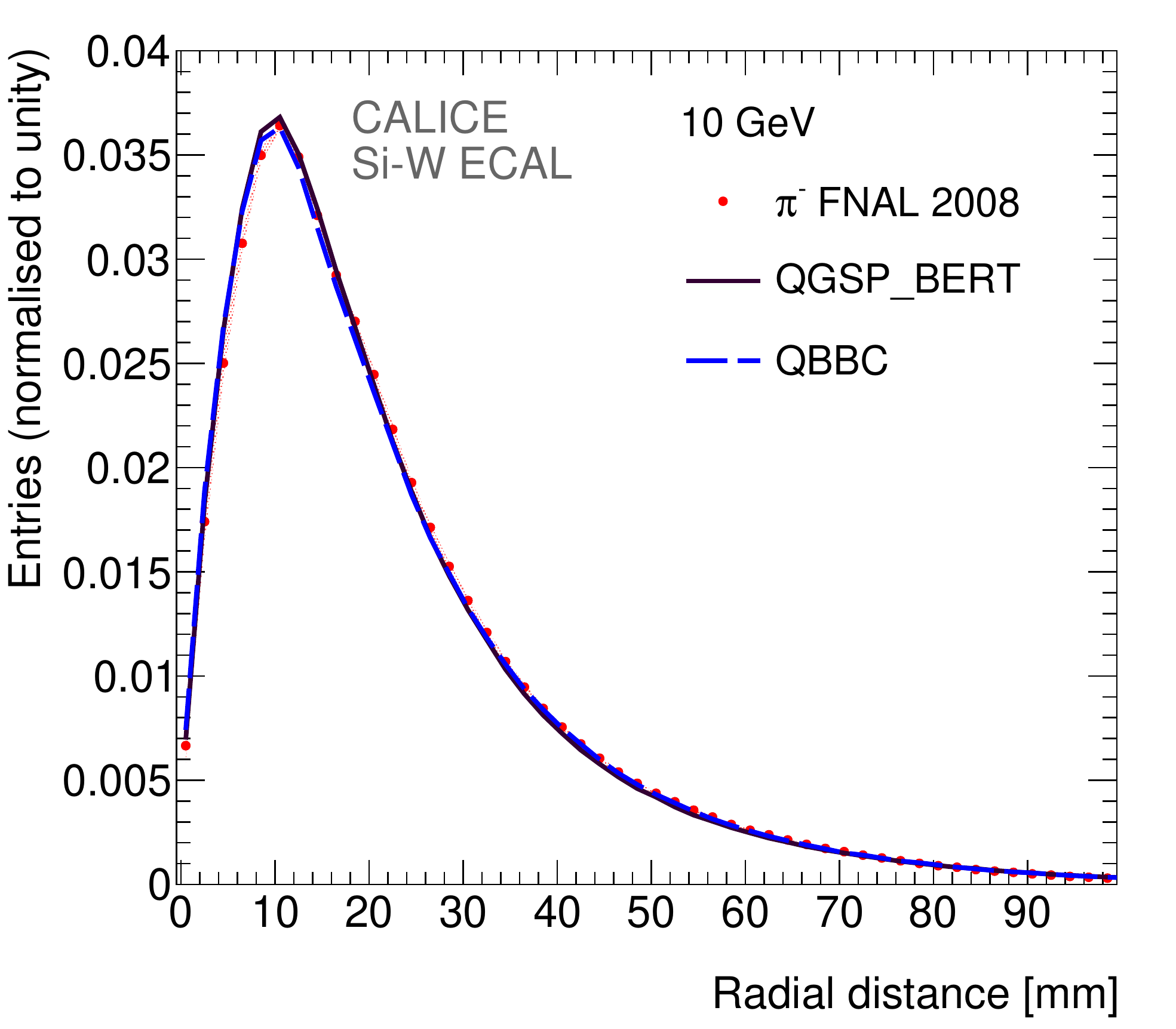}}
    \caption{\sl The radial distance from the shower centre of hits in the shower for interacting events at 2, 6 and 10\,GeV, for data and the Monte Carlo physics lists {\sc qgsp\_bert} and {\sc qbbc}. $\Delta r$ is 2 mm.}
    \label{figure:radialdistanceqgsp}
  }
\end{figure}

The radial distribution of hits in the shower and the radial energy profile can be used as a measure of the lateral extension of the shower formed as a result of an interaction.
Figures~\ref{figure:radialdistanceftfp} and \ref{figure:radialdistanceqgsp} show the radial distance of shower hits with respect to the lateral shower barycentre for beam energies of 2, 6 and 10\,GeV. 
The bin size, $\Delta r$, is 2 mm.
Only hits in the interaction layer and subsequent layers are taken into account.
The histograms are normalised to unity in order to compare the shape of the distributions.
In Fig.~\ref{figure:radialdistanceftfp} the data are compared to {\sc ftfp\_bert} and {\sc ftfp\_bert\_hp} while in Fig.~\ref{figure:radialdistanceqgsp} they are compared to {\sc qgsp\_bert} and {\sc qbbc}.
The data are shown with their statistical and total uncertainties.
The predictions of all physics lists are within 6\% of the data. 
 
\begin{figure}[h]
  {\centering
    \subfloat[]{\includegraphics[width=0.5\textwidth]{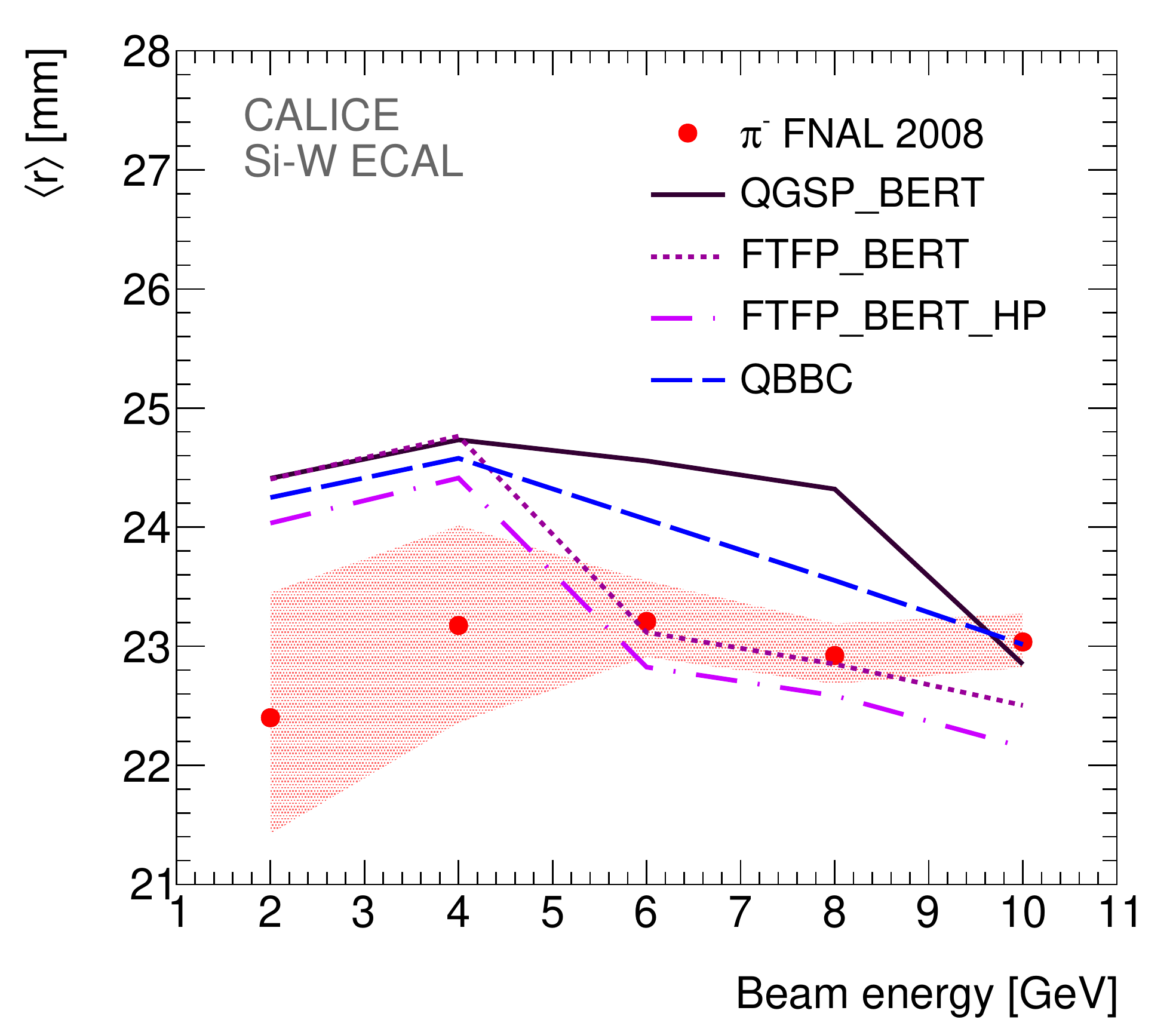}}
    \subfloat[]{\includegraphics[width=0.5\textwidth]{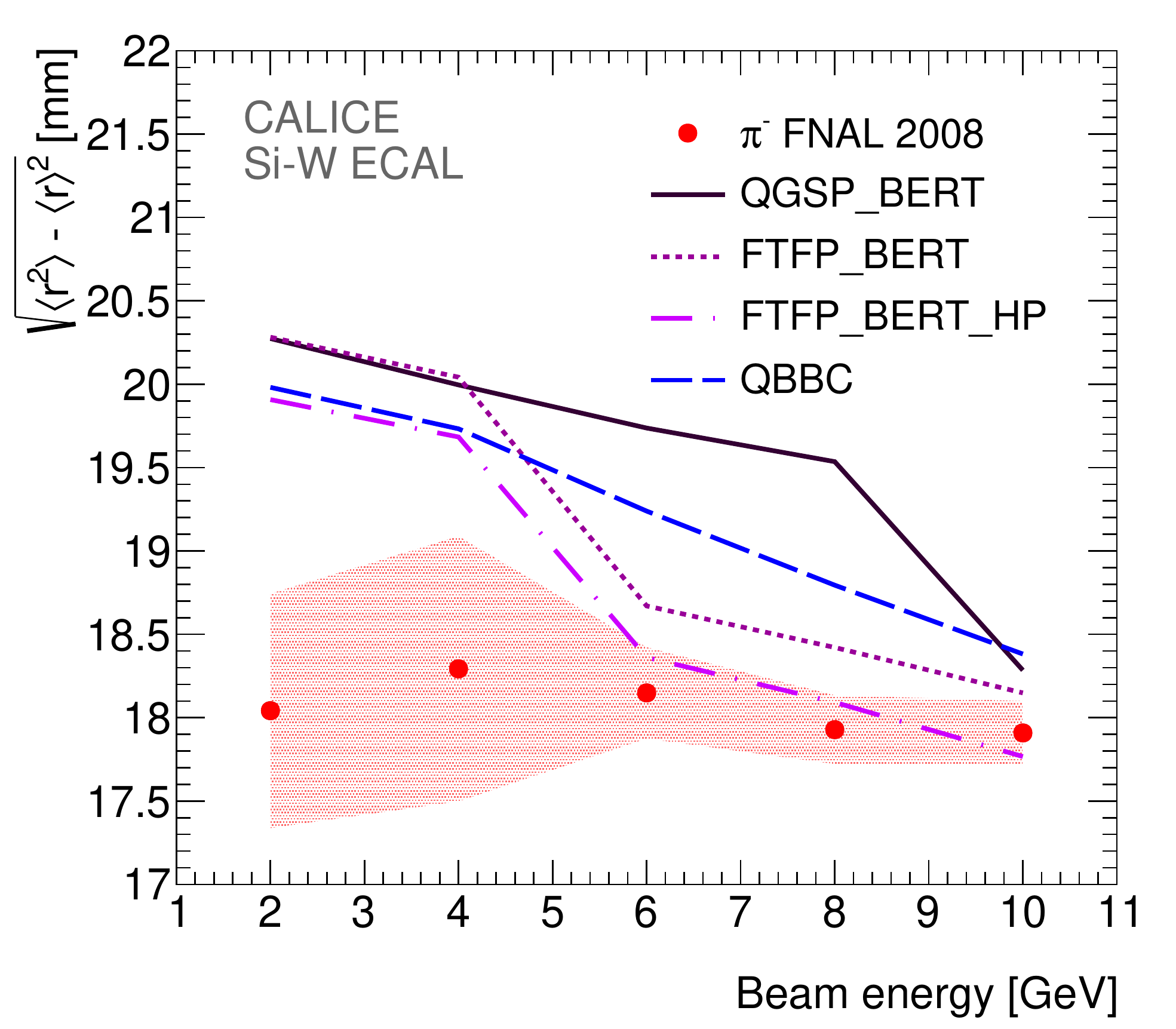}}
    \caption{\sl Mean (a) and standard deviation (b) of the radial distance of hits for interacting events as a function of beam energy (2\,GeV to 10\,GeV) for data and various Monte Carlo physics lists.}
    \label{figure:meansigmaradialdistance}
  }
\end{figure}

Figure~\ref{figure:meansigmaradialdistance} shows the mean, $\langle r \rangle$, and standard deviation, $\sqrt{\langle r^2 \rangle - \langle r \rangle^2}$, of the radial distance.
While for data they are constant within 4\%, in the simulation the mean decreases by 7\% between 2 and 10\,GeV and the standard deviation decreases by 10\%.
The Monte Carlo models agree with the data within 7\%, but the {\sc qgsp\_bert} and {\sc qbbc} physics lists overestimate the mean radial distance for almost all energies, while {\sc ftfp\_bert} and {\sc ftfp\_bert\_hp} overestimate for 2 and 4\,GeV after which there is a very abrupt transition to a smaller mean and standard deviation.
Between 4 and 6\,GeV these physics lists change from the Bertini cascade to the Fritiof string model.
The transition from the Bertini cascade to the Low Energy Parametrized model in {\sc qgsp\_bert} is also visible.
For the energy range between 4 and 10 GeV the {\sc qbbc} physics list is in the transition region from the Bertini cascade to the Fritiof string model and is thus in between {\sc qgsp\_bert} and {\sc ftfp\_bert}.
The Bertini model generates too wide showers while the Fritiof model seems to agree better with the data.
Additionally the high precision treatment of low energy neutrons in {\sc ftfp\_bert\_hp} gives a systematically smaller mean and standard deviation compared to {\sc ftfp\_bert}.

\FloatBarrier

\begin{figure}[h]
  {\centering
    \subfloat[]{\includegraphics[width=0.33\textwidth]{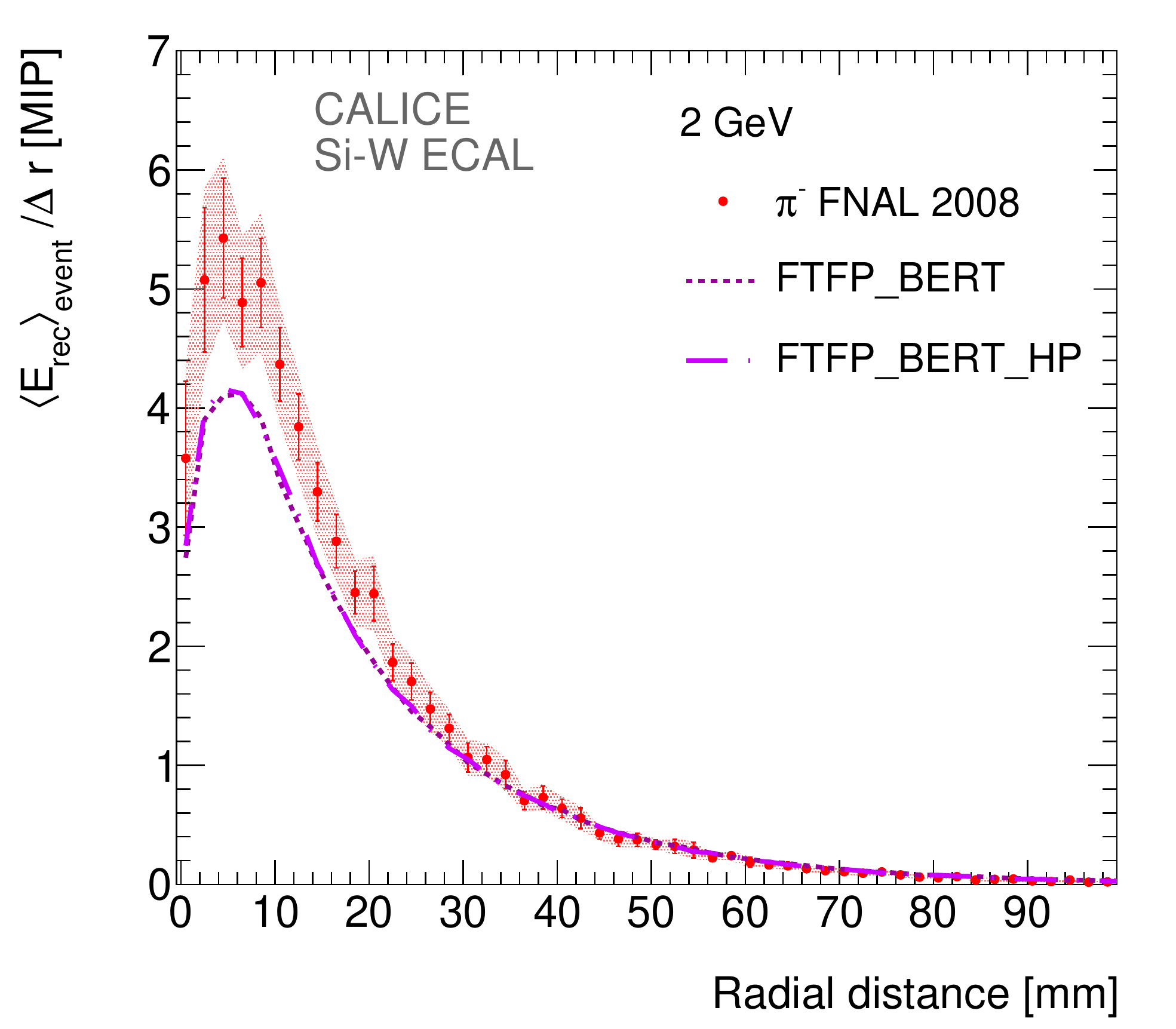}}
    \subfloat[]{\includegraphics[width=0.33\textwidth]{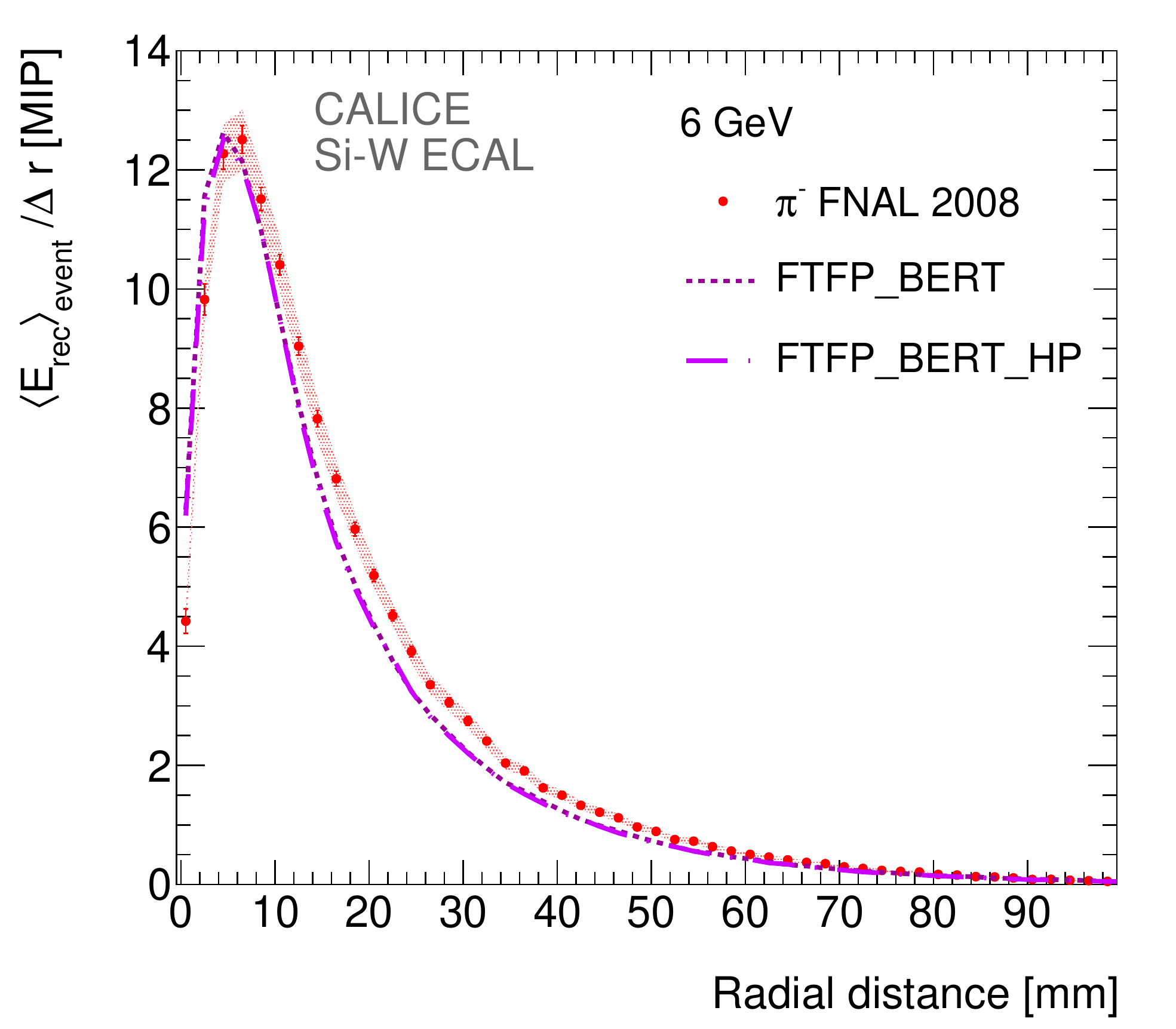}}
    \subfloat[]{\includegraphics[width=0.33\textwidth]{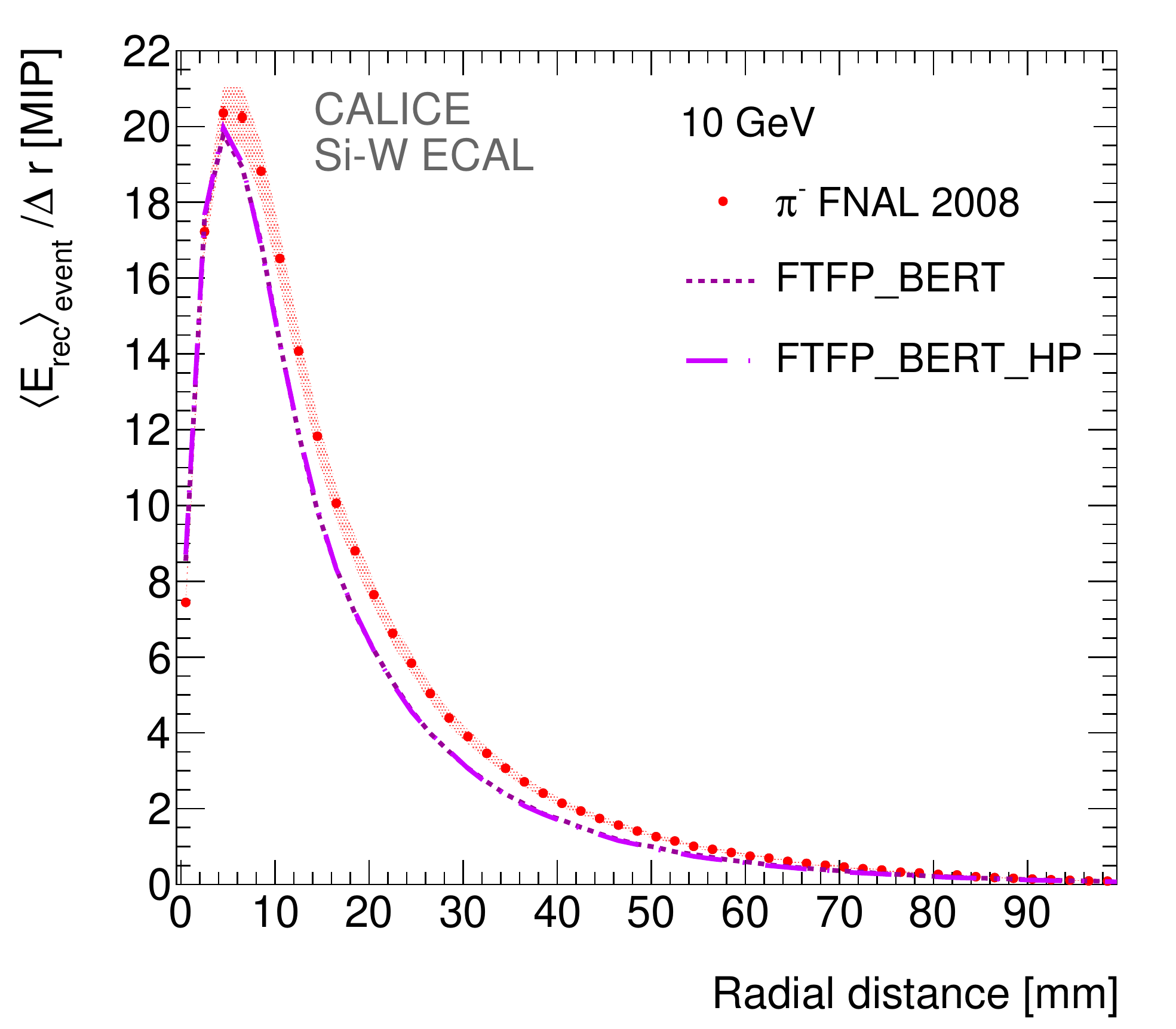}}
    \caption{\sl The radial energy profile for interacting events at 2, 6 and 10\,GeV, for data and the Monte Carlo physics lists {\sc ftfp\_bert} and {\sc ftfp\_bert\_hp}. $\Delta r$ is 2 mm.}
    \label{figure:radialprofileftfp}
  }
\end{figure}

\begin{figure}[h]
  {\centering
    \subfloat[]{\includegraphics[width=0.33\textwidth]{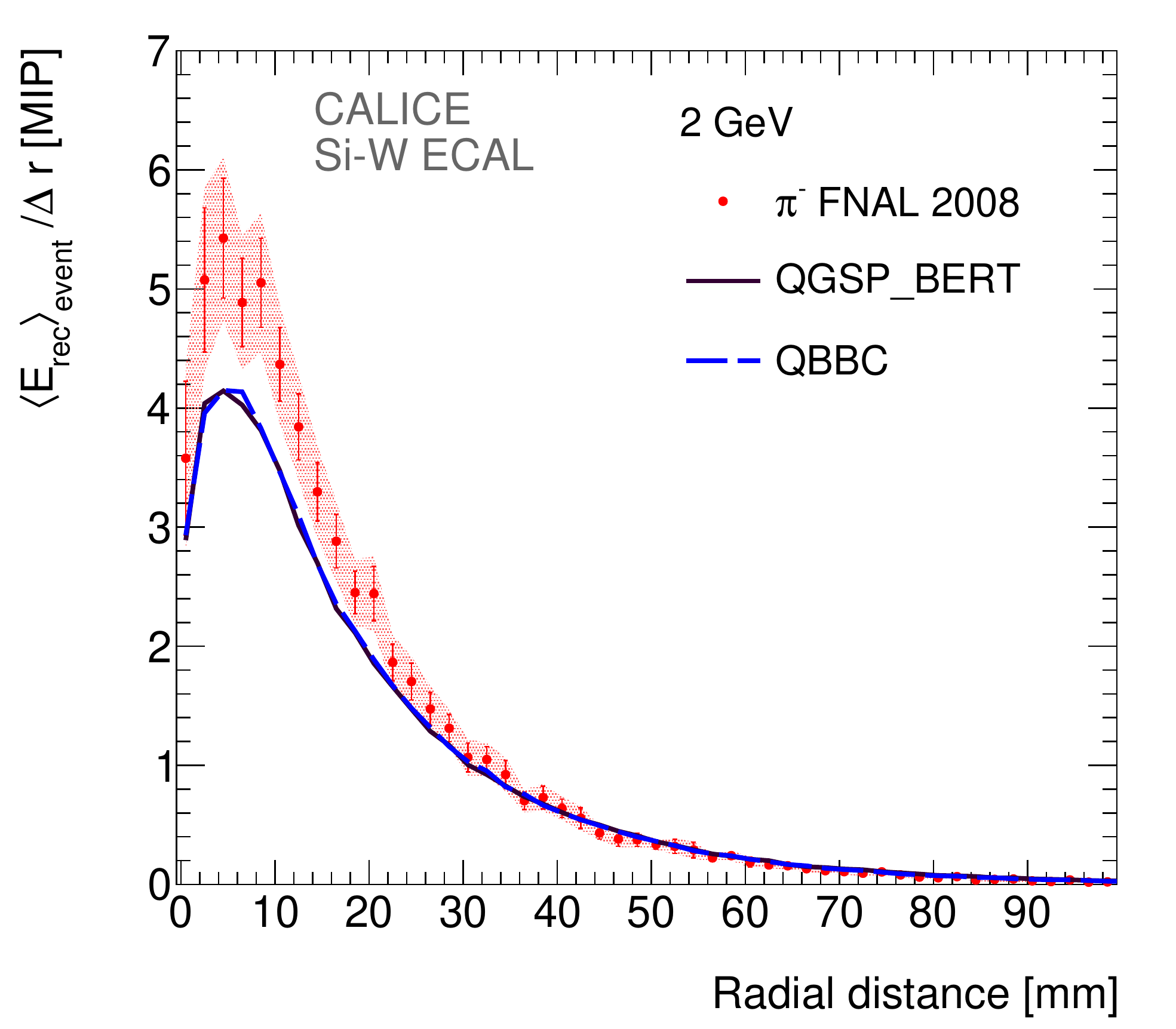}}
    \subfloat[]{\includegraphics[width=0.33\textwidth]{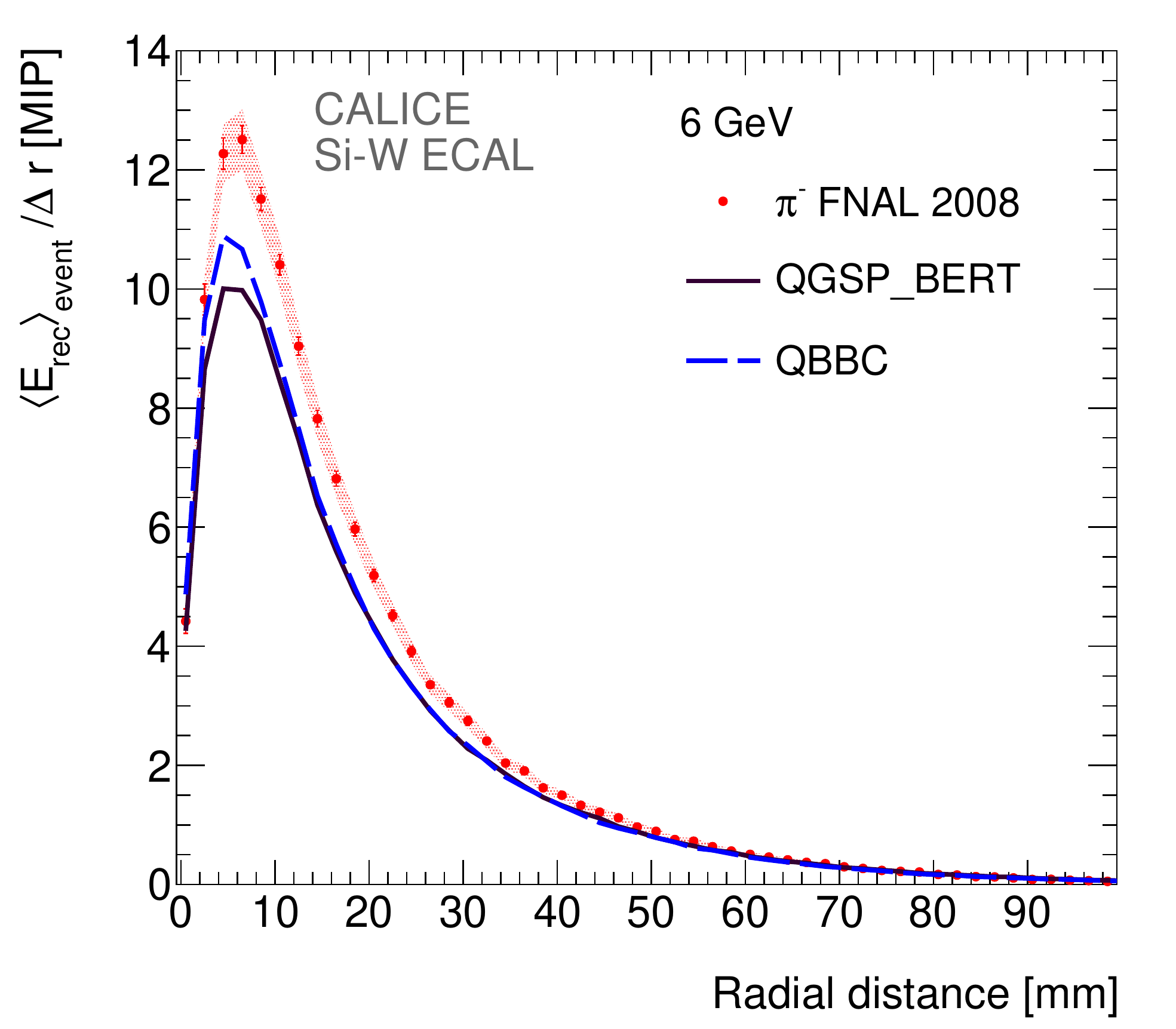}}
    \subfloat[]{\includegraphics[width=0.33\textwidth]{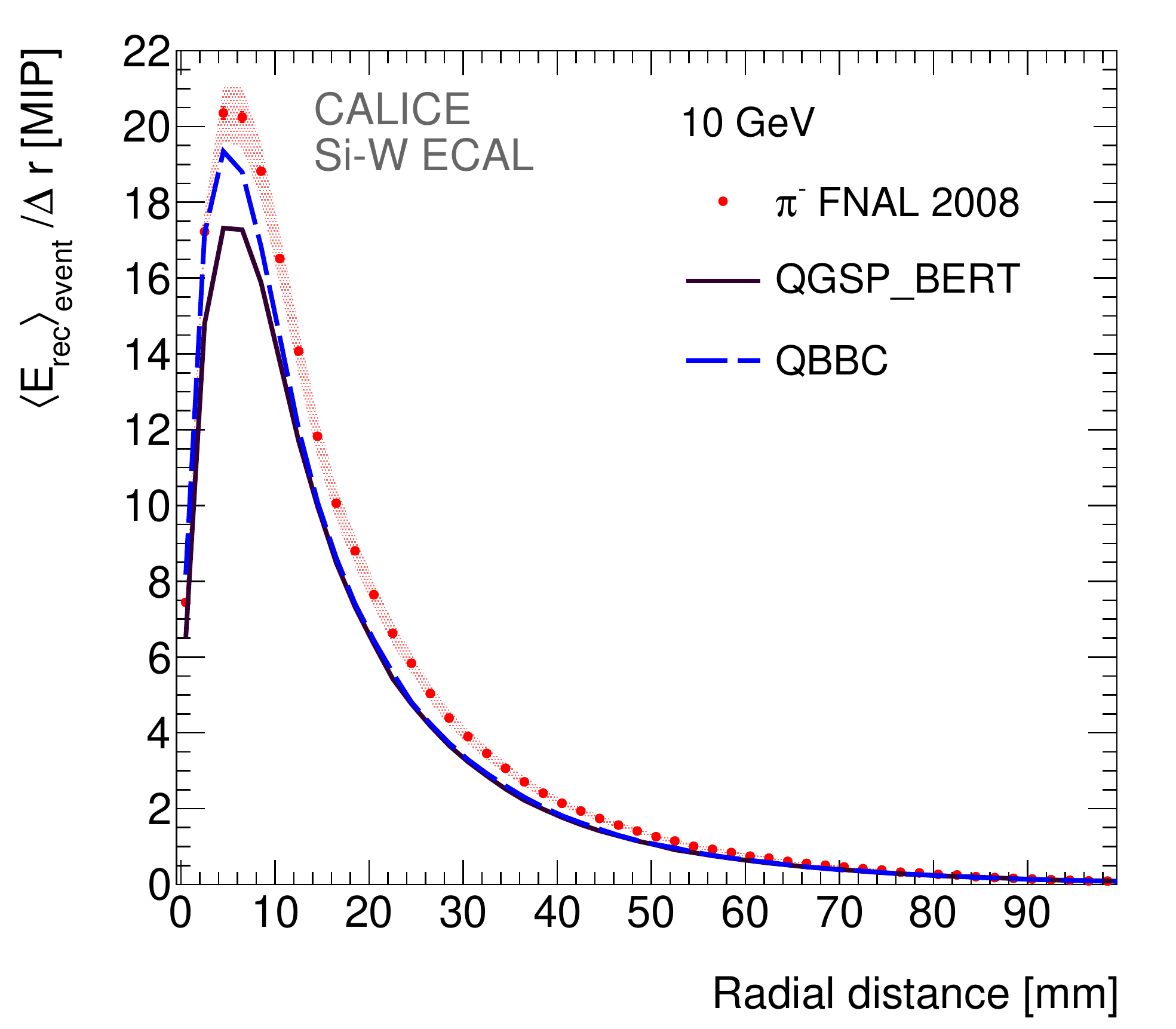}}
    \caption{\sl The radial energy profile for interacting events at 2, 6 and 10\,GeV, for data and the Monte Carlo physics lists {\sc qgsp\_bert} and {\sc qbbc}. $\Delta r$ is 2 mm.}
    \label{figure:radialprofileqgsp}
  }
\end{figure}

Figures~\ref{figure:radialprofileftfp} and \ref{figure:radialprofileqgsp} show the radial energy profile, defined here as the reconstructed energy per event as a function of the radial distance to the shower barycentre, at 2, 6 and 10\,GeV. 
In Fig.~\ref{figure:radialprofileftfp} the data are compared to {\sc ftfp\_bert} and {\sc ftfp\_bert\_hp}, in Fig.~\ref{figure:radialprofileqgsp} they are compared to {\sc qgsp\_bert} and {\sc qbbc}.
Overall, the reconstructed energy is underestimated by all the physics lists, especially {\sc qgsp\_bert}, which is compatible with Fig.~\ref{figure:showerenergy}.
This underestimation of the reconstructed energy is caused by a smaller number of hits in the simulation compared to the data, as the mean energy per hit is comparable.

\begin{figure}[h]
  {\centering
    \subfloat[]{\includegraphics[width=0.33\textwidth]{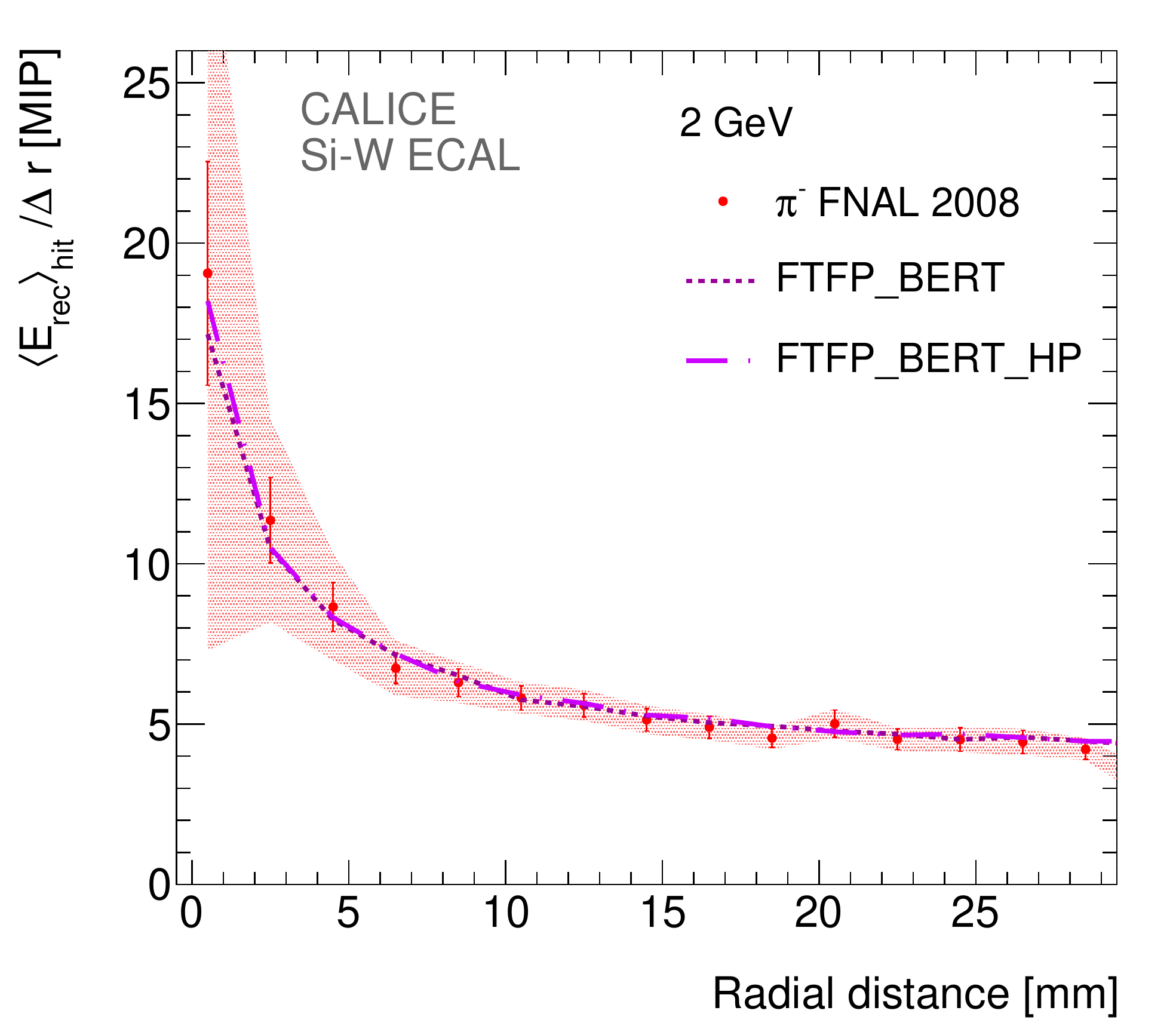}}
    \subfloat[]{\includegraphics[width=0.33\textwidth]{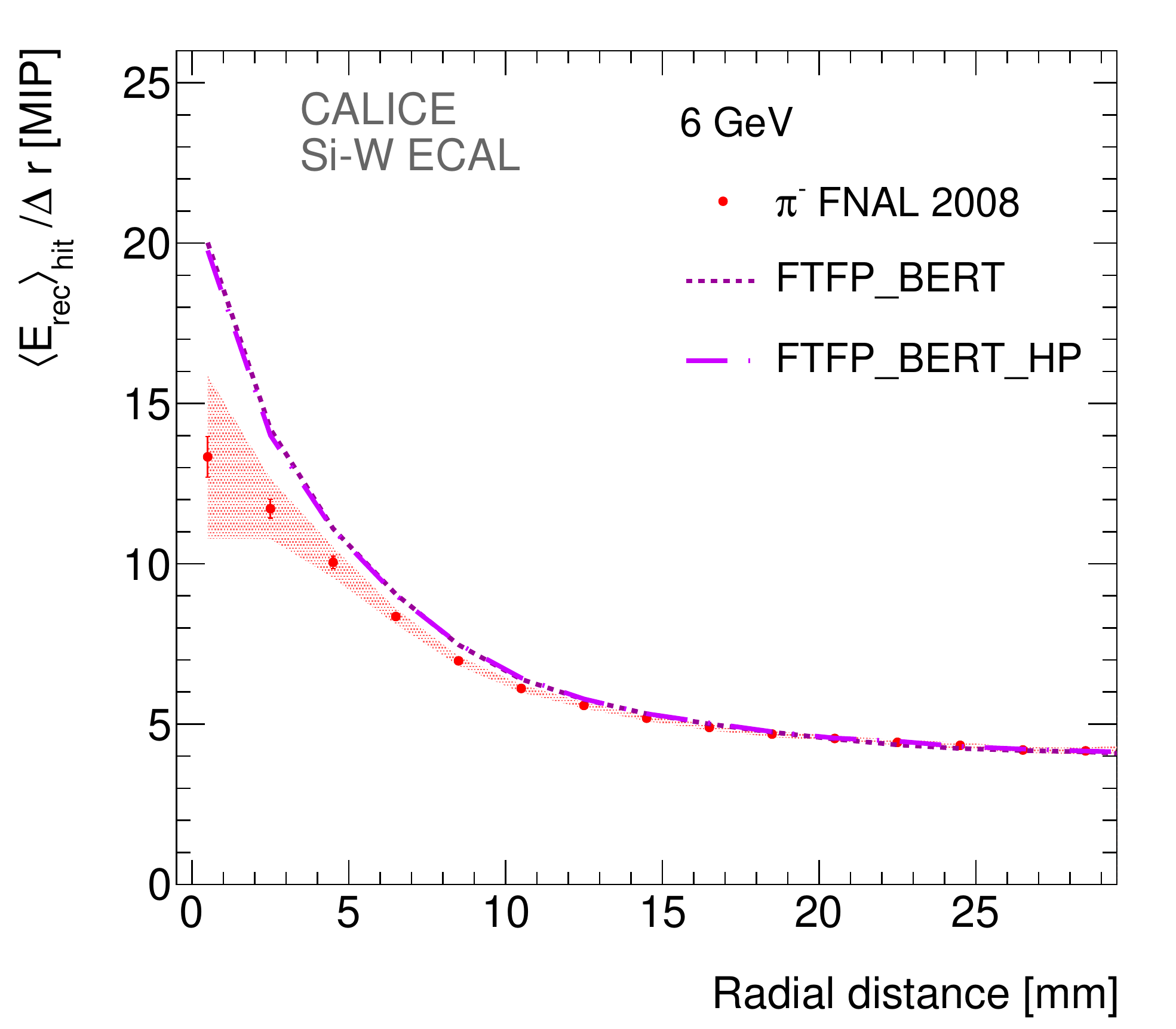}}
    \subfloat[]{\includegraphics[width=0.33\textwidth]{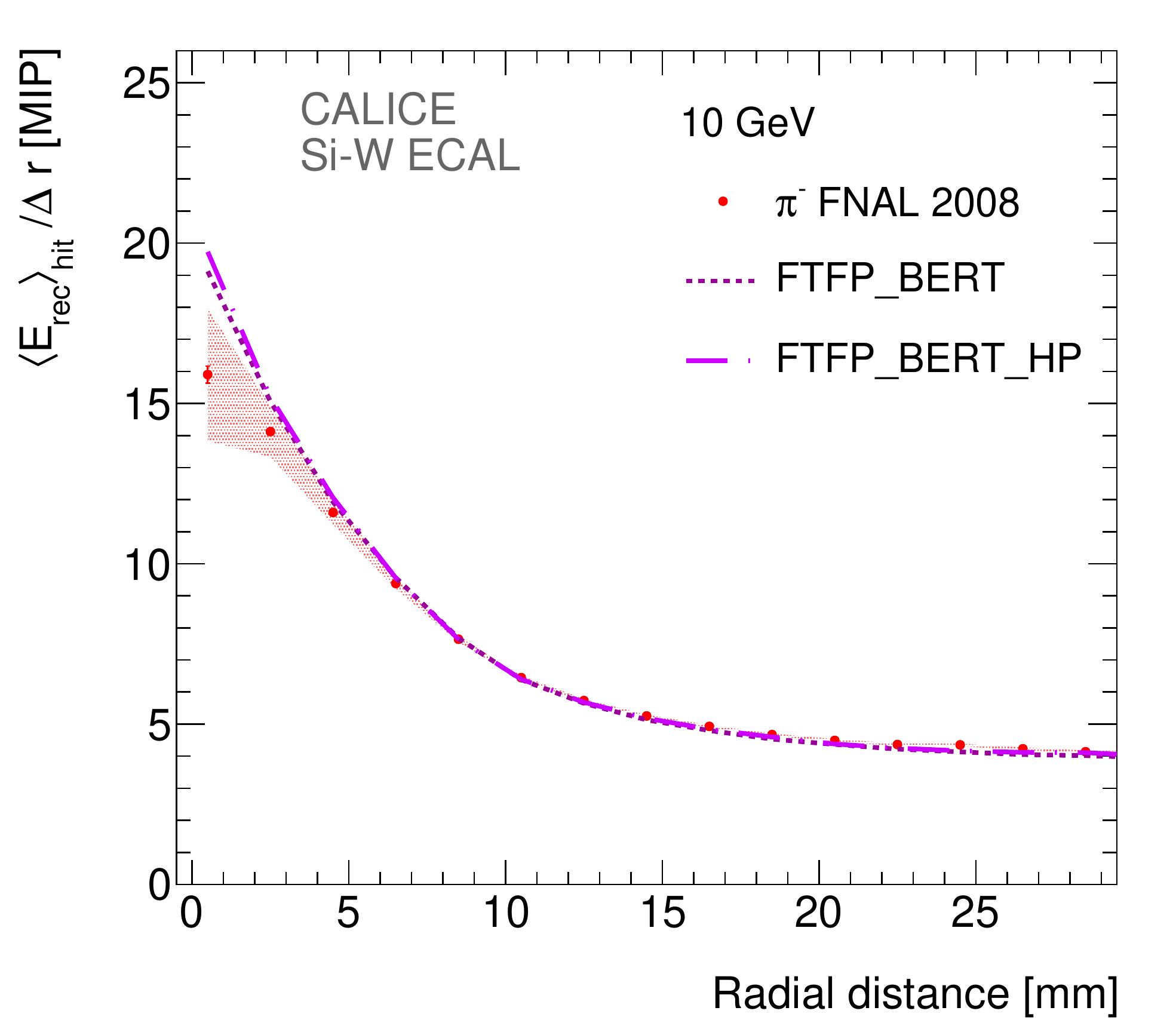}}
    \caption{\sl The radial mean hit energy for interacting events at 2, 6 and 10\,GeV, for data and the Monte Carlo physics lists {\sc ftfp\_bert} and {\sc ftfp\_bert\_hp}. $\Delta r$ is 2 mm.}
    \label{figure:radialmeanhitenergyftfp}
  }
\end{figure}

\begin{figure}[h]
  {\centering
    \subfloat[]{\includegraphics[width=0.33\textwidth]{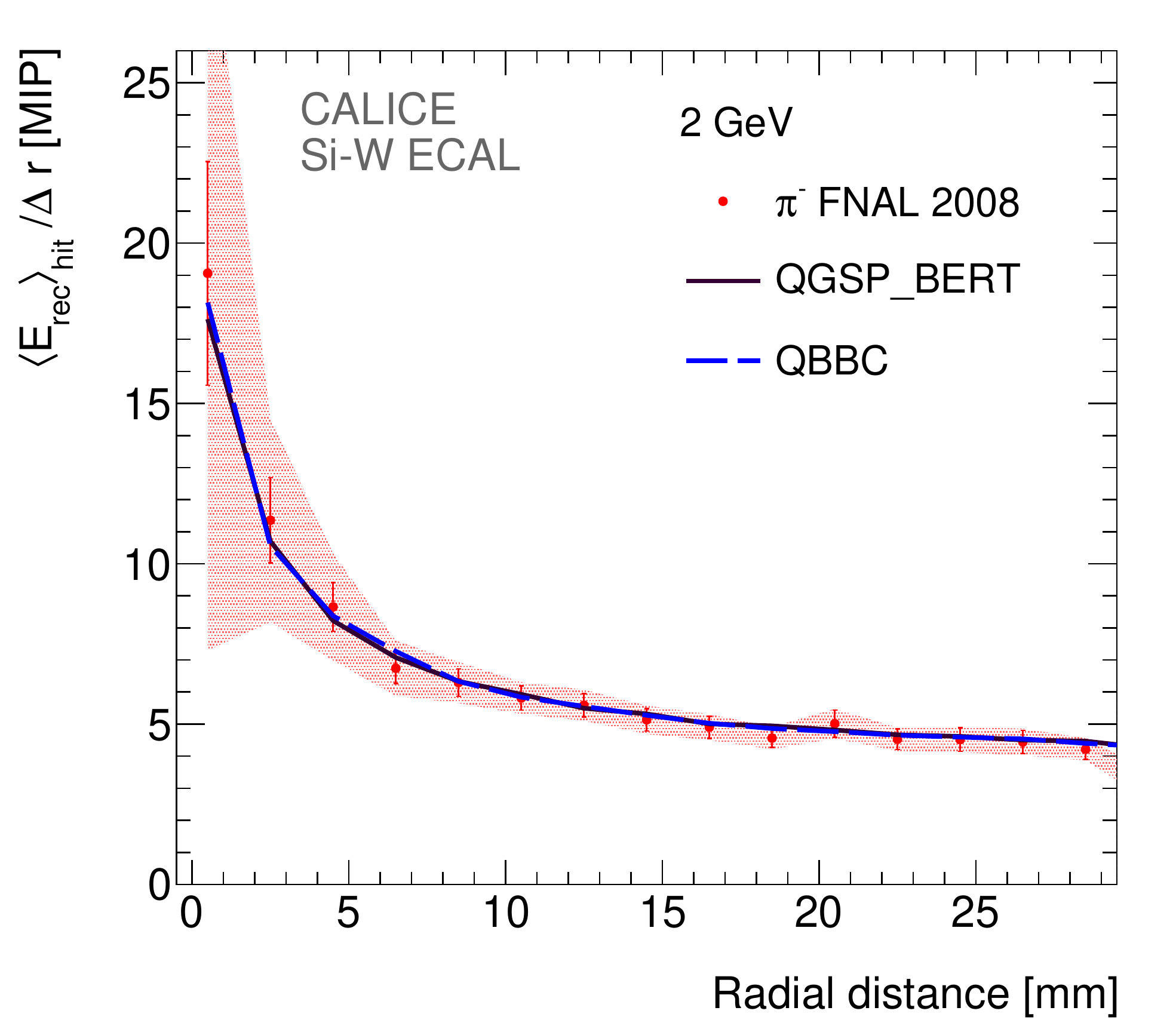}}
    \subfloat[]{\includegraphics[width=0.33\textwidth]{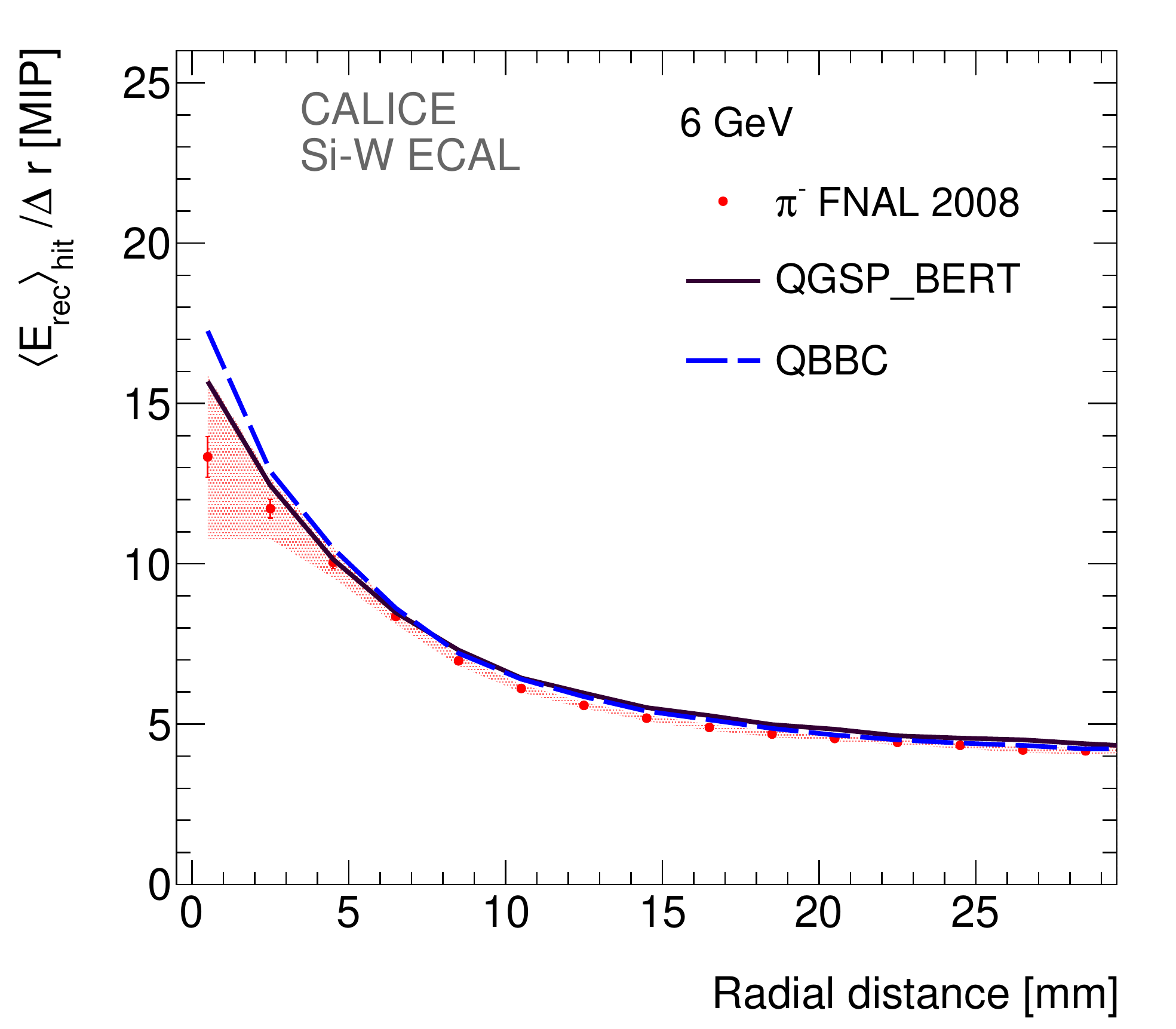}}
    \subfloat[]{\includegraphics[width=0.33\textwidth]{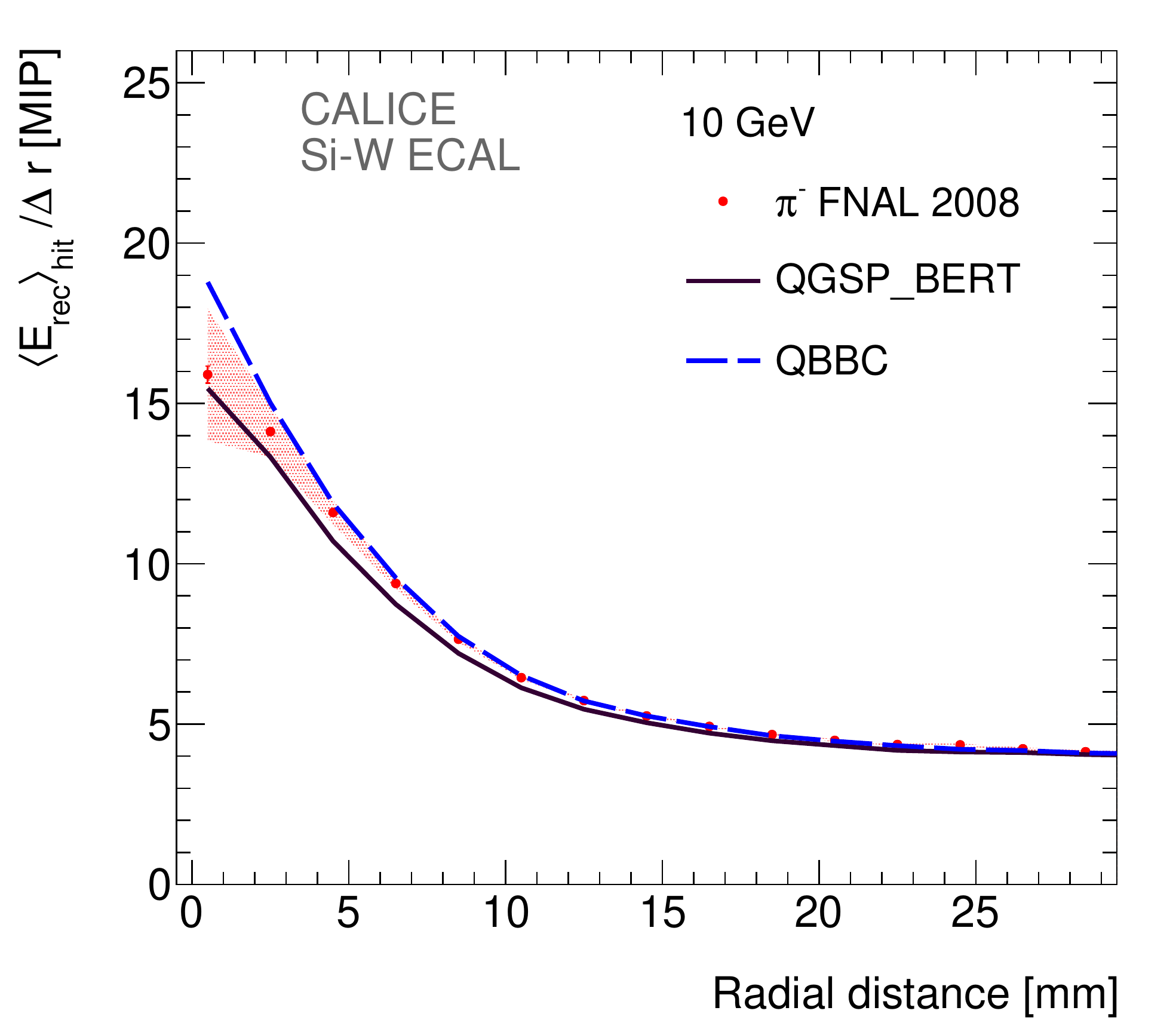}}
    \caption{\sl The radial  mean hit energy for interacting events at 2, 6 and 10\,GeV, for data and the Monte Carlo physics lists {\sc qgsp\_bert} and {\sc qbbc}. $\Delta r$ is 2 mm.}
    \label{figure:radialmeanhitenergyqgsp}
  }
\end{figure}

Only at small radii do the physics lists {\sc ftfp\_bert}, {\sc ftfp\_bert\_hp} and {\sc qbbc} have a higher mean hit energy for higher energies, as can be seen in Fig.~\ref{figure:radialmeanhitenergyftfp} and \ref{figure:radialmeanhitenergyqgsp}.
A higher energy can also be seen in Fig.~\ref{figure:radialprofileftfp} for small radii at 6 and 10\,GeV.
In Fig.~\ref{figure:radialmeanhitenergyftfp} the mean energy per hit in {\sc ftfp\_bert} and {\sc ftfp\_bert\_hp} are compared to the data.
This comparison suggests that too much energy is deposited close to the shower axis in the Fritiof model.
The effect is smaller for {\sc qbbc} and especially {\sc qgsp\_bert}, as can be seen in Fig.~\ref{figure:radialmeanhitenergyqgsp}.
At 10\,GeV {\sc qgsp\_bert} even slightly underestimates, due to the admixture of the Low Energy Parametrized model.

\begin{figure}[h]
  {\centering
    \subfloat[]{\includegraphics[width=0.5\textwidth]{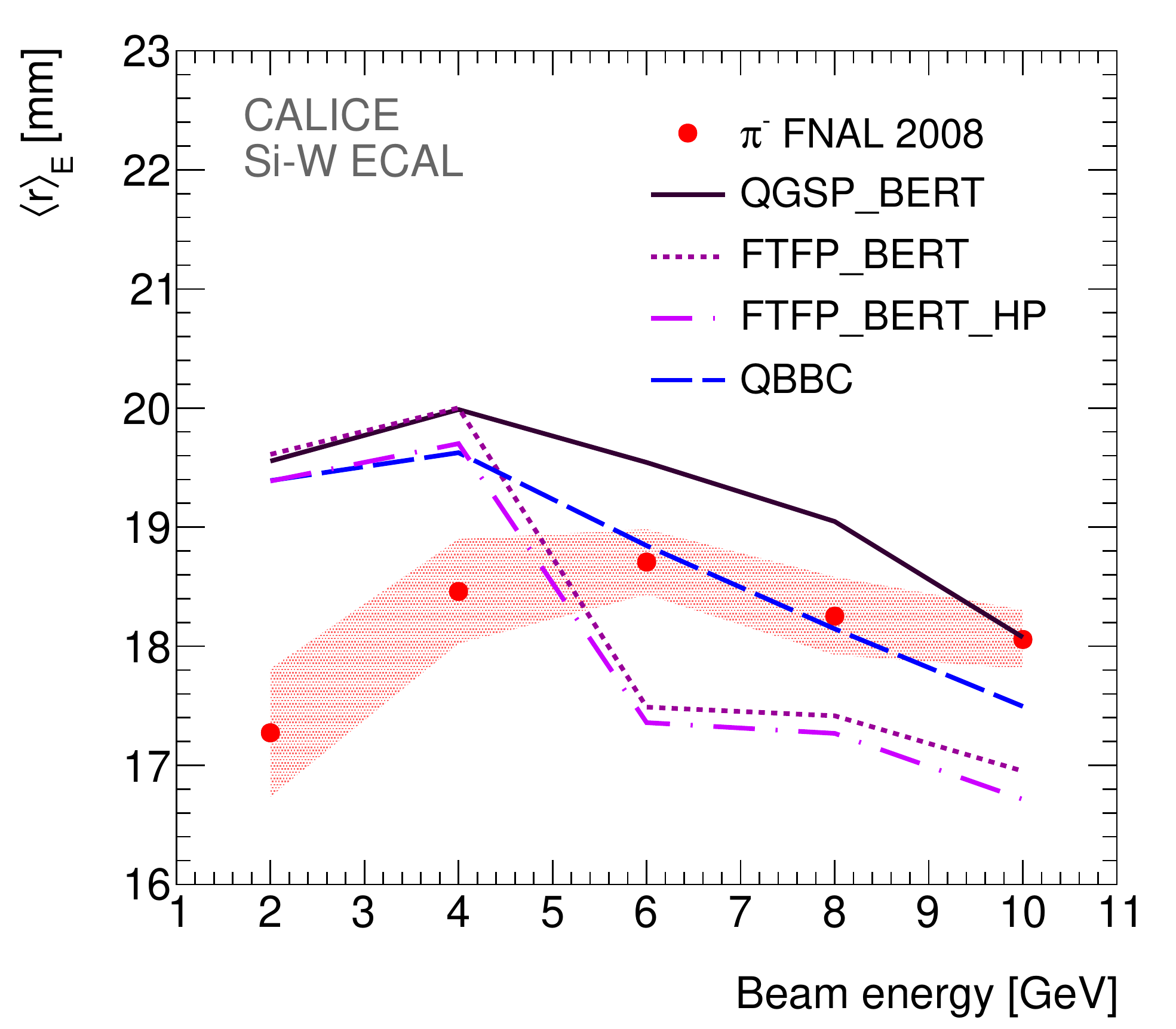}}
    \subfloat[]{\includegraphics[width=0.5\textwidth]{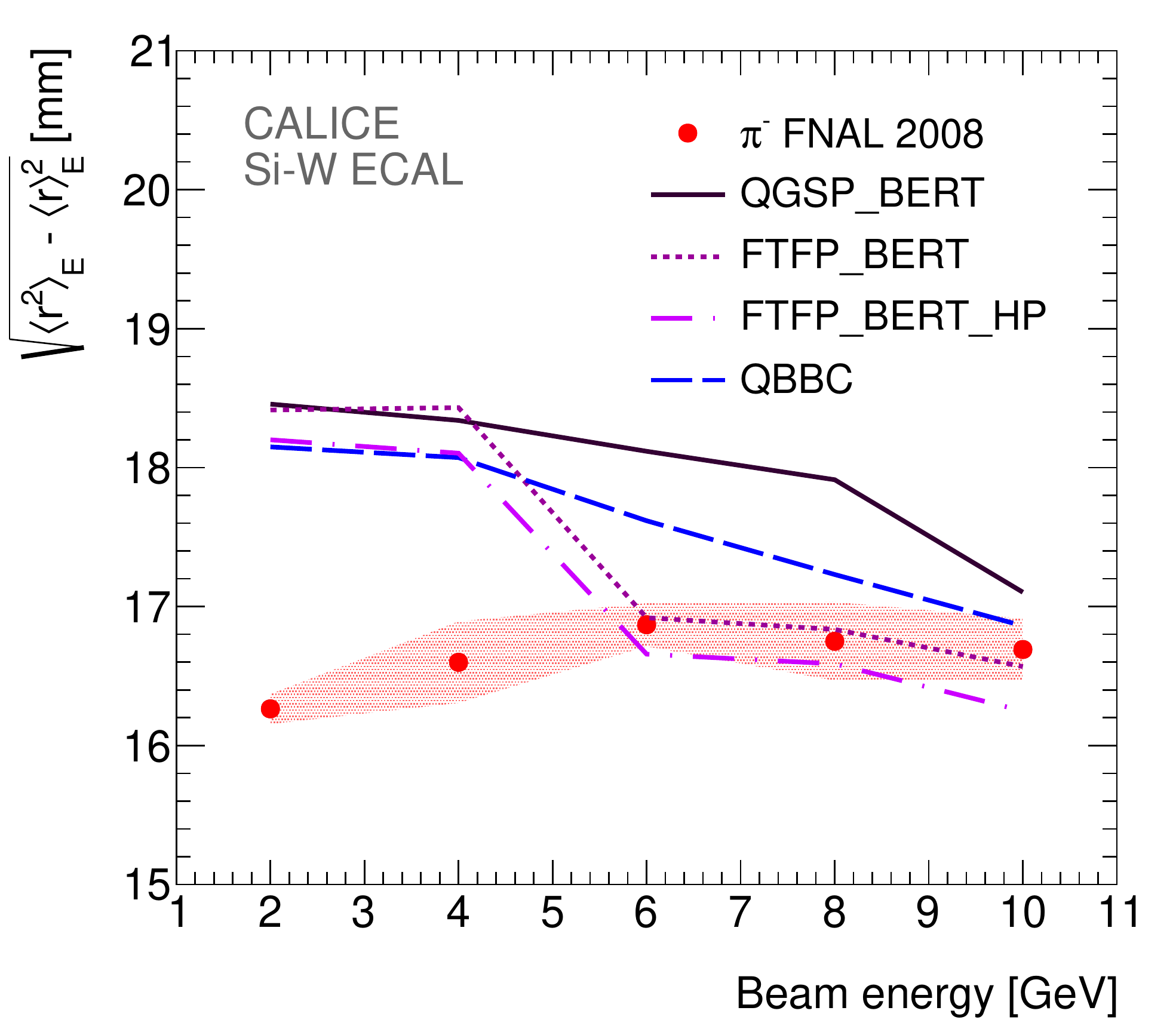}}
    \caption{\sl Mean (a) and standard deviation (b) of the radial energy profile for interacting events as a function of beam energy (2\,GeV to 10\,GeV) for data and various Monte Carlo physics lists.}
    \label{figure:meansigmaradialprofile}
  }
\end{figure}

Figure~\ref{figure:meansigmaradialprofile} shows the mean and standard deviation of the radial energy profiles as a function of the beam energy.
Again the model transition between 4 and 6\,GeV in {\sc ftfp\_bert} and {\sc ftfp\_bert\_hp} is very distinct.
The model transition in {\sc qgsp\_bert} that falls in between 8 and 10\,GeV has less influence. 
The {\sc qbbc} physics list is again in between {\sc qgsp\_bert} and {\sc ftfp\_bert}.
The Bertini cascade model generates too wide an energy distribution for all energies except for 10\,GeV, while the Fritiof model clearly deposits the energy too close to the shower axis, but simulates the standard deviation better.
The {\sc qbbc} physics list describes the mean best where it combines the Bertini and Fritiof models.

\subsection{Longitudinal shower distributions}

\begin{figure}[h]
  {\centering
    \subfloat[]{\includegraphics[width=0.33\textwidth]{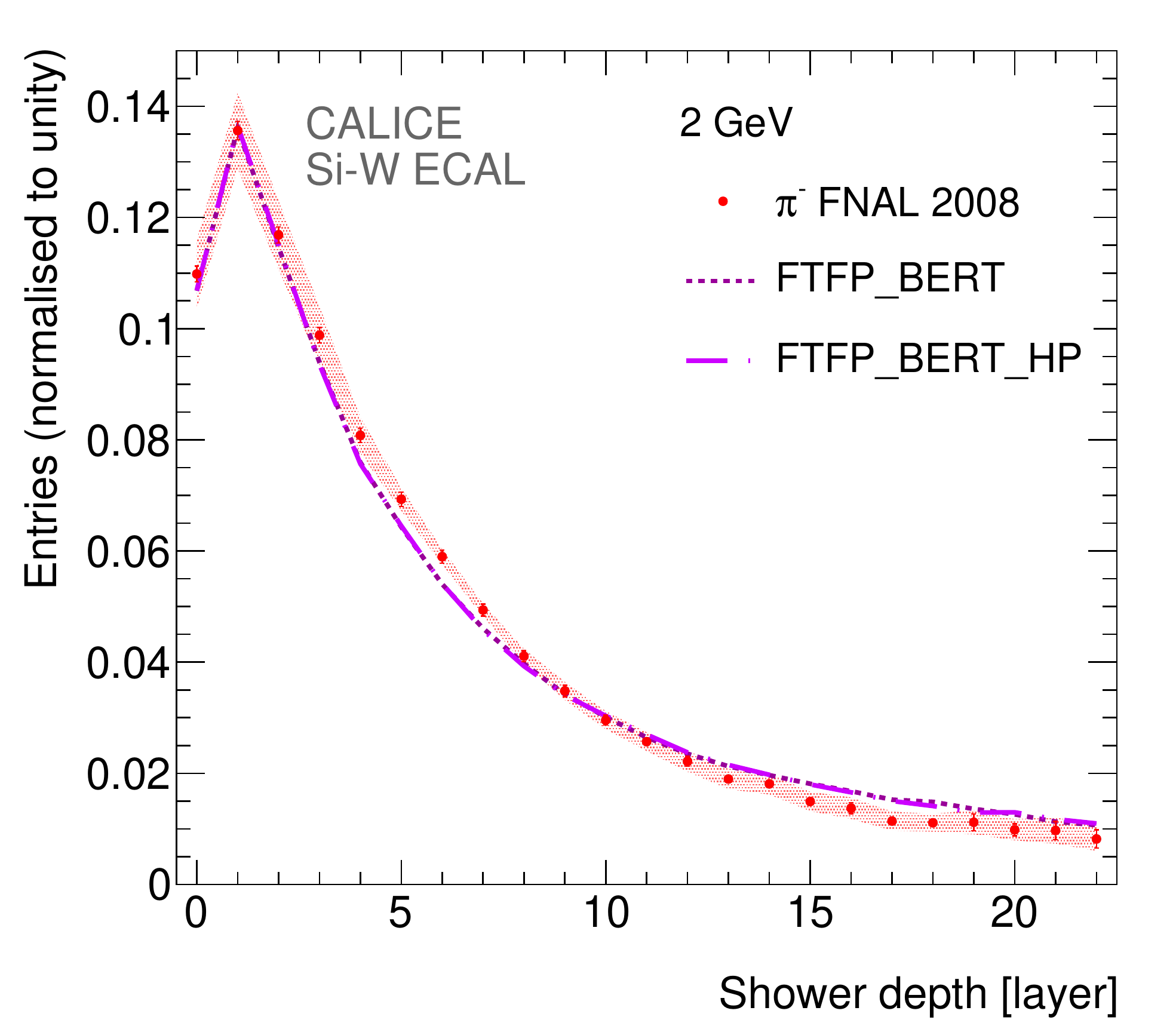}}
    \subfloat[]{\includegraphics[width=0.33\textwidth]{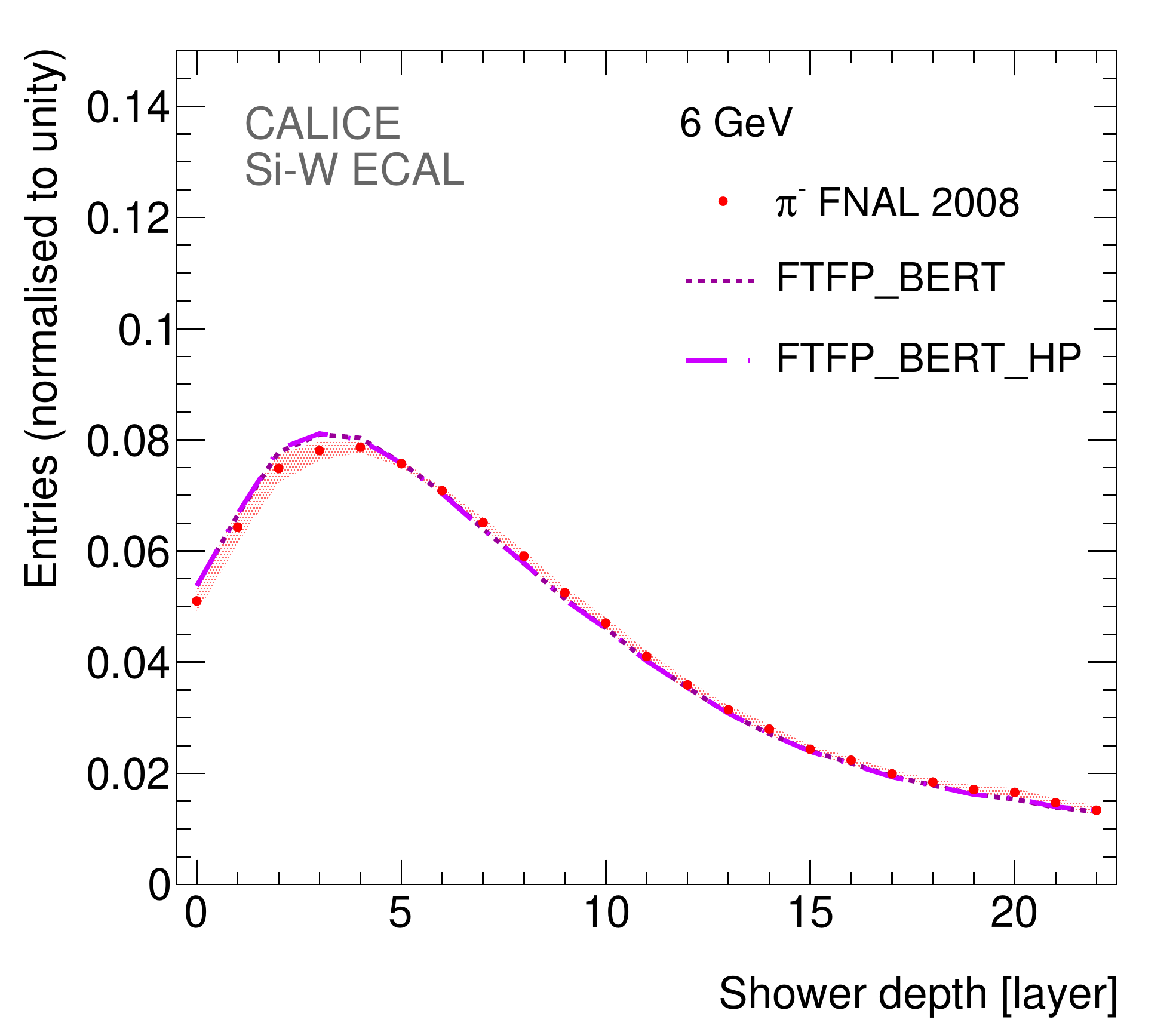}}
    \subfloat[]{\includegraphics[width=0.33\textwidth]{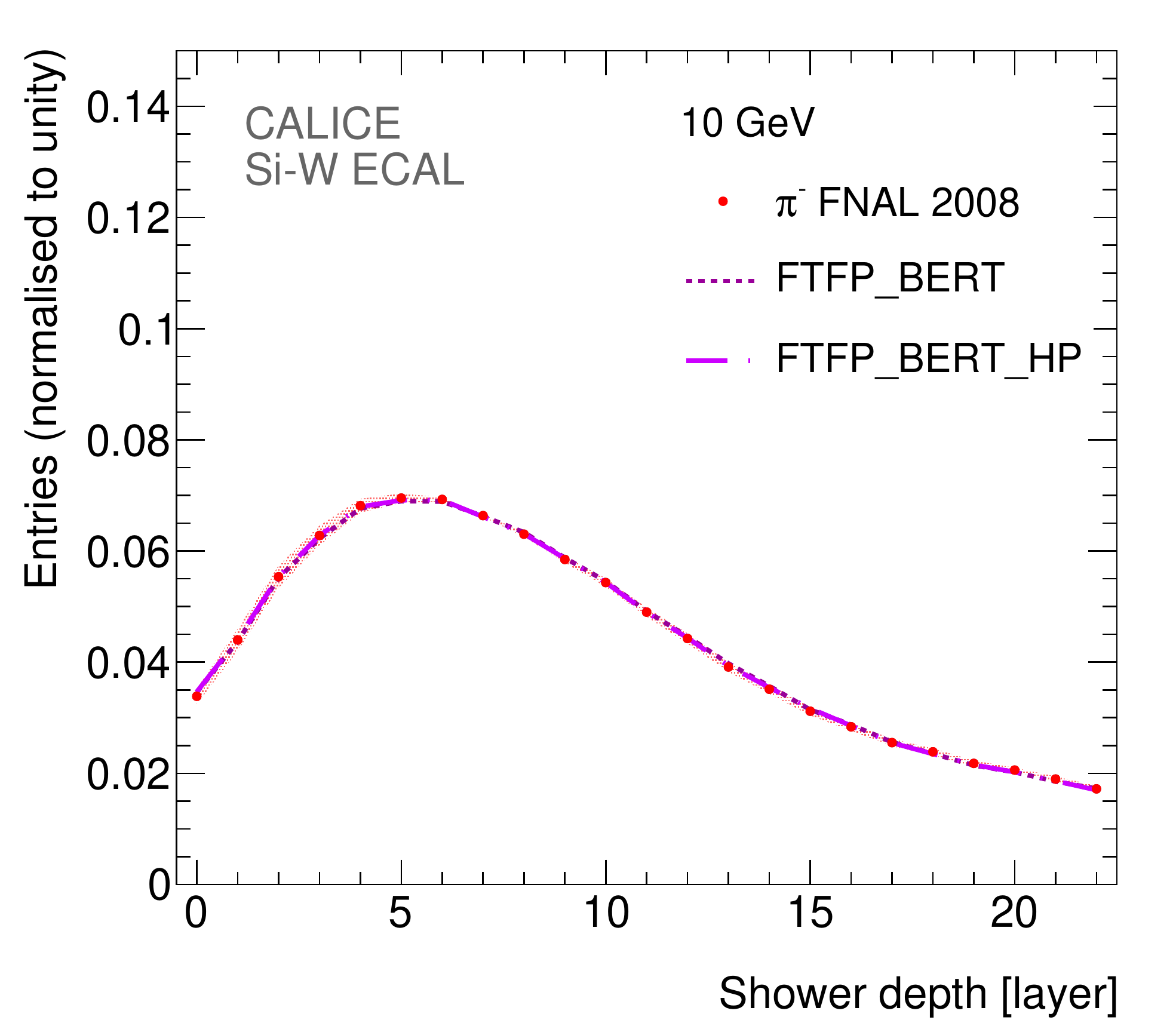}}
    \caption{\sl The longitudinal shower hit distribution for interacting events at 2, 6 and 10\,GeV, for data and the Monte Carlo physics lists {\sc ftfp\_bert} and {\sc ftfp\_bert\_hp}.}
    \label{figure:zftfp}
  }
\end{figure}

\begin{figure}[h]
  {\centering
    \subfloat[]{\includegraphics[width=0.33\textwidth]{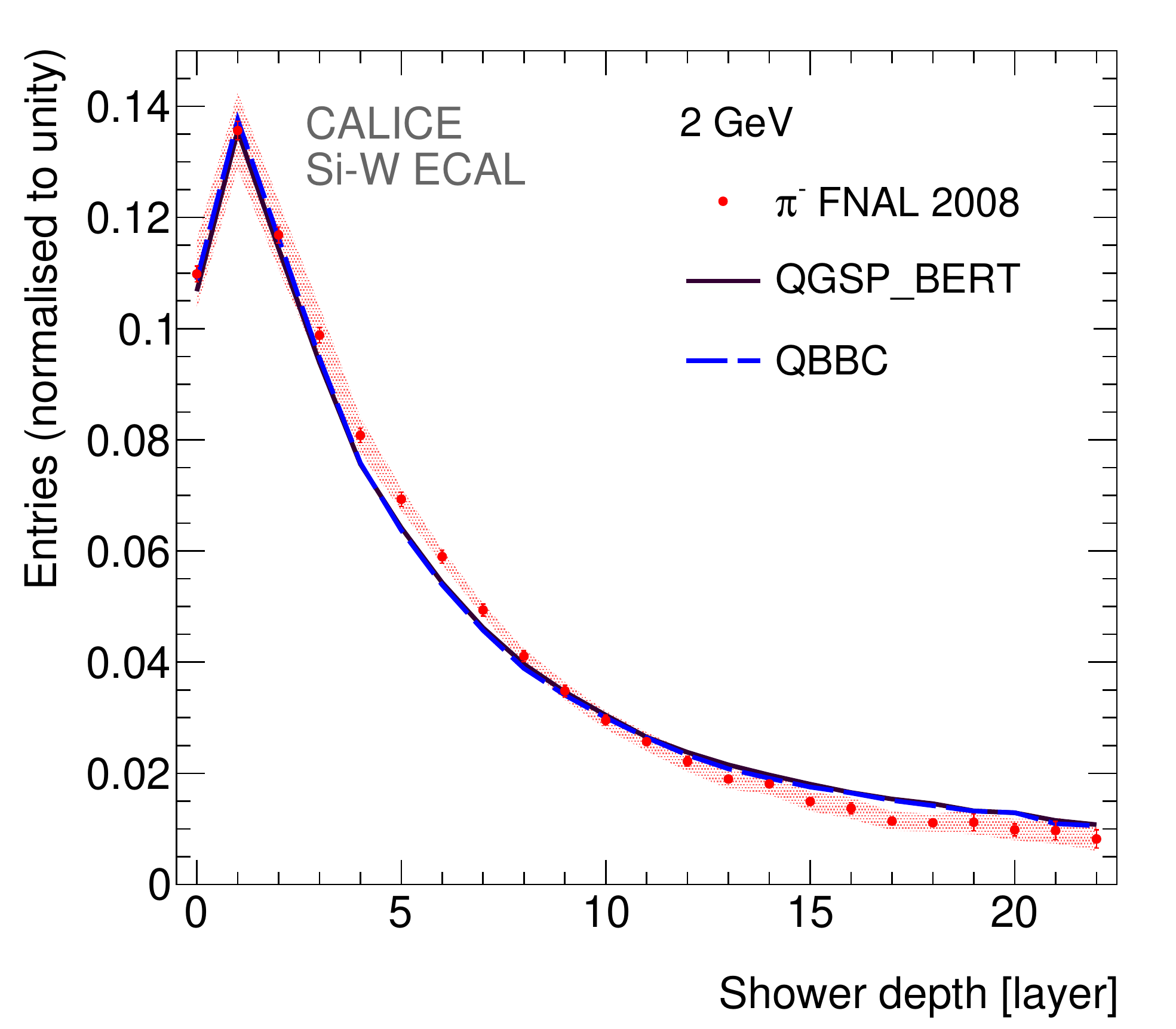}}
    \subfloat[]{\includegraphics[width=0.33\textwidth]{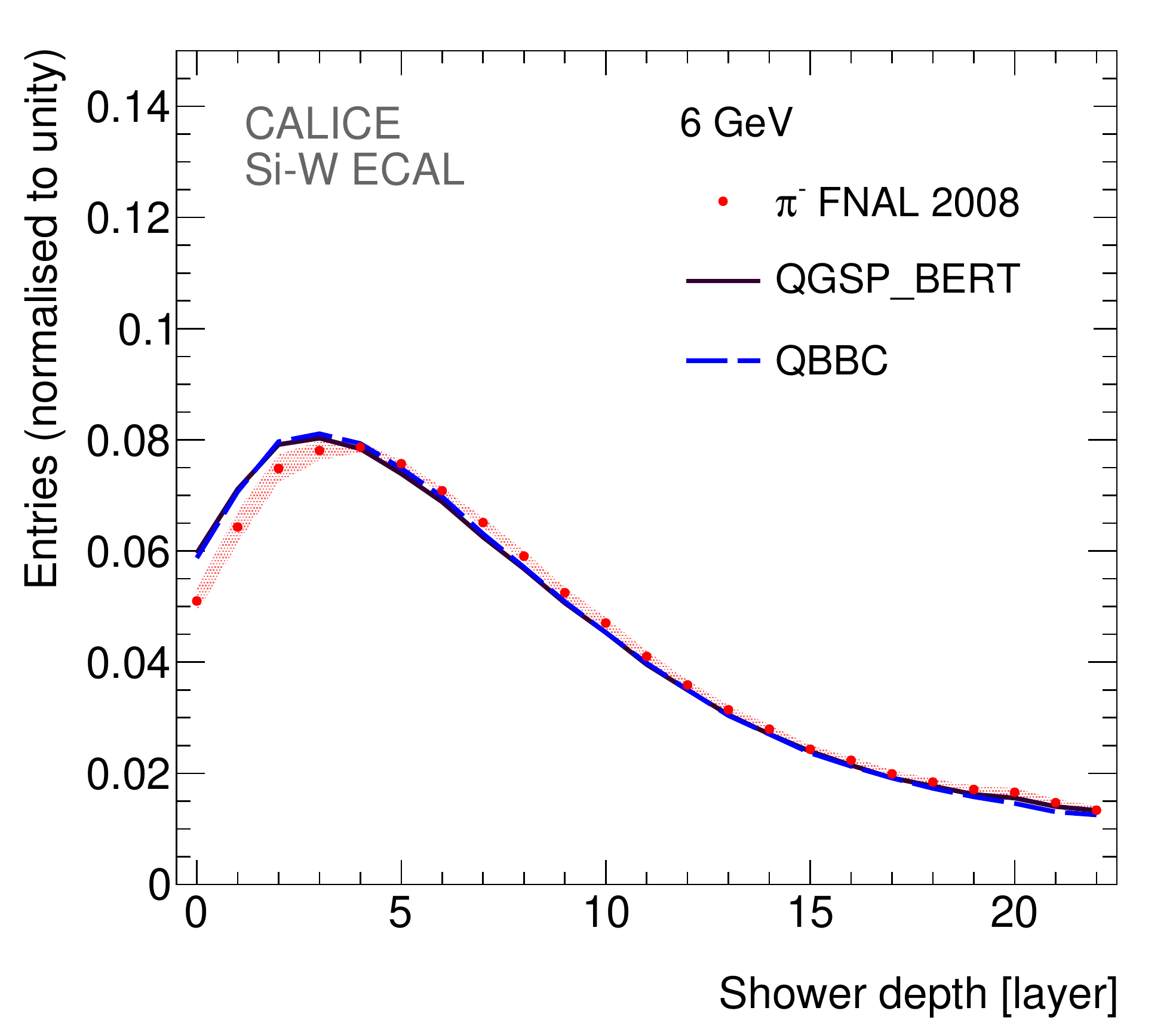}}
    \subfloat[]{\includegraphics[width=0.33\textwidth]{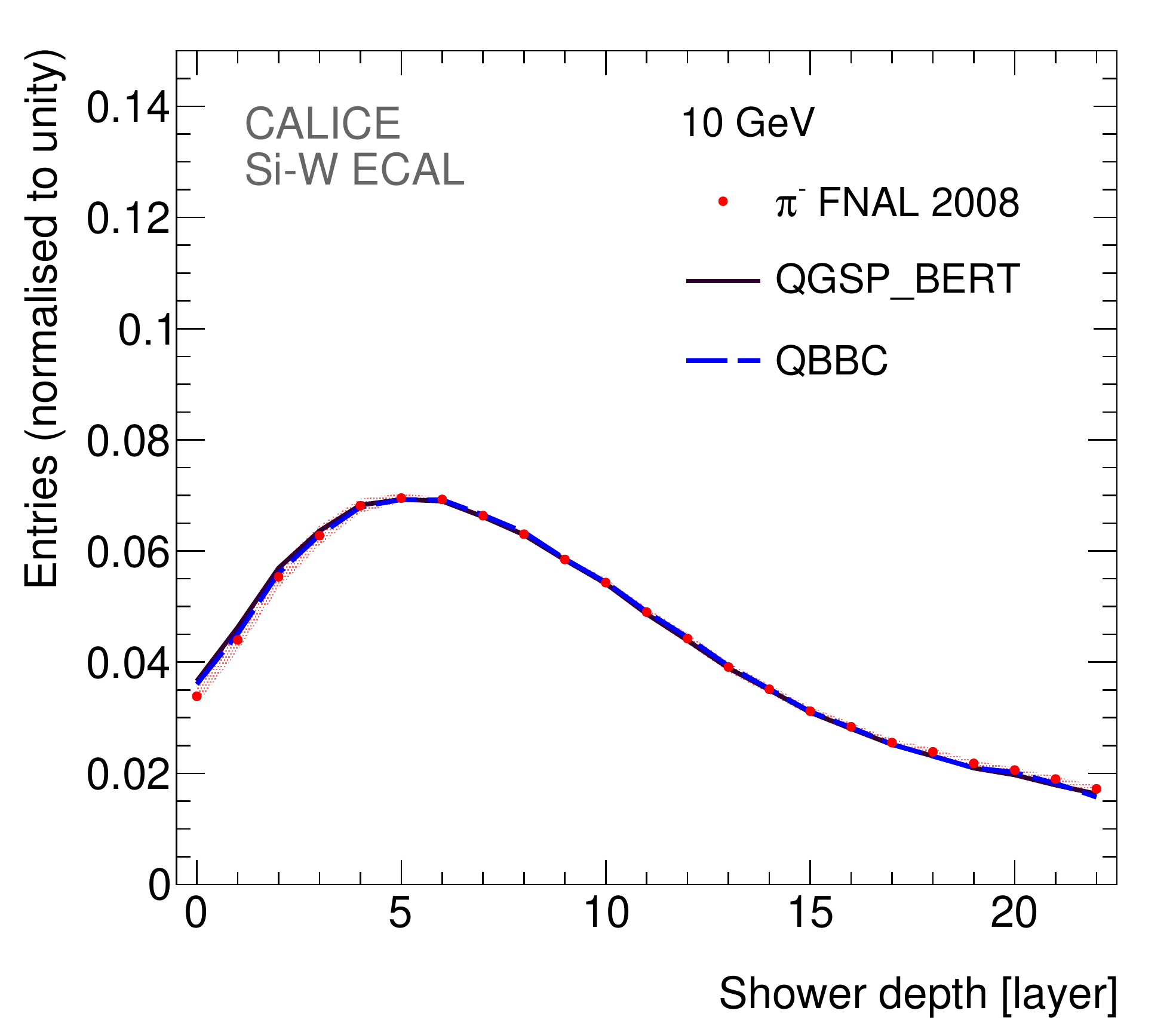}}
    \caption{\sl The longitudinal shower hit distribution for interacting events at 2, 6 and 10\,GeV, for data and the Monte Carlo physics lists {\sc qgsp\_bert} and {\sc qbbc}.}
    \label{figure:zqgsp}
  }
\end{figure}

The next global observable considered is the longitudinal distribution of hits and that of the reconstructed energy.
Figures~\ref{figure:zftfp} and~\ref{figure:zqgsp} show the hit distribution in the shower as a function of layer number where the first layer is taken to be the identified interaction layer, so the x-axis represents the shower depth in layers.
To take into account showers which extend beyond the physical dimensions of the prototype, the average in a given bin is determined by considering only events which contribute energy in the corresponding layer.
Figure~\ref{figure:zftfp} shows the distributions at 2, 6 and 10\,GeV for the physics lists {\sc ftfp\_bert} and {\sc ftfp\_bert\_hp} compared to the data while Fig.~\ref{figure:zqgsp} shows the same for {\sc qgsp\_bert} and {\sc qbbc}.
The distributions are normalised to unity in order to compare the shape of the distributions.
The longitudinal hit distribution in showers (shower shape) is reasonably well modelled by all physics lists.
At 10\,GeV the desctription is best, while at 6\,GeV {\sc ftfp\_bert} overestimates at the peak by 4\% while {\sc qgsp\_bert} and {\sc qbbc} are too high for the first few layers by at most 16\%, at 2\,GeV the shape of the simulated distributions deviates from that of the data.

\begin{figure}[h]
  {\centering
    \subfloat[]{\includegraphics[width=0.5\textwidth]{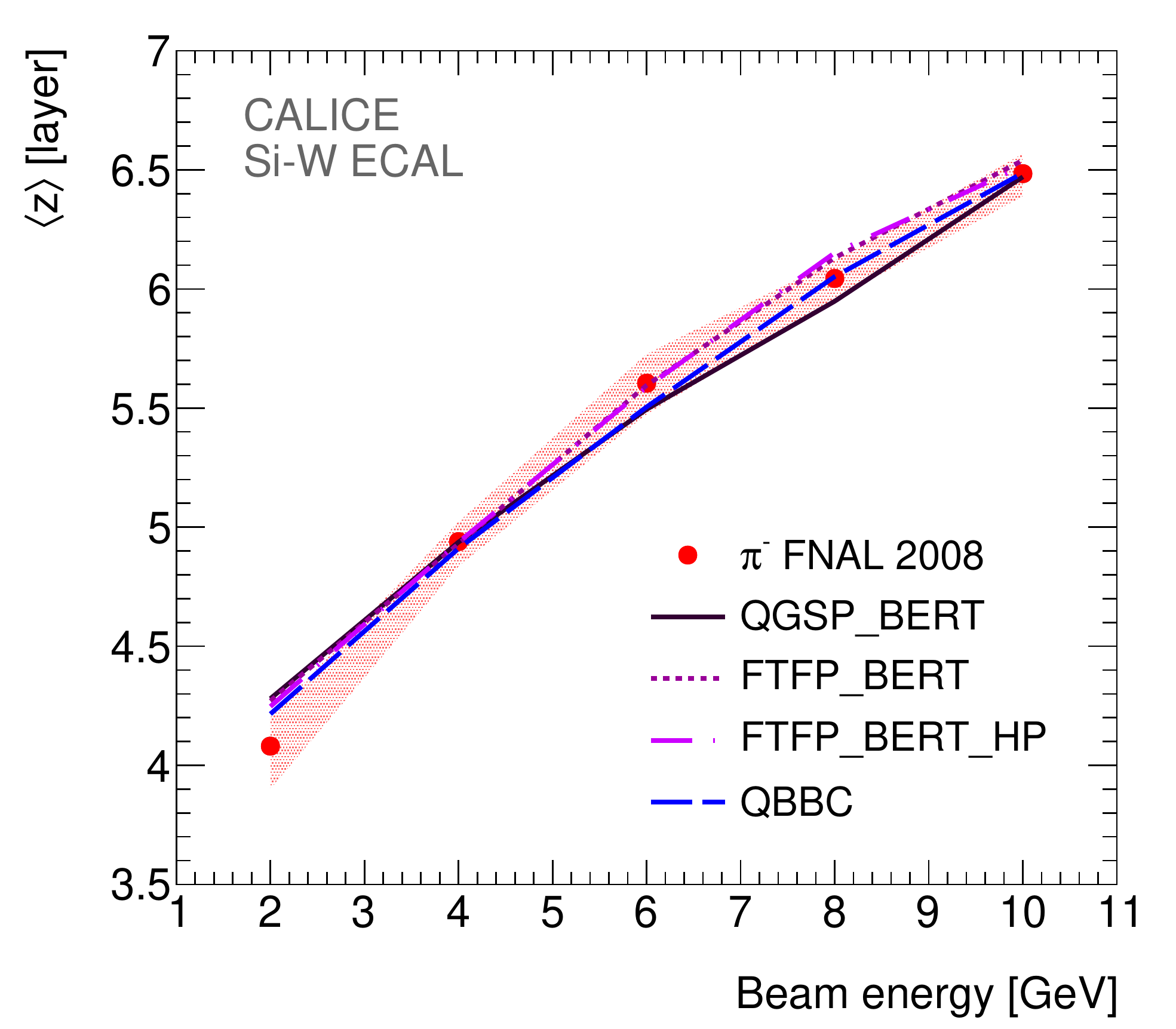}}
    \subfloat[]{\includegraphics[width=0.5\textwidth]{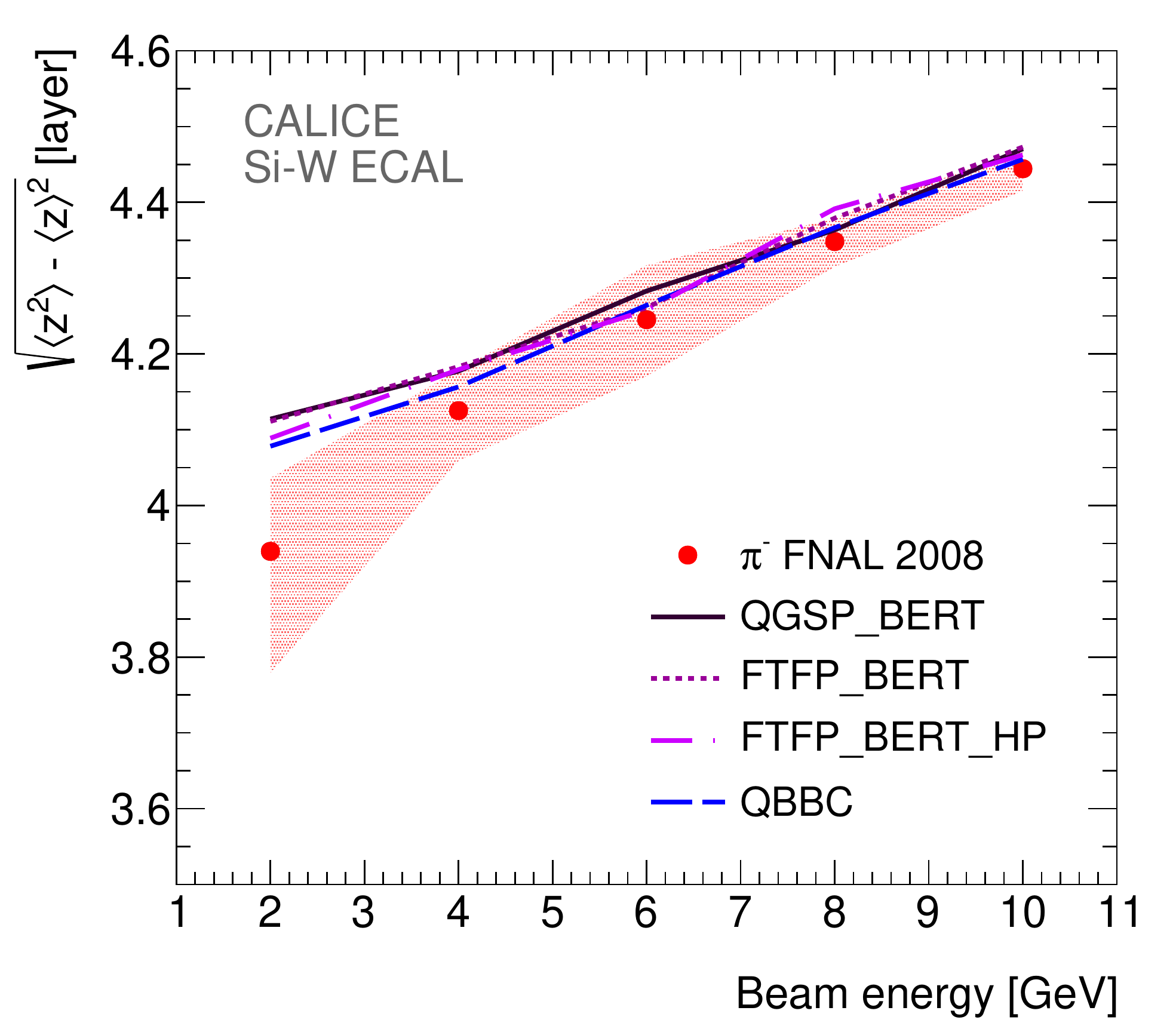}}
    \caption{\sl Mean (a) and standard deviation (b) of the longitudinal shower hit distribution for interacting events as a function of beam energy (2\,GeV to 10\,GeV) for data and various Monte Carlo physics lists.}
    \label{figure:meansigmaz}
  }
\end{figure}

Figure~\ref{figure:meansigmaz} shows the mean, $\langle z \rangle$, and standard deviation, $\sqrt{\langle z^2 \rangle - \langle z \rangle^2}$, of the longitudinal hit distribution for the data and all four physics lists.
The mean increases with beam energy and is very well described by all physics lists, the standard deviation increases less strongly and is compatible with the data except at 2\,GeV, where the data is at most 4.5\% smaller than the Monte Carlo.

\begin{figure}[h]
  {\centering
    \subfloat[]{\includegraphics[width=0.33\textwidth]{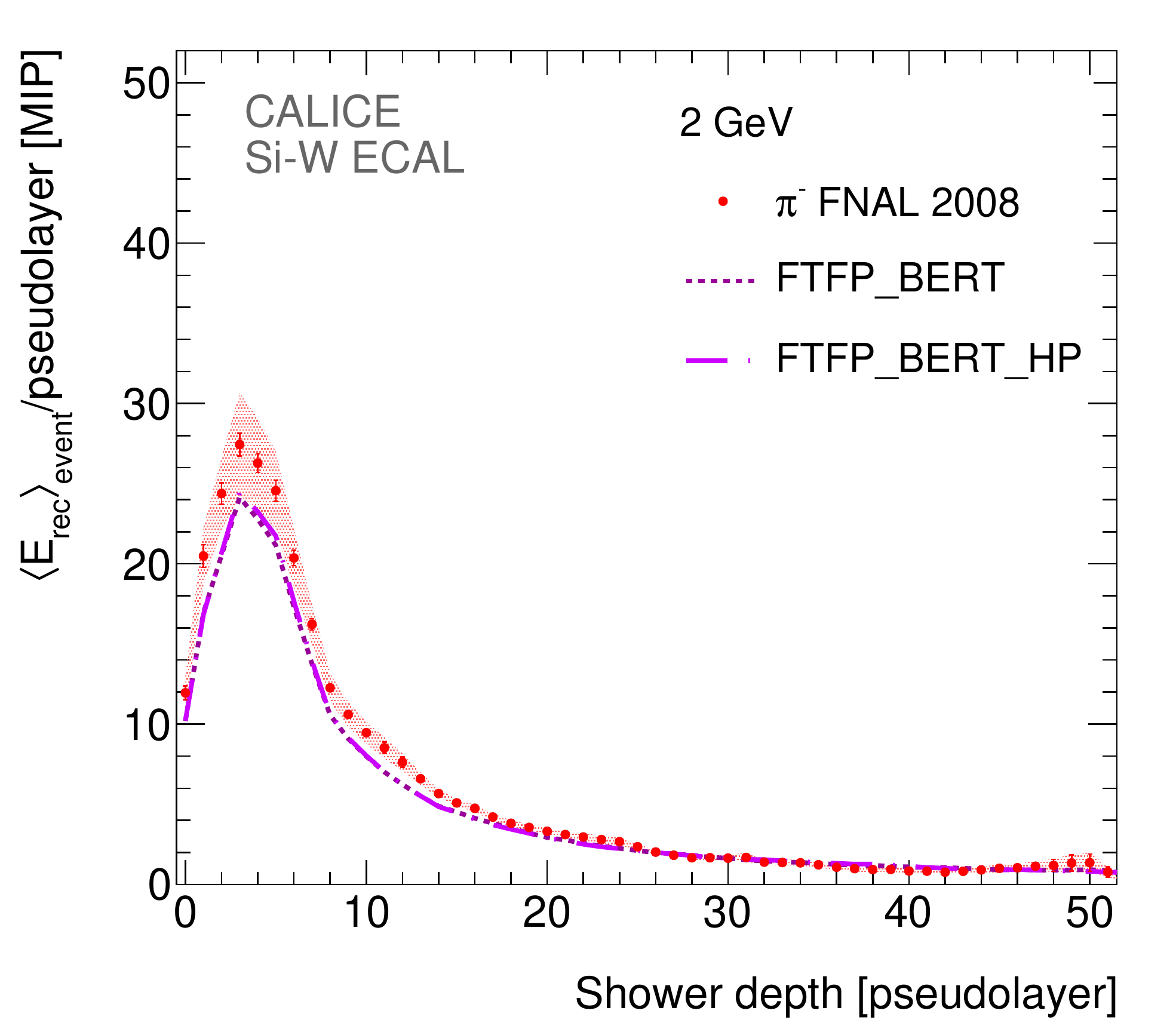}}
    \subfloat[]{\includegraphics[width=0.33\textwidth]{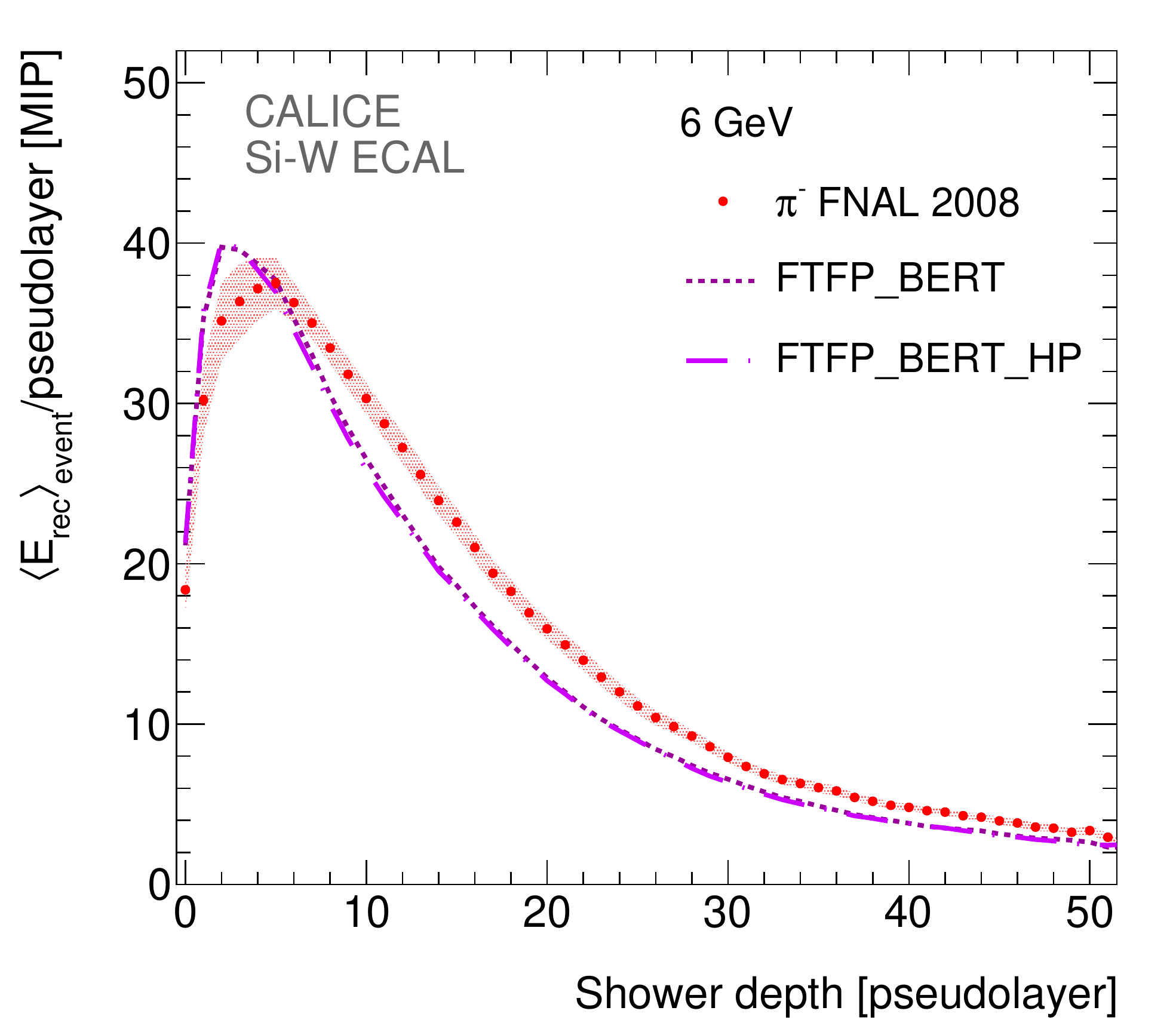}}
    \subfloat[]{\includegraphics[width=0.33\textwidth]{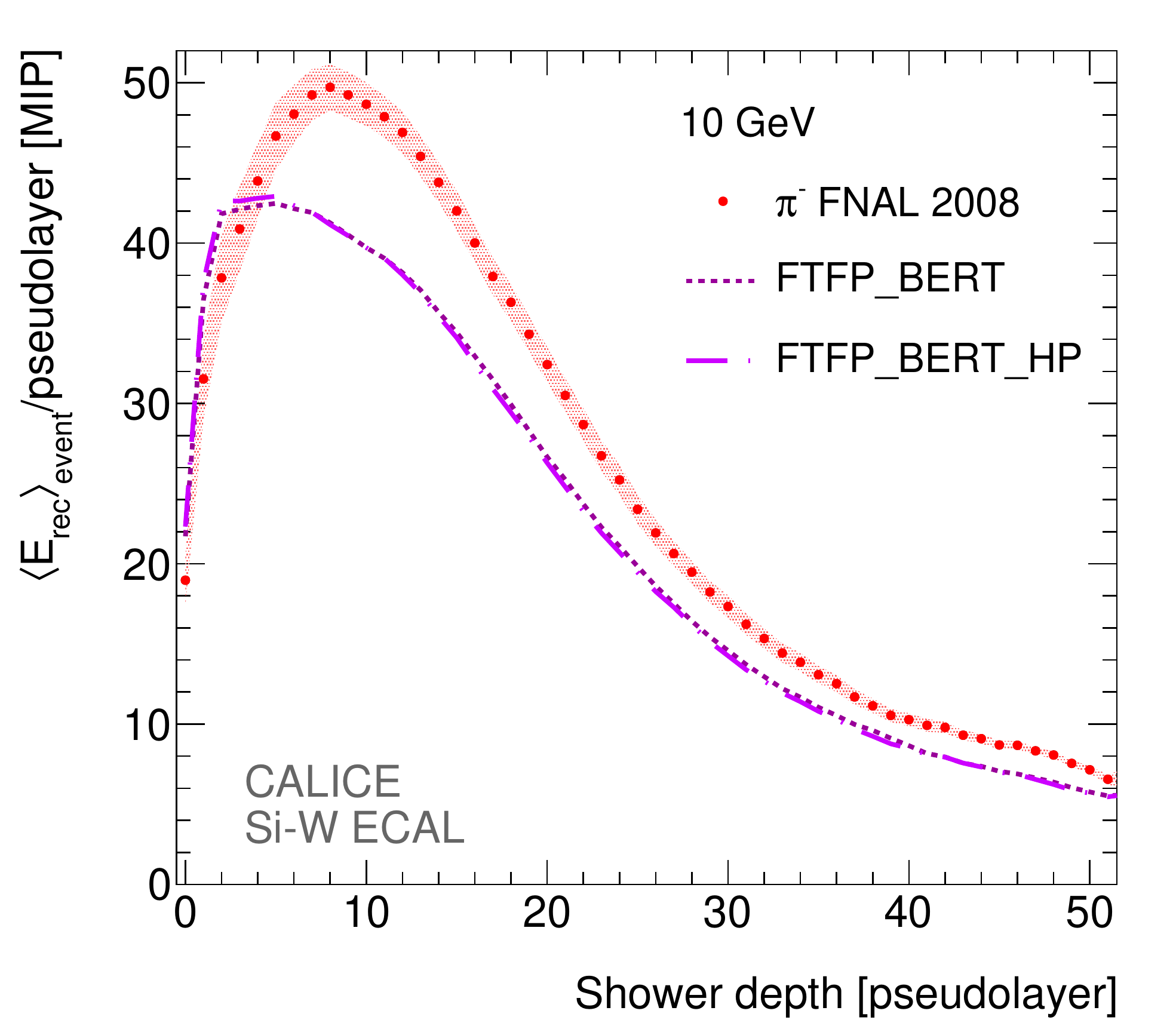}}
    \caption{\sl The longitudinal energy profile for interacting events at 2, 6 and 10\,GeV, for data and the Monte Carlo physics lists {\sc ftfp\_bert} and {\sc ftfp\_bert\_hp}.}
    \label{figure:longitudinalprofileftfp}
  }
\end{figure}

\begin{figure}[h]
  {\centering
    \subfloat[]{\includegraphics[width=0.33\textwidth]{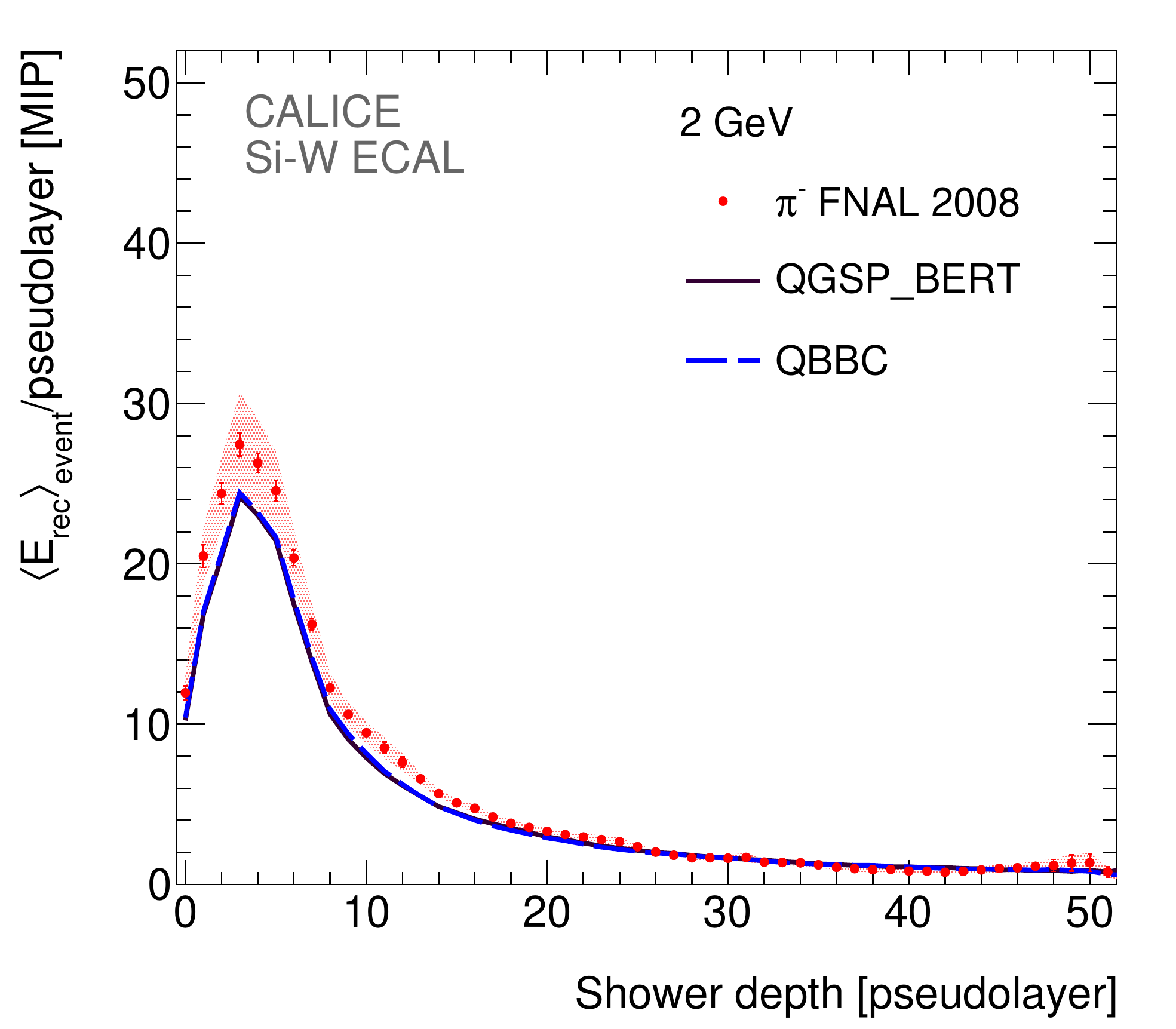}}
    \subfloat[]{\includegraphics[width=0.33\textwidth]{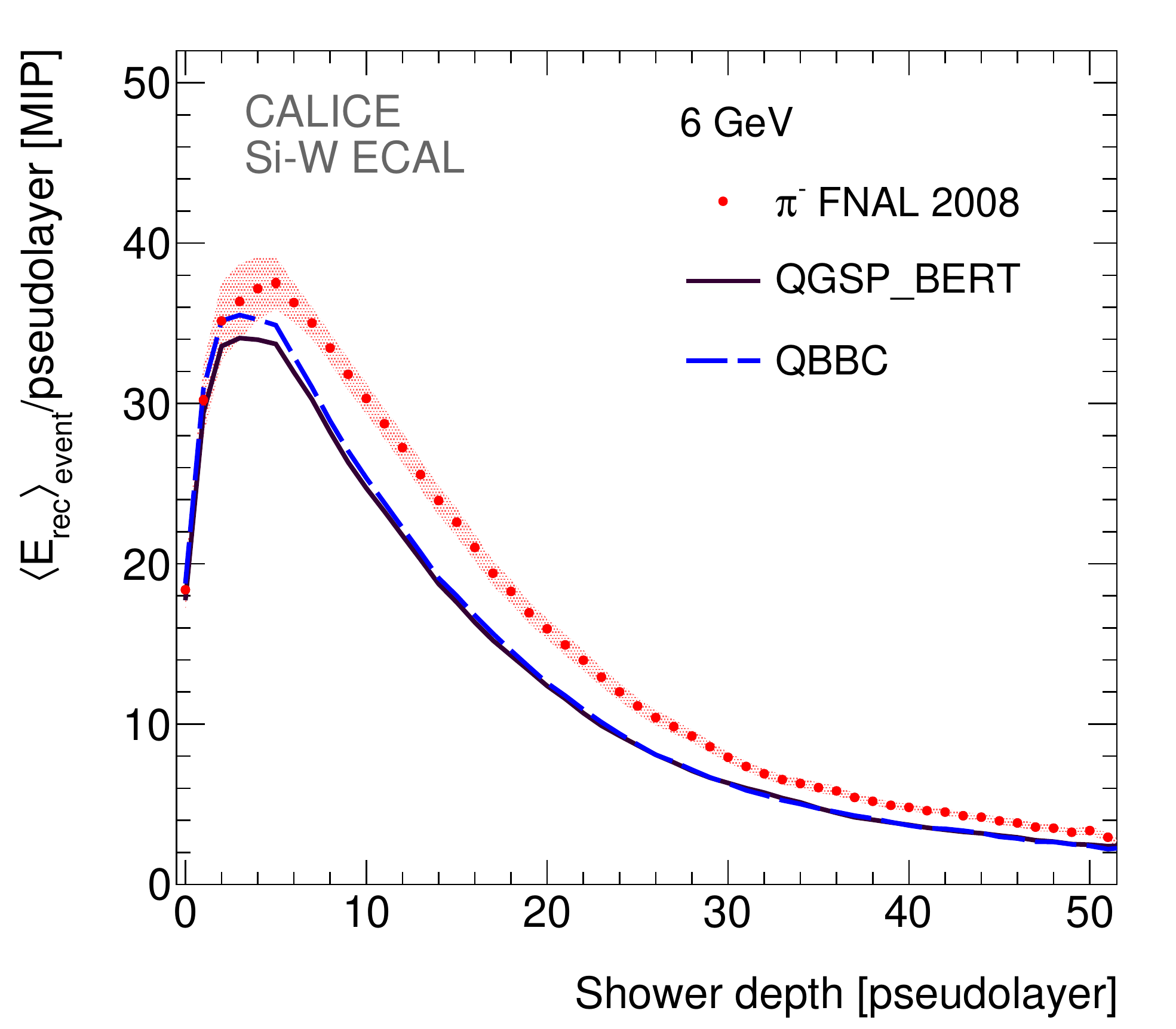}}
    \subfloat[]{\includegraphics[width=0.33\textwidth]{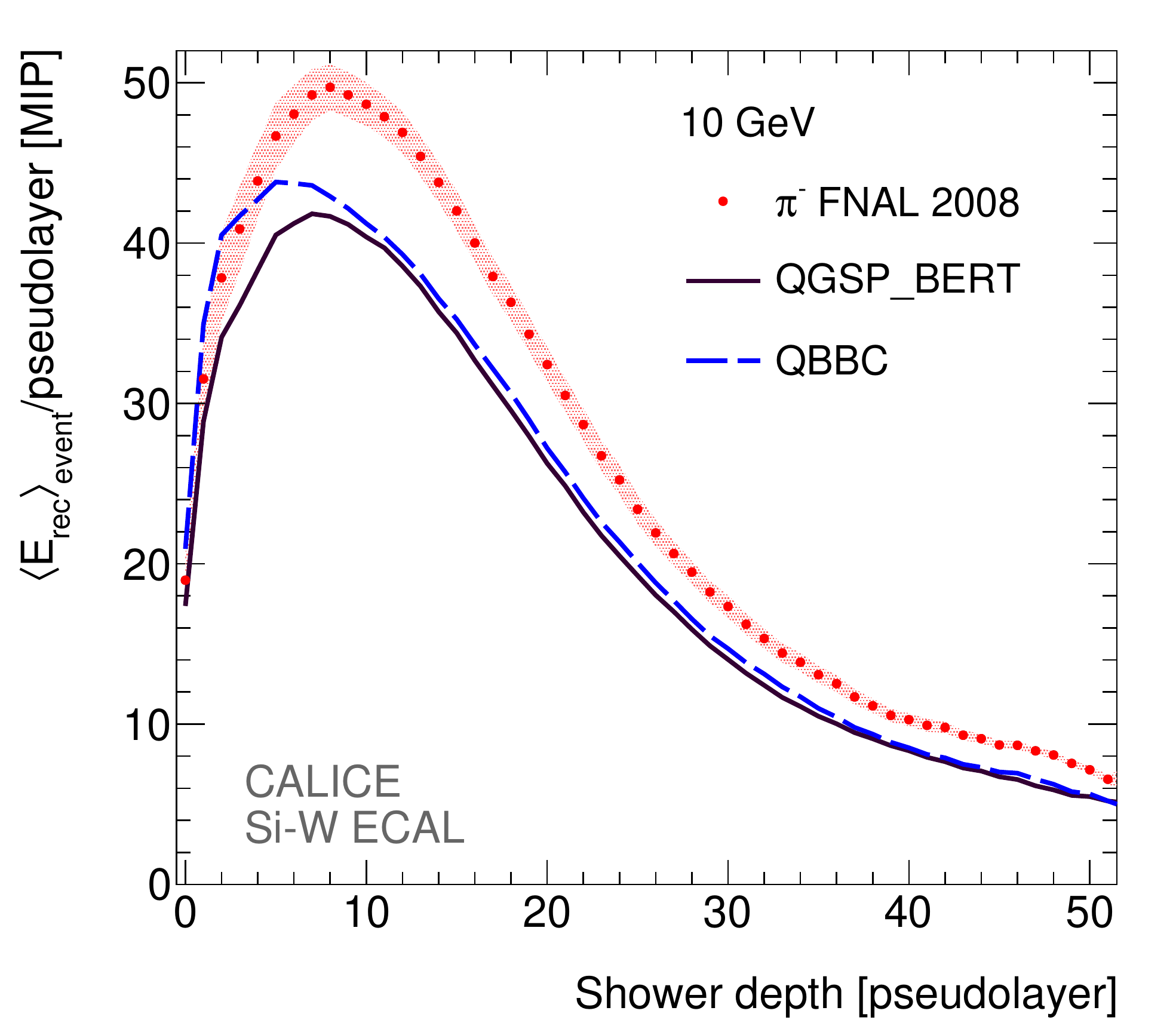}}
    \caption{\sl The longitudinal energy profile for interacting events at 2, 6 and 10\,GeV, for data and the Monte Carlo physics lists {\sc qgsp\_bert} and {\sc qbbc}.}
    \label{figure:longitudinalprofileqgsp}
  }
\end{figure}

The longitudinal energy profiles are defined as introduced in~\cite{2010_CALICE} and also start from the reconstructed interaction layer. 
They give the energy in MIPs per {\em pseudolayer}. 
Pseudolayers are introduced in order to account for the different sampling fractions in the Si-W ECAL. 
There is a one to one correspondence between physical layers and pseudolayers in the first module, 
while each layer in the second module has been subdivided in two pseudolayers 
and those in the third module have been subdivided into three pseudolayers.
The energy in the added pseudolayers is calculated by interpolating between the reconstructed energy in the considered physical layer and the reconstructed energy in the previous physical layer.
Figures~\ref{figure:longitudinalprofileftfp} and \ref{figure:longitudinalprofileqgsp} show the longitudinal energy profiles for 2, 6 and 10\,GeV. 
The Monte Carlo physics lists are again divided over the two figures. 
The profiles are averaged for each bin separately by considering only events which have contributed energy in the corresponding pseudolayer, in order to reduce the influence of showers which extend beyond the physical dimensions of the prototype.
   
\begin{figure}[h]
  {\centering
    \subfloat[]{\includegraphics[width=0.33\textwidth]{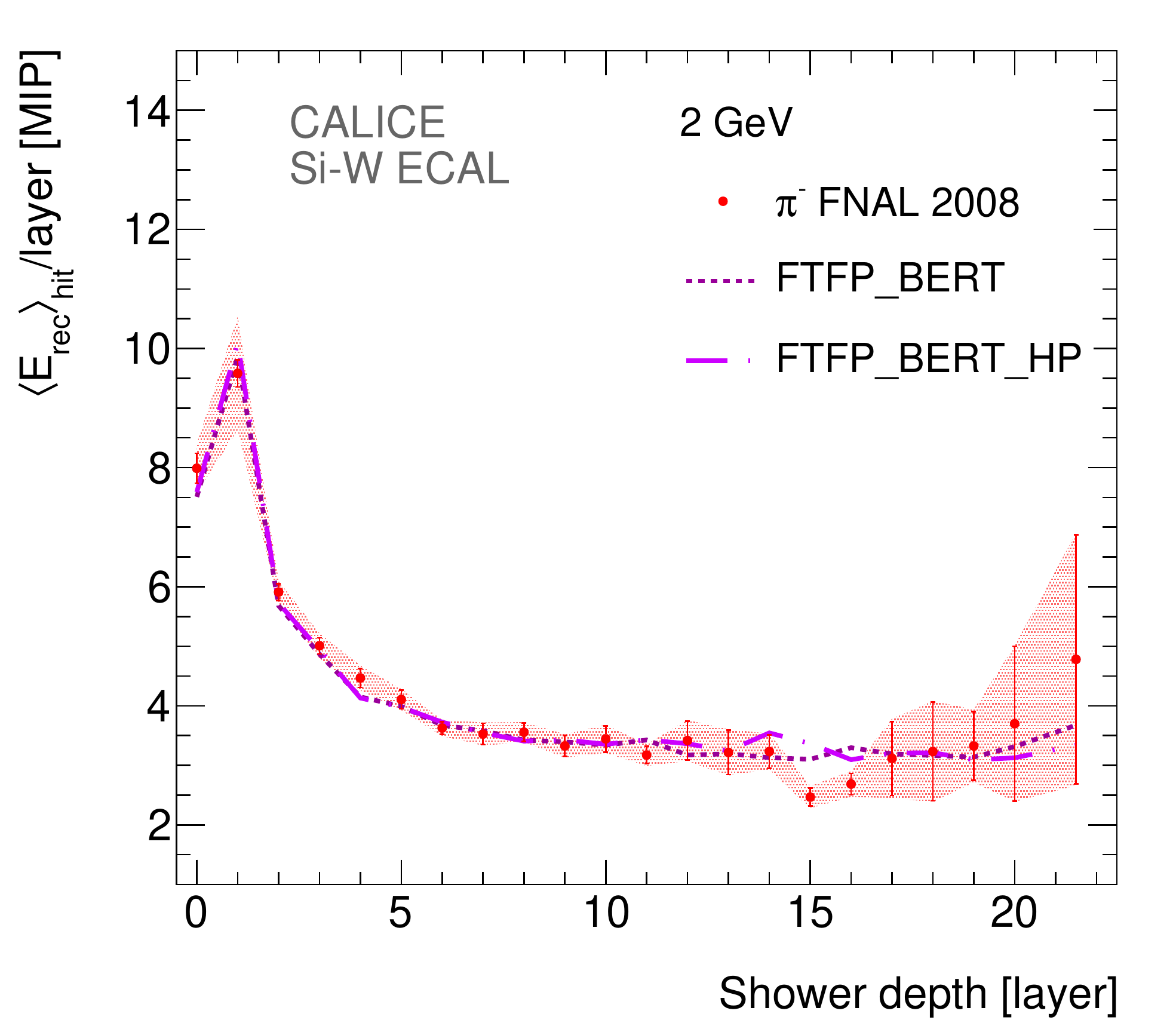}}
    \subfloat[]{\includegraphics[width=0.33\textwidth]{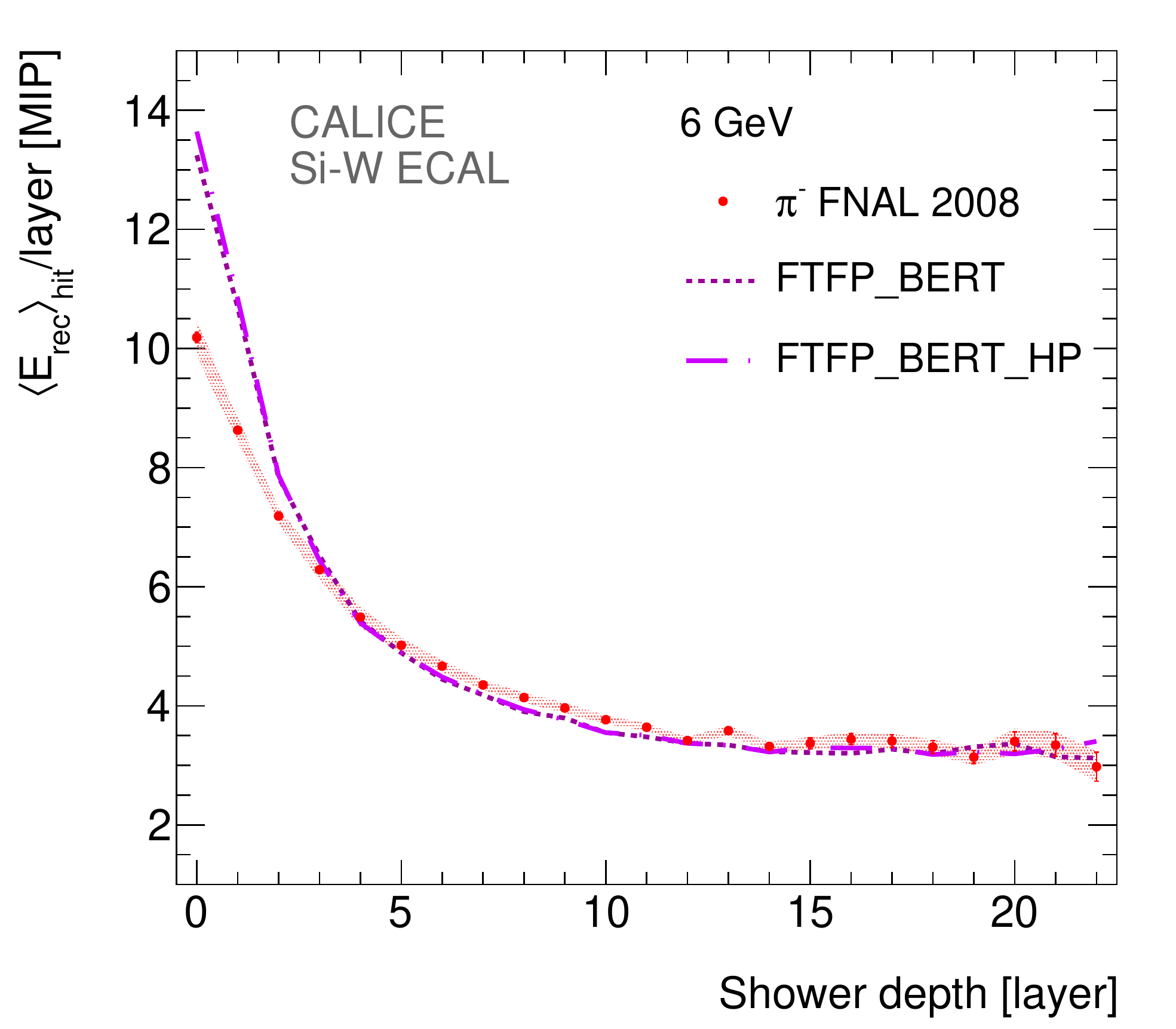}}
    \subfloat[]{\includegraphics[width=0.33\textwidth]{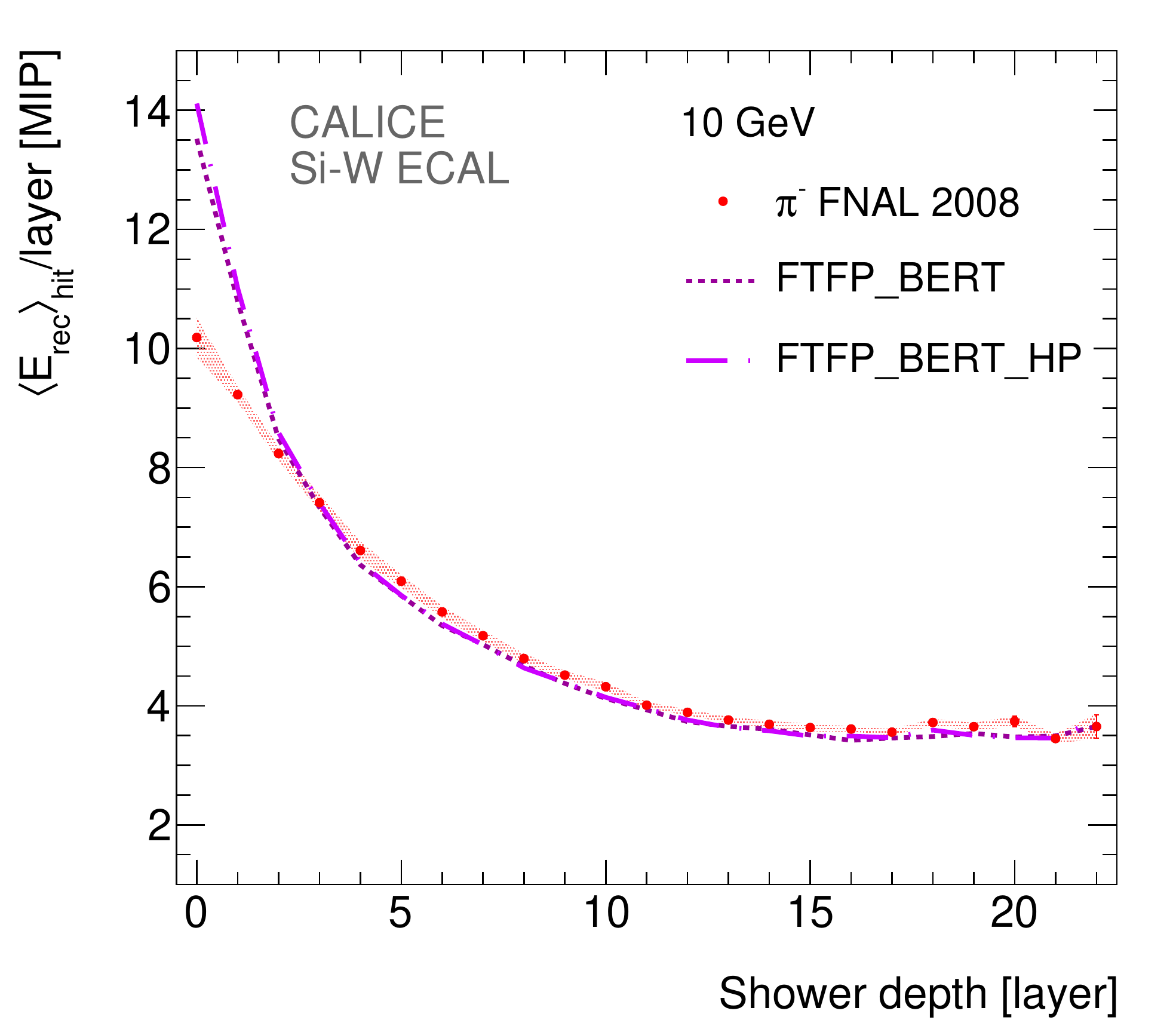}}
    \caption{\sl The longitudinal mean hit energy for interacting events at 2, 6 and 10\,GeV, for data and the Monte Carlo physics lists {\sc ftfp\_bert} and {\sc ftfp\_bert\_hp}. For 2\,GeV the last two layers have been combined into one data point because of their low number of entries.}
    \label{figure:longitudinalmeanhitenergyftfp}
  }
\end{figure}

\begin{figure}[h]
  {\centering
    \subfloat[]{\includegraphics[width=0.33\textwidth]{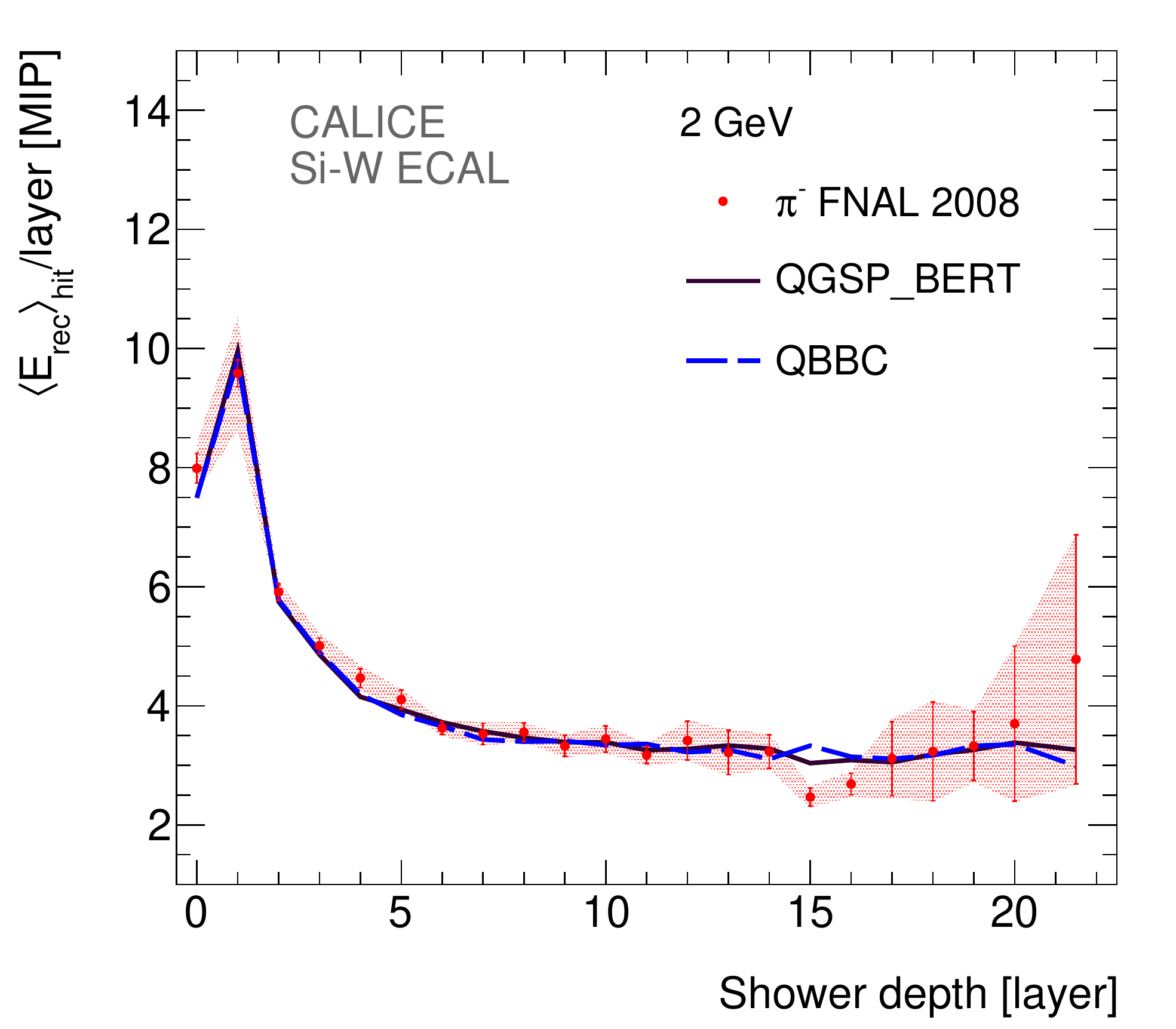}}
    \subfloat[]{\includegraphics[width=0.33\textwidth]{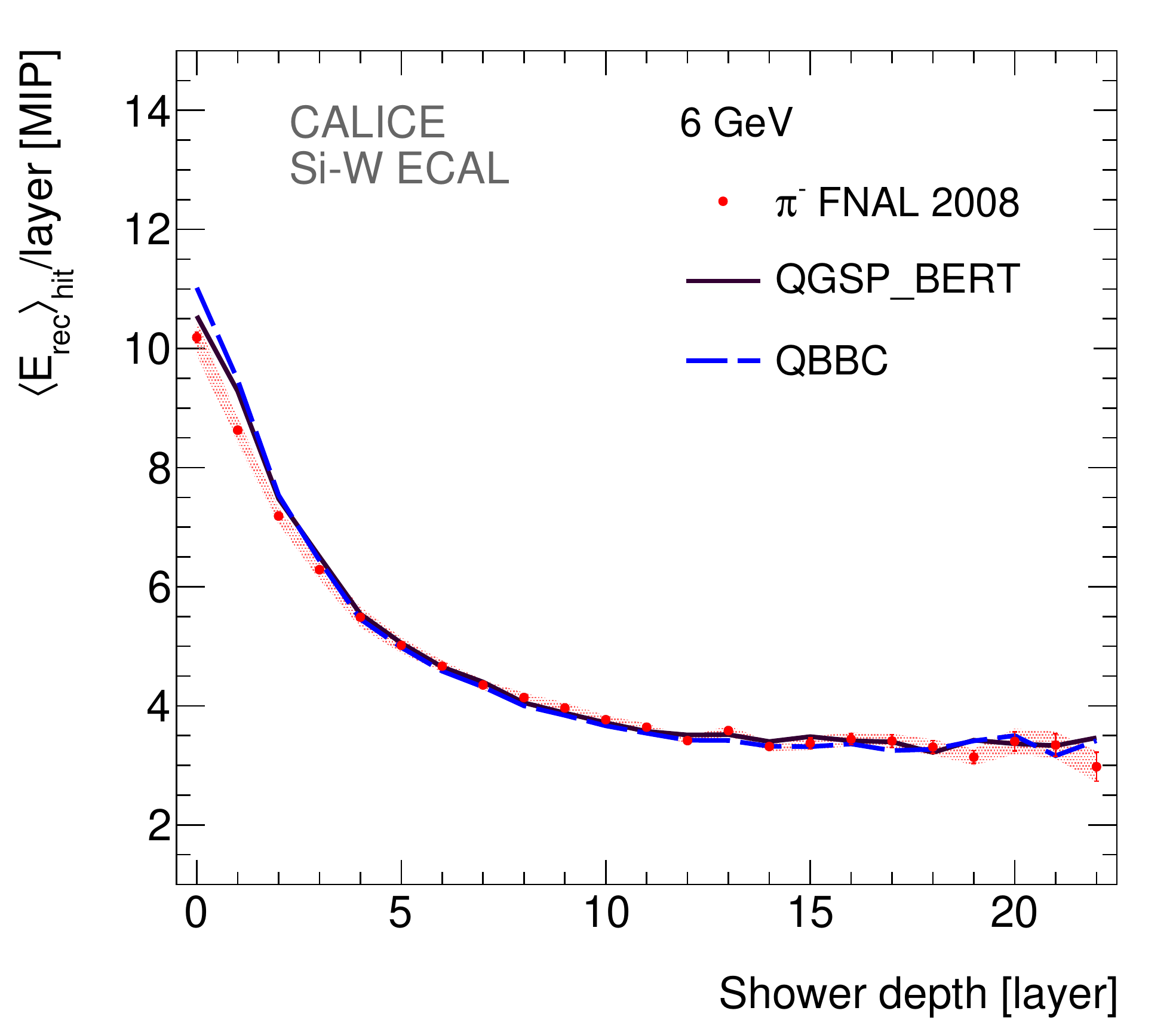}}
    \subfloat[]{\includegraphics[width=0.33\textwidth]{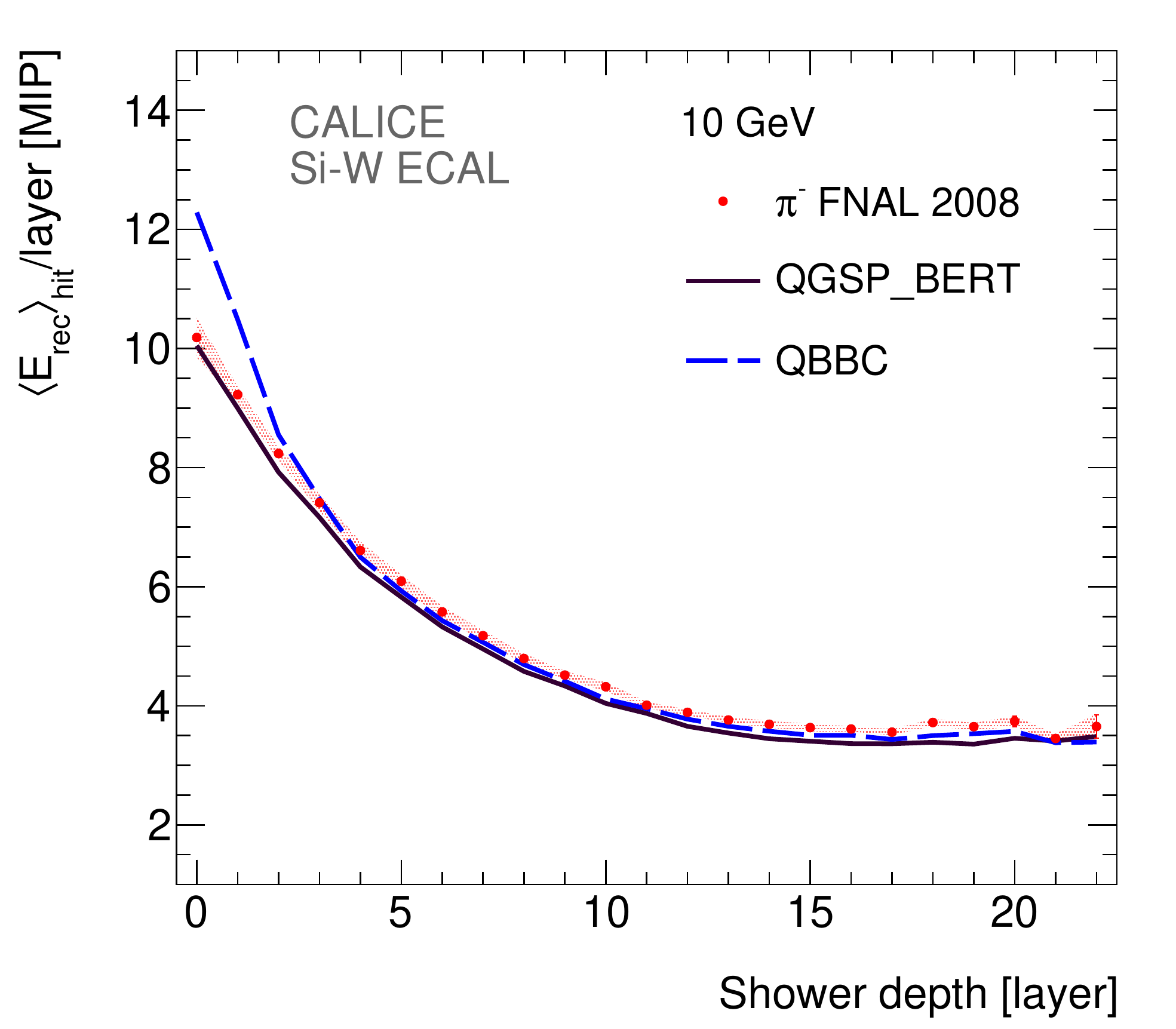}}
    \caption{\sl The longitudinal mean hit energy for interacting events at 2, 6 and 10\,GeV, for data and the Monte Carlo physics lists {\sc qgsp\_bert} and {\sc qbbc}. For 2\,GeV the last two layers have been combined into one data point because of their low number of entries.}
    \label{figure:longitudinalmeanhitenergyqgsp}
  }
\end{figure}

The longitudinal energy profile descriptions are progressively worse with increasing energy and overall the energy deposition is underestimated.
Just like for the radial distributions, the mean energy per hit is similar in data and simulations, which can be seen in Fig.~\ref{figure:longitudinalmeanhitenergyftfp} and ~\ref{figure:longitudinalmeanhitenergyqgsp}.
These figures show the mean energy per hit for each physical layer in the shower.
While at higher energies the mean hit energy in the data is a little higher than in the Monte Carlo, this does not explain the deficit in the deposited energy as seen in Fig.~\ref{figure:longitudinalprofileftfp} and~\ref{figure:longitudinalprofileqgsp}.
This means the lower energy in the simulations can be attributed to a lower number of hits.
Near the shower start the mean hit energy for beam energies above 4\,GeV is overestimated in {\sc ftfp\_bert}, {\sc ftfp\_bert\_hp} and {\sc qbbc}.
This overestimation results in a small excess in the deposited energy near the shower start (Fig.~\ref{figure:longitudinalprofileftfp} and~\ref{figure:longitudinalprofileqgsp}).
Too much energy is being deposited close to the interaction layer by the Fritiof model.

\begin{figure}[h]
  {\centering
    \subfloat[]{\includegraphics[width=0.5\textwidth]{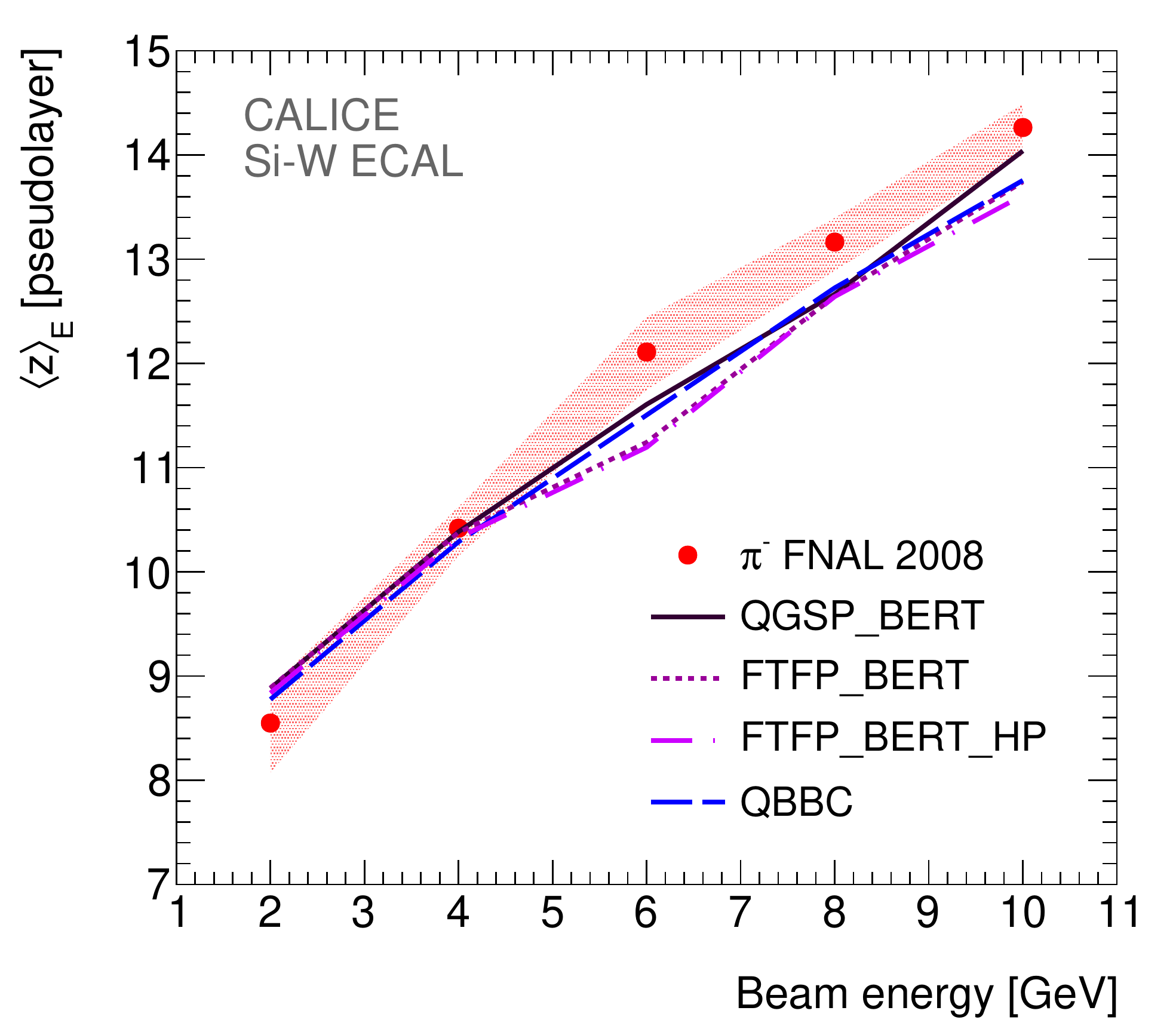}}
    \subfloat[]{\includegraphics[width=0.5\textwidth]{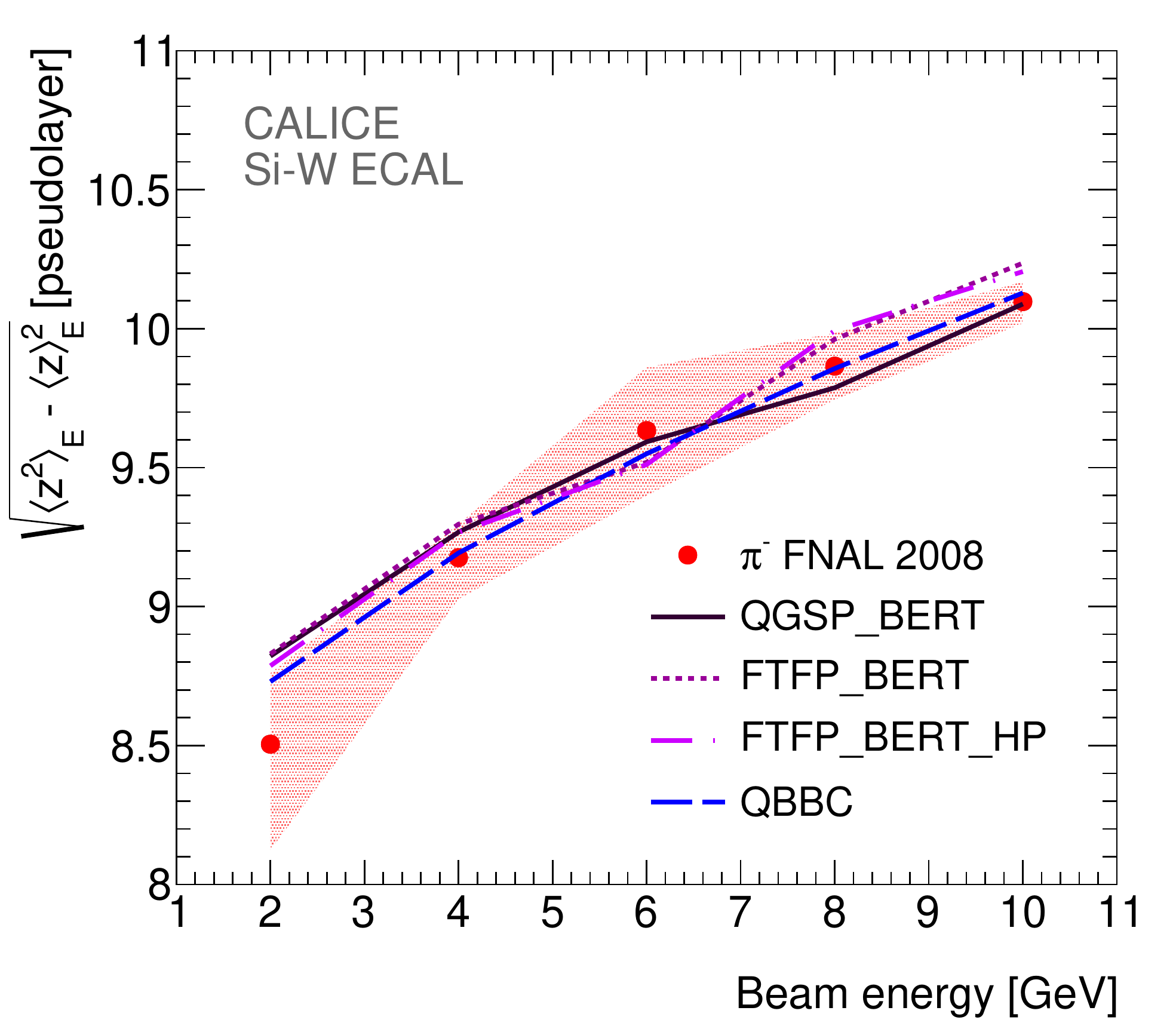}}
    \caption{\sl Mean (a) and standard deviation (b) of the longitudinal energy profile for interacting events as a function of beam energy (2\,GeV to 10\,GeV) for data and various Monte Carlo physics lists.}
    \label{figure:meansigmalongitudinalprofile}
  }
\end{figure}

Figure~\ref{figure:meansigmalongitudinalprofile} shows the mean, $\langle z \rangle_{E}$, and standard deviation, $\sqrt{\langle z^2 \rangle_{E} - \langle z \rangle^2_{E}}$, of the longitudinal profiles for all four physics lists compared to the data. 
The mean is underestimated at higher energies which supports the observation of too much deposited energy near the interaction layer.
The standard deviation is compatible with the data within the uncertainties only for {\sc qbbc}.
The difference between the physics lists is maximally 4\%.

\begin{figure}[h]
  {\centering
    \subfloat[]{\includegraphics[width=0.5\textwidth]{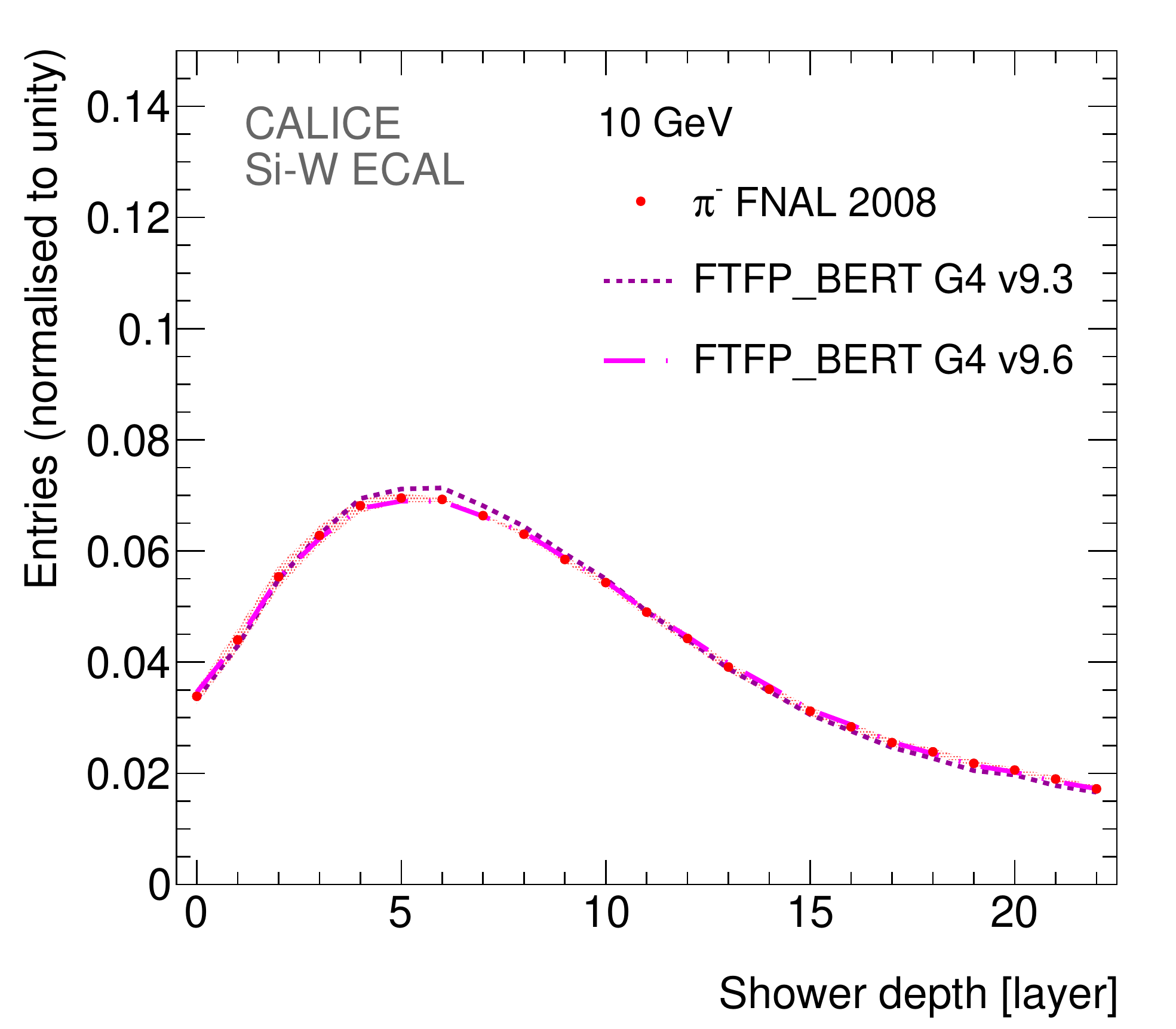}}
    \subfloat[]{\includegraphics[width=0.5\textwidth]{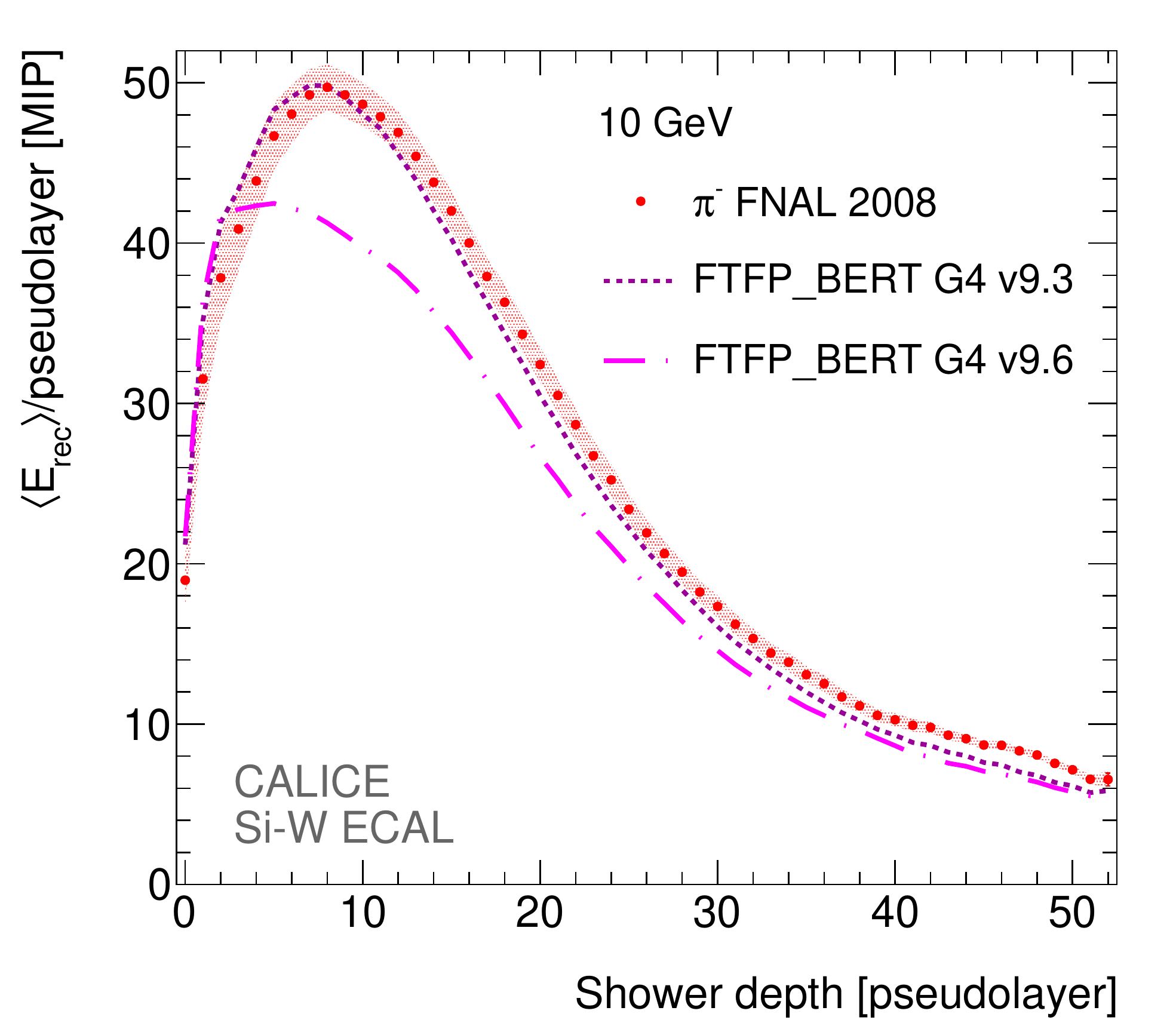}}
    \caption{\sl The longitudinal hit distribution (a) and energy profile (b) for interacting events at 10\,GeV, for data and the Monte Carlo physics list {\sc ftfp\_bert} for two different {\sc Geant4} versions.}
    \label{figure:longitudinalprofilecompared}
  }
\end{figure}

The hadronic models implemented in {\sc Geant4} are constantly being revised and improved to best describe the available data.
The analysis presented in this paper initially used {\sc Geant4} version 9.3~\cite{2012_Doublet} and was later updated to version 9.6.
Between these two versions the Fritiof string model has been significantly revised and tuned based on thin target data and LHC test beam data, while the Bertini cascade model has undergone only minor revisions.
The changes in the Fritiof model have led to a larger mismatch between the data and the physics list {\sc ftfp\_bert} in the longitudinal energy profile, as is illustrated in Fig.~\ref{figure:longitudinalprofilecompared}. 
{\sc ftfp\_bert} in version 9.3 describes the data reasonably well at 10\,GeV, while in version 9.6 it clearly does not.
On the other hand, the longitudinal hit distribution is well modelled and, while the change between the versions is small, the description is better in version 9.6.
For {\sc qgsp\_bert} such a change in the longitudinal energy profile is not seen and in both versions the energy is underestimated.
This kind of discrepancy has not been observed in other detector configurations; in a recent CALICE publication~\cite{2014_CALICE} the longitudinal energy profile of $\pi^-$s in a scintillator-tungsten hadronic calorimeter prototype is well described by {\sc ftfp\_bert} in {\sc Geant4} version 9.6.
The observed discrepancy with the Si-W ECAL data could be related to the sensitive material of the prototype, silicon for the Si-W ECAL, as the optimisation of the Fritiof model has been mostly done with data obtained from detectors with scintillator as sensitive material.
Recently a bug has been identified in the implementation of the Fritiof String model, which could be responsible for the discrepancy\footnote{Geant4 10.1-beta-01 Release Notes: http://geant4.web.cern.ch/geant4/support/Beta4.10.1-1.txt}. 
Corrections are being implemented in the next release of {\sc Geant4} ({\sc Geant4} 10.1). 
This possible origin of the energy discrepancy will be verified or excluded once this release is available in the CALICE analysis software. 

\section{Summary, Conclusions and Outlook}

This study demonstrates the large potential of the CALICE Si-W ECAL to obtain a detailed image of the early part of hadronic cascades. 
Data obtained in test beams with negatively charged pions ($\pi^-$) with an energy between 2 and 10\,GeV are compared to Monte Carlo predictions employing different physics lists of the {\sc Geant4} simulation tool kit.

If a hadronic interaction takes place within the Si-W ECAL volume, the start of the shower can be reconstructed with an accuracy of $\pm$ 2 layers at an efficiency of at least 50\% at 2\,GeV and 87\% at 10\,GeV.
This interaction finding efficiency is found from simulated events.
At the low beam energies studied here interactions are selected using not only the absolute energy increase in subsequent layers but also the relative energy increase.

The accuracy with which the Monte Carlo describes the data varies with the beam energy and the chosen physics observable.
None of the physics lists describe the entire set of data, but overall the Monte Carlo are within 20\% of the data and for most observables much closer.
The longitudinal hit distribution is very well described, while the mean is shifted for the radial hit distribution.
On the other hand the physics observables which take into account the energy deposition are not reproduced well by the Monte Carlo.
The reconstructed energy is too low due to a lower number of hits.
Combining the longitudinal and radial energy profiles it seems that especially the Fritiof model deposits too much energy near the interaction region. 

The radial distributions prove to be sensitive to the different hadronic models implemented in the physics lists.
The transition between the Bertini cascade and Fritiof string model in {\sc ftfp\_bert} and {\sc ftfp\_bert\_hp} is much more pronounced 
in the mean and standard deviation of the radial observables than the longitudinal observables.
Additionally the deviations of the physics lists from the data and each other are larger.
The precision treatment of neutrons in {\sc ftfp\_bert\_hp} gives smaller mean and standard deviations.
The results for {\sc qbbc} tend to be between {\sc qgsp\_bert} and {\sc ftfp\_bert}, as expected.

In conclusion, no preference for a hadronic model is seen as none of the physics lists reliably reproduce the data in detail.
The main deficiencies are in the longitudinal and radial energy profiles.
The observables that are well described show 3 -- 7\% difference between physics lists. 
The level of agreement between the data and simulations depends also on the version of {\sc Geant4}. 

Future analysis into hadronic showers will attempt to classify inelastic reactions in terms of shower topology. 
This comprises the determination of size and energy density of the interaction region as well as the measurements of tracks emerging from the interaction region. 
These steps will further exploit the lateral granularity of the Si-W ECAL which will be even higher, $5\times5$ mm$^2$, in the baseline design for the International Large Detector (ILD) at the ILC.  
They may form a solid base for the development and improvement of particle flow algorithms.

\section{Acknowledgements}

We gratefully acknowledge the DESY, CERN and FNAL managements for their support and hospitality, 
and their accelerator staff for the reliable and efficient beam operation.
The authors would like to thank the RIMST (Zelenograd) group for their help and sensors manufacturing.
This work was supported
by the P2IO LabEx (ANR-10-LABX-0038) in the framework `Investissements d'Avenir' (ANR-11-IDEX-0003-01) managed by the French National Research Agency (ANR);
by the Quarks and Leptons Programme of CNRS/IN2P3 France;
by the Bundesministerium f\"{u}r Bildung und Forschung, Germany;
by the  the DFG cluster of excellence `Origin and Structure of the Universe' of Germany ; 
by the Helmholtz-Nachwuchsgruppen grant VH-NG-206;
by the BMBF, grant no. 05HS6VH1;
by the Alexander von Humboldt Foundation (including Research Award IV, RUS1066839 GSA);
by joint Helmholtz Foundation and RFBR grant HRJRG-002, SC Rosatom;
by the Russian Ministry of Education and Science contracts 4465.2014.2 and 14.A12.31.0006 and the Russian
Foundation for Basic Research grant 14-02-00873A;
by MICINN and CPAN, Spain;
by CRI(MST) of MOST/KOSEF in Korea;
by the US Department of Energy and the US National Science Foundation;
by the Ministry of Education, Youth and Sports of the Czech Republic under the projects AV0 Z3407391, AV0 Z10100502, LC527  and LA09042  and by the Grant Agency of the Czech Republic under the project 202/05/0653;  
and by the Science and Technology Facilities Council, UK.

\bibliographystyle{elsarticle-num.bst}
\bibliography{paper}

\end{document}